\documentclass[12pt,showpacs,superscriptaddress,preprint,unsortedaddress]{revtex4}
\usepackage{graphicx}
\usepackage{comment}
\usepackage{amstext}
\usepackage{amsmath}
\usepackage{amsthm}
\usepackage{amssymb}
\usepackage{color}
\DeclareMathOperator{\imag}{Im}

\newtheorem{thr}{\textbf{Theorem}}[section]

\renewcommand{\eqref}[1]{(\ref{eq:#1})}
\newcommand{\figref}[1]{Fig.~\ref{fig:#1}}
\newcommand{\appref}[1]{Appendix~\ref{app:#1}}
\newcommand{\thrref}[1]{Theorem~\ref{thm:#1}}
\newcommand{\secref}[1]{Sec.~\ref{sec:#1}}

\newcommand{\tabref}[1]{Tab.~\ref{tab:#1}}

\begin{document}

\title{Extreme Value Statistics of the Total Energy
  in an Intermediate Complexity Model of the Mid-latitude Atmospheric Jet.\\
  Part I: Stationary case.}

\author{Mara Felici}
\email{mara.felici@math.unifi.it}
\homepage{www.unicam.it/matinf/pasef} \affiliation{
  PASEF -- Physics and Applied Statistics of Earth Fluids,
  Dipartimento di Matematica ed Informatica,
  Universit\`{a} di Camerino,
  Via Madonna delle Carceri, 62032 Camerino (MC), Italy}
\affiliation{
  Dipartimento di Matematica U. Dini,
  Universit\`a di Firenze,
  viale Morgagni 67/A --  50134 Firenze, Italy}
\author{Valerio Lucarini}
\author{Antonio Speranza}
\author{Renato Vitolo}
\affiliation{
  PASEF -- Physics and Applied Statistics of Earth Fluids,
  Dipartimento di Matematica ed Informatica,
  Universit\`{a} di Camerino,
  Via Madonna delle Carceri, 62032 Camerino (MC), Italy}


\begin{abstract}
  An intermediate complexity baroclinic model for the atmospheric
  jet at middle-latitudes is used as a stochastic generator
  of earth-like time series: in the present case
  the total energy of the system.
  Statistical inference of extreme values is applied to
  yearly maxima sequences of the time series,
  in the rigorous setting provided by extreme value theory.
  In particular, the Generalized Extreme Value (GEV) family
  of distributions is used here as a fundamental model
  for its simplicity and generality.
  Several physically realistic values of the parameter $T_E$,
  descriptive of the forced equator-to-pole temperature gradient and
  responsible for setting the average baroclinicity in the
  atmospheric model, are examined.
  Stationary time series of the total energy are generated and
  the estimates of the three GEV parameters -- location, scale and shape --
  are inferred by maximum likelihood methods.
  Standard statistical diagnostics, such as return level and
  quantile-quantile plots, are systematically
  applied to asses goodness-of-fit.
  The location and scale GEV parameters are found to have
  a piecewise smooth, monotonically increasing dependence on $T_E$.
  This is in agreement with the similar dependence on $T_E$ observed
  in the same system when other dynamically and physically relevant
  observables are considered. The shape parameter also increases
  with $T_E$ but is always negative, as \textit{a priori} required
  by the boundedness of the total energy of the system.
  The sensitivity of the statistical inference process is studied
  with respect to the selection procedure of the maxima:
  the roles of both the length of maxima sequences and of the length of
  data blocks over which the maxima are computed are critically analyzed.
  Issues related to model sensitivity are also explored by
  varying the resolution of the system.
\end{abstract}

\maketitle

\tableofcontents

\clearpage
\section{Introduction}

The study of climatic extreme events is of paramount importance
for society, particularly in the fields of engineering and
environmental and territorial planning. Indeed, temporal
variations in the statistics of extreme events may have more acute
and disruptive effects than changes in the mean
climate~\cite{KB92}. In works of economical nature (see \emph{e.g.}
Nordhaus~\cite{Nor94}), the special role played by the extreme
events in terms of impacts is included with the hypothesis that
the costs associated with climatic change can be represented as
strongly nonlinear functions of the observed variations in surface
temperature. This constitutes a clear motivation for which, when
the impacts of climatic change are analyzed, the interest for
variations in the statistics of extreme events plays a strategic
role \cite{IPCC01,Luc02}.

In the scientific literature, some recent papers in which the
existence of trends in the frequency of extreme (precipitation)
events was pushed forward in quantitative terms are those by Karl
\emph{et al.}~\cite{KKEQ96, KK98}. Here the authors stated that
the \textit{percentage of the U.S.A. with a much above normal
proportion of total annual precipitation from extreme
precipitation events (daily events at or above 2 inches)} showed
an increase from 9\% in 1910-1920 to about 11\% in the '90s.
Despite severe scientific criticism to these papers
by many other researchers in the field, the basic idea that the
frequency of extreme events may change together with average
surface temperature was discussed more and more and, eventually,
it became one of the issues of analysis for the Intergovernmental
Panel for Climate Change: a specific a specific report on
\textit{Changes in extreme weather and climate events} was issued
in 2002 ~\cite{IPCC02}. Basic questions, when dealing with
extremes of complex processes, is: what is the correct way of
measuring extremes? Are we concentrating on \emph{local} or \emph{global}
fluctuations of the system in question? How do we measure local extremes?
Extremes of wind speeds, of rainfall amounts, of economical damage?
Moreover, the enhancement in the extreme events might be quantified
either in terms of number of events, or in size of the average extreme event,
or a combination thereof. Several other ambiguities make it often
difficult to follow literature on the subject.

Overall, two important weaknesses of much work on the subject of
extreme meteo-climatic events and of their trends are:
\begin{itemize}
\item
  the lack of interpretation of the dynamical mechanisms
  that should cause the hypothesized changes in the frequency
  of extremes of various nature;
  often such mechanisms are just alluded to instead of being explicitely
  formulated and analyzed;
\item
  the lack of a common and theoretically founded definition of ``extremes''.
\end{itemize}
The deficit in the first point above may negatively affect both
deterministic and statistical studies of the phenomena in
question. One major example on global processes is that, despite
the great attention attracted by the subject, very few researchers
have investigated in detail the basic mechanisms that should
associate an increased $CO_2$ concentration to enhanced extreme
weather events.
The chain of mechanisms possibly linking $CO_2$ concentration and
weather extremes is too long even for an adequate qualitative
discussion here, but we shall concentrate on the basic sequence:
enhanced surface temperature $\longrightarrow$ enhanced
baroclinicity $\longrightarrow$ changes in the upper tail of the
probability distribution function of the baroclinic disturbances.
But no robust analysis of this complex dynamical ``chain'' has
been offered so far.

As for the second point above, the lack of a common rigorous
framework for the statistical analysis of extremes (with few
exceptions such as \emph{e.g.}~\cite{KPN02,ZK98,ZK00}) provides a
serious drawback for the interpretation and comparison of results
from different studies. Moreover, this problem is not even
justified, since mathematical theories of extreme events are
well-developed~\cite{Cas88,Col01,EKM97,FT28,Gal78,Gne43,LLR83}
and the derived methods are quite successful
in many applications~\cite{KPN02,PRT05,ZK98,ZK00}.
One basic ingredient of the theory relies on Gnedenko's
theorem~\cite{Gne43}, which states that, under fairly mild assumptions,
the distribution of the block-maxima of a sample
of independent identically distributed variables converges to a
family of three distinct distributions, the so-called Generalized
Extreme Value (GEV) distributions.
See \appref{theory} for a brief description.
Notice that one of the earliest applications of this theory
in the natural sciences occurred specifically in a meteorologic-climatic
setting~\cite{Jen55}.
Other statistical models for extreme events include the $r$-largest
statistics, threshold exceedance models such as the generalized
Pareto distribution, and point processes, see~\cite{Col01}.

The reliability of parametric estimates for extreme value models
strongly depends on the asymptotic nature of extreme value theory.
In particular, at least the following issues should be checked or
addressed~\cite{Col01}:
  \begin{enumerate}
  \item
    \emph{independence} of the selected extreme values;
  \item
    using a \emph{sufficiently large} number of extremes;
  \item
    using values that are \emph{genuinely} extreme.
  \end{enumerate}
Despite the importance of the third requirement, many studies
actually deal with so-called \emph{soft extremes}~\cite{KtK03},
which are maxima of too short data blocks or with too small return
periods for the basic assumptions of the theory to hold.
This is often the consequence of the limited amount of available data:
on one hand, one has to restrict to maxima of data blocks,
thereby discarding \emph{most} available data;
on the other hand one would like to have a \emph{long}
sequence of extreme values.
The net result is that the assumptions of the extreme value theorems
often go unchecked and are sometimes plainly impossible to check,
since the available climatic records cover at best the last century.
Therefore, thinking in terms of annual maxima, in such cases we
only have 100 extremes. The inevitable consequence of adapting the
definition of extremes to the needs of the work is a serious
reduction of reliability of the resulting estimates.

The goal of this paper is to infer and critically quality-check
the statistical description of extreme values in the GEV
distributions framework on the ``earth-like'' time series
produced by a dynamical system descriptive of the mid-latitude
atmospheric circulation featuring a chaotic regime.
Such system has internally generated noise and can be effectively
considered as a \textit{stochastic generator} of data.
Time series of the system's total energy $E(t)$ are used,
which is a relevant global physical quantity.
We analyze how the GEV distribution inferred from block
maxima of $E(t)$ depends on the value of the most important
parameter of the system, namely the forced equator-to-pole
temperature difference $T_E$, which controls the baroclinicity of
the model.
The reliability of the GEV fits is studied, by
considering both shorter sequences of extremes and \emph{soft extremes}.
Moreover, issues related to model error and sensitivity are
briefly examined by analyzing the effects of variations
in model resolution.
The use of numerically generated data allows us to avoid all the
difficulties related to shortness of the available climatic
records, such as missing observations and low-quality data.
In particular, we do not need to worry about the wastage
of data caused by the selection of annual maxima,
which is a serious limitation when considering observational data.
In such methodological sense, our approach is similar to that
of~\cite{ZZL03} as far as statistical inference is concerned.
However, an important difference is that the statistics
of the time series $E(t)$ generated by the atmospheric model
\emph{cannot be directly chosen}: there is no explicit formula
relating the probability density function of the adopted
observable and the parameter $T_E$. This problem we analyze
elsewhere~\cite{LSV05}.

The structure of the paper is now outlined.
In \secref{setup} we first describe the set-up
of the numerical experiments performed with the atmospheric model
and then the methods of statistical analysis of extreme values
adopted for the total energy time series.
The results for the considered reference case of 1000 yearly
maxima are presented in \secref{GEVinference}. Assessment of the
sensitivity of the inferences is studied in \secref{GEVsensitivity},
by varying the length of yearly maxima sequences, the block length
over which maxima are taken, and the model resolution. The
dependence of the GEV parameters with respect to $T_E$ is also
analyzed in this section. \secref{conclusions} summarizes the
results and their relation with the above discussion.
The theory and the methods of Extreme Value distributions,
as far as needed in the present work, are briefly reviewed in \appref{theory}.
The model of the baroclinic jet used as a stochastic generator is
described in \appref{model}, referring to~\cite{LSV05} for a
thorough discussion.

\newpage
\section{Data and Methods}
\label{sec:setup}

\subsection{Total Energy of the Atmospheric Model}
\label{sec:data}

We consider a quasi-geostrophic intermediate complexity
model~\cite{SM88,MTS90,LSV05} (also see \appref{model}),
providing a basic re\-pre\-sen\-tation of the turbulent jet
and of the baroclinic conversion and barotropic stabilization
processes which characterize the physics of the mid-latitudes
atmospheric circulation.
The model is relaxed towards a given equator-to-pole
temperature profile which acts as baroclinic forcing. It features
several degrees of freedom in the latitudinal direction and two
layers in the vertical - the minimum for baroclinic conversion to
take place~\cite{Ped87,Phi54}. The system's statistical properties
radically change when the parameter $T_E$, determining the
forced equator-to-pole temperature gradient, is changed.
In particular, as $T_E$ increases a transition occurs from a
stationary to an earth-like chaotic regime with internally
generated noise. By chaotic, we mean that the system possesses a
strange attractor in phase space~\cite{ER}.
For a detailed description of the model physics and dynamics
see~\cite{LSV05}.

In the present setting, the model is used as a stochastic generator
of earth-like time series for testing the reliability of different
statistical approaches~\cite{ZZL03} and studying the dynamics
of extremes~\cite{LSV05}.
A uniformly spaced grid of 21 values of the parameter
$T_E$ is fixed in the range $[10,50]$, starting from 10 and
increasing with step 2. The baroclinic model is run for $T_E$
fixed at each of these values, producing 21 simulations of length
1000 years (preceded by an initial transient of five years) where
the total energy $E(t)$ is written every 6 hours.
The formula of the total energy is given in \appref{model},
equation~\eqref{energydensity}.
We recall that, in the non-dimensionalization of the system,
$T_E=1$ corresponds to $3.5\,K$, 1 unit of total energy corresponds
to roughly $5\times10^{17} J$, and $t=0.864$ is one day,
see~\cite{LSV05} for details.

\begin{figure}[p]
  \centering
  \includegraphics[width=0.32\textwidth]{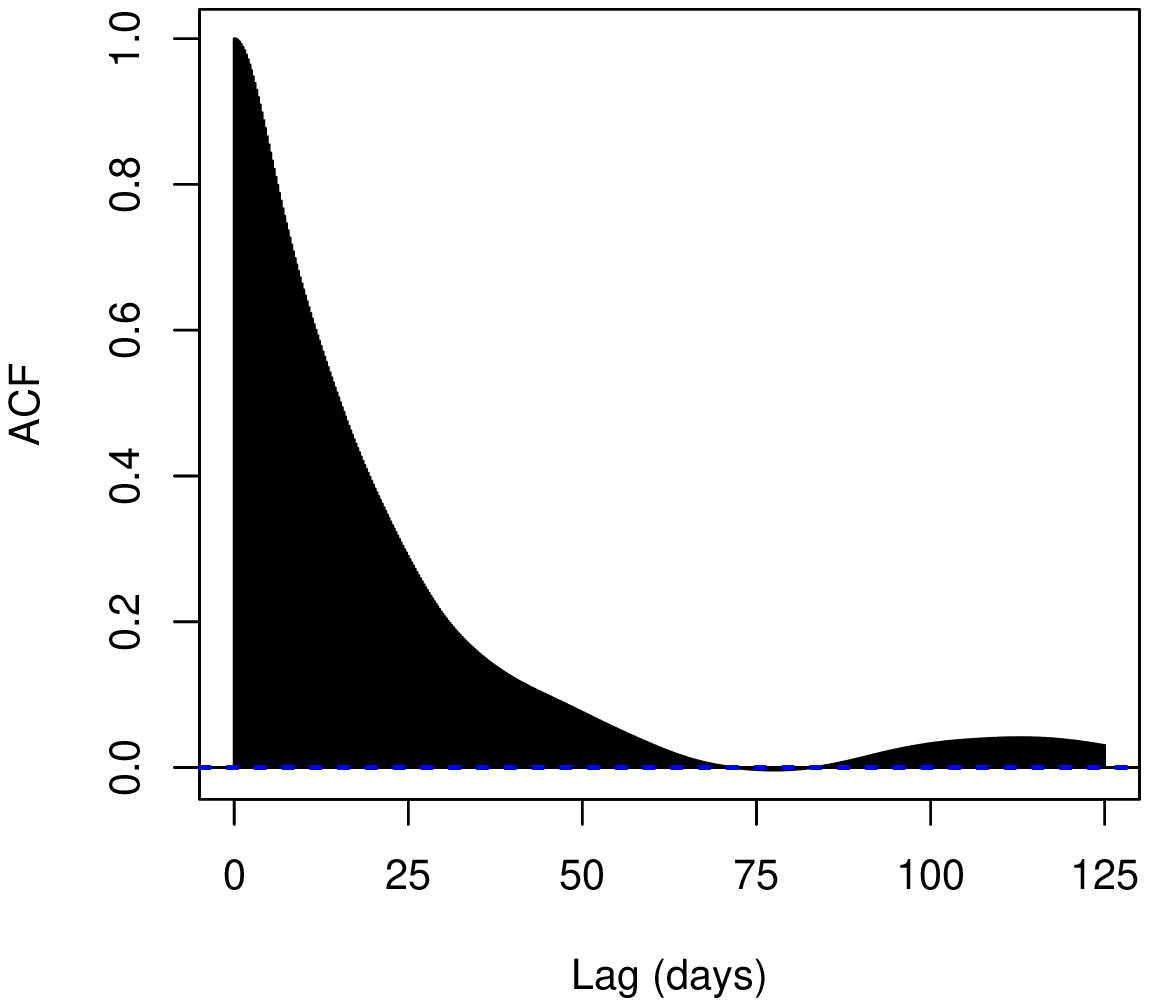}
  \includegraphics[width=0.32\textwidth]{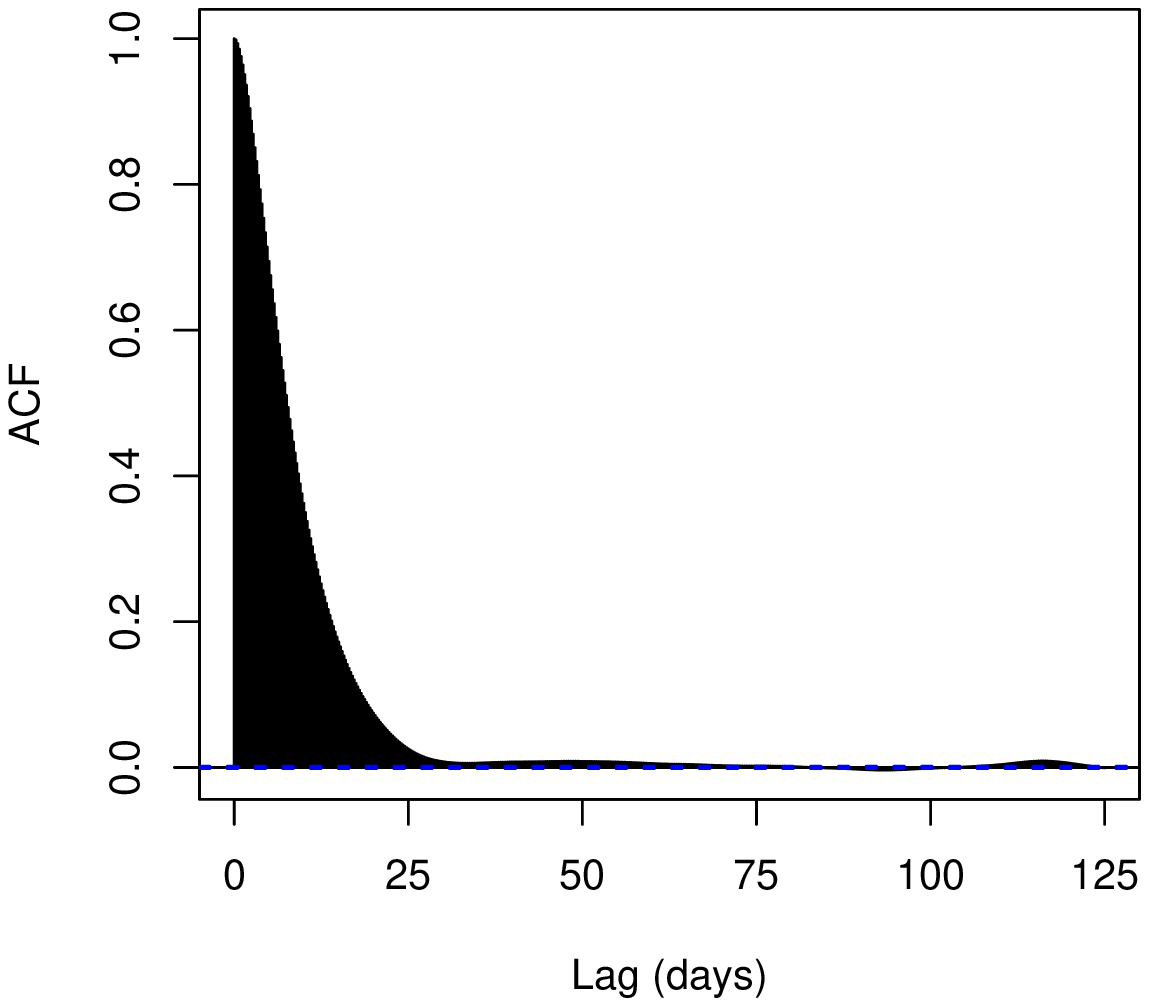}
  \includegraphics[width=0.32\textwidth]{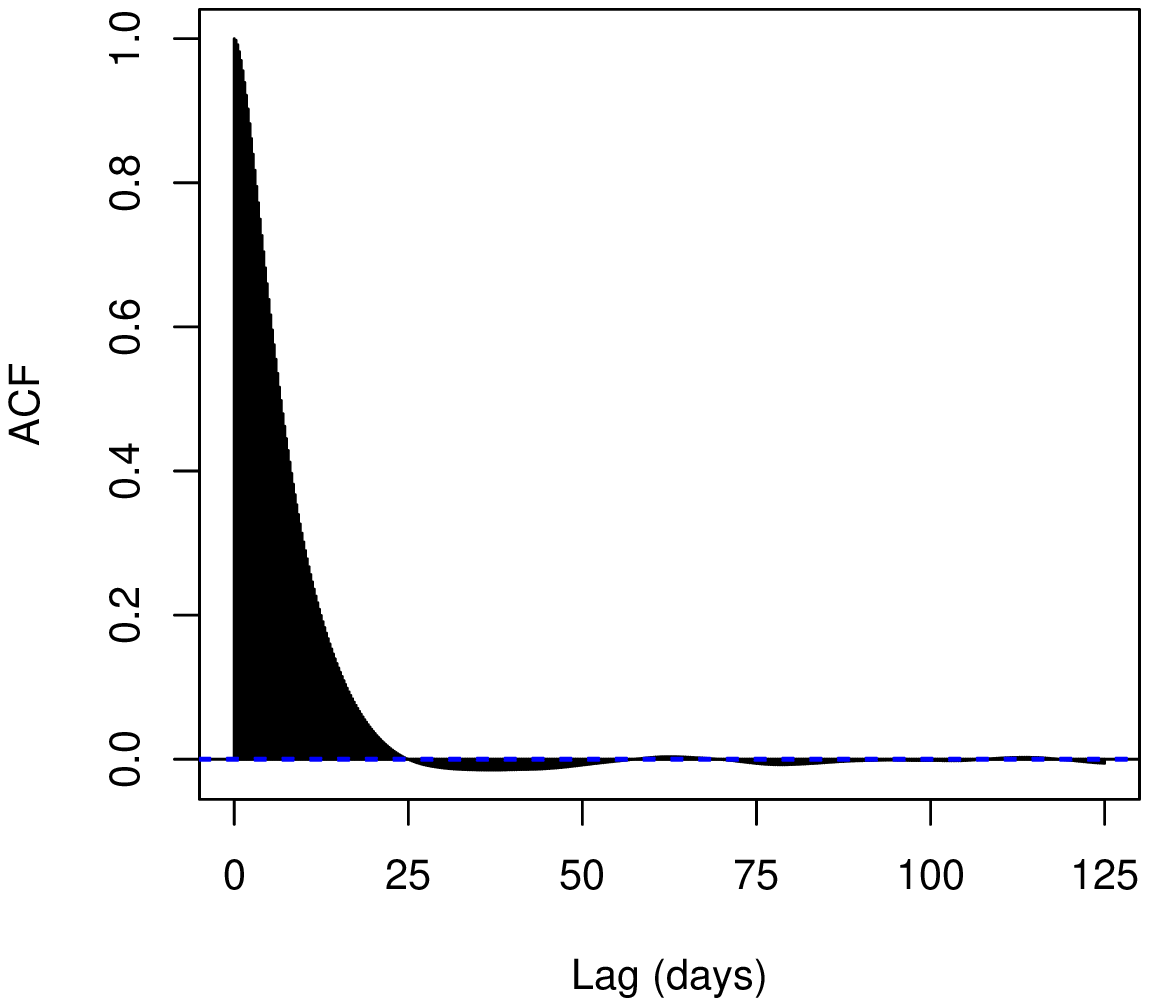}
  \caption{
    Autocorrelations of the total energy time series for
    $T_E=10,30,50$ (left, center, right, respectively),
    time-lag in days on the horizontal axis.
    The full 6-hourly time-series of 1000 years have been
    used, see \secref{data}.
  }
  \label{fig:acf}
\end{figure}

For each of the selected values of $T_E$, a chaotic attractor
is numerically detected in the phase space of the model.
This is illustrated by the autocorrelations of the time
series of the total energy $E(t)$ (\figref{acf}),
which decay to zero on a time scale that is comparable with that
of the atmospheric system (roughly 10-15 days~\cite{Lor67}).
Since all parameters of the model are kept fixed in each simulation,
by discarding the initial transient, the time series of $E(t)$
may be considered a realization of a stationary stochastic process.

\begin{figure}[p]
  \centering
  \includegraphics[width=0.32\textwidth]{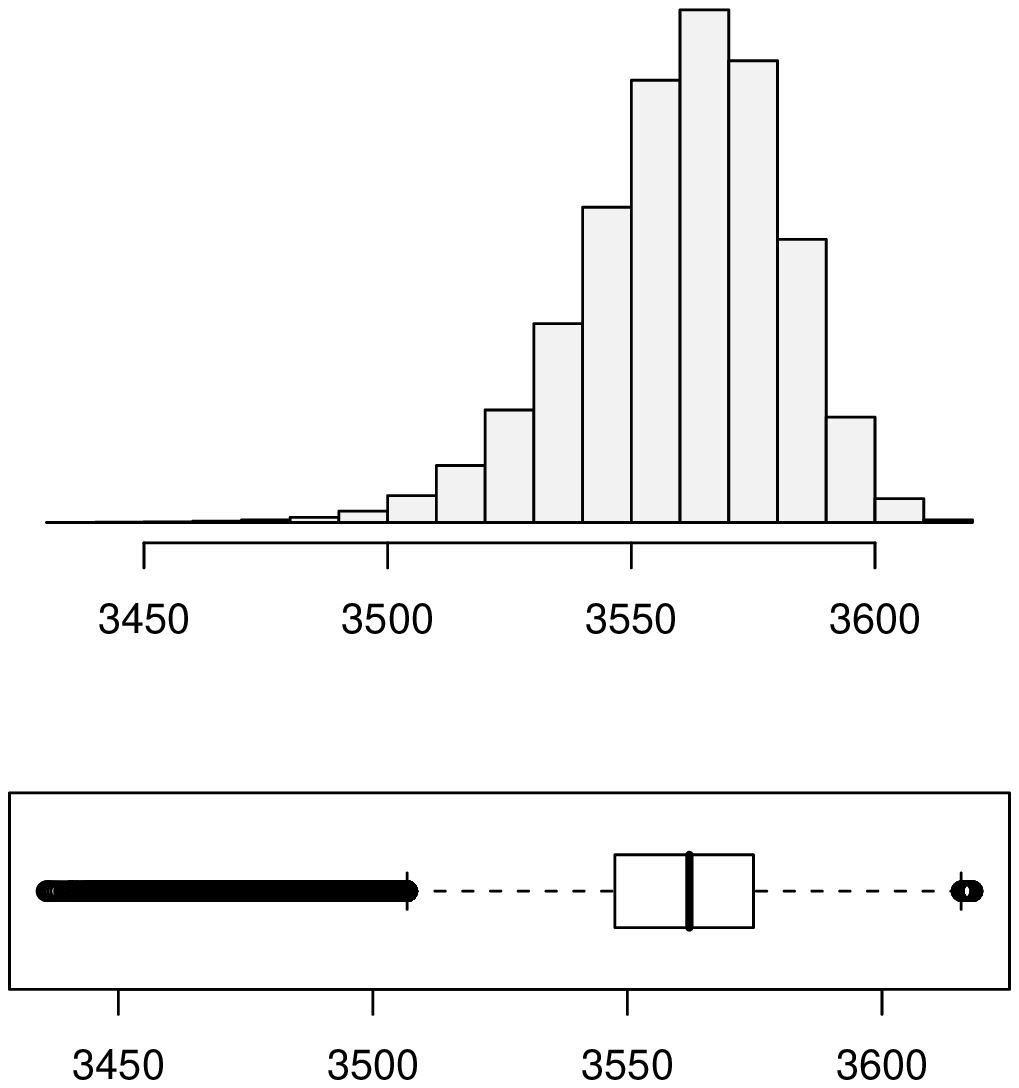}
  \includegraphics[width=0.32\textwidth]{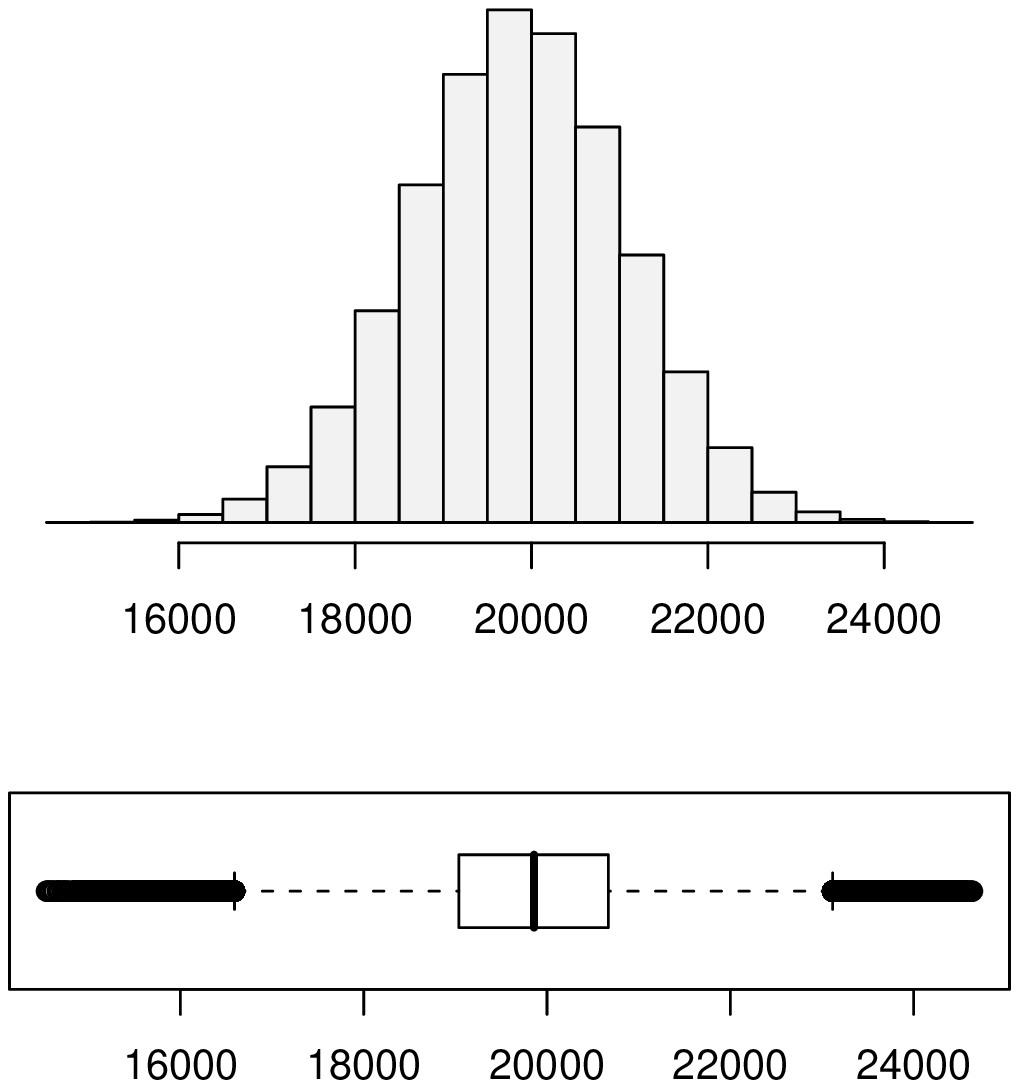}
  \includegraphics[width=0.32\textwidth]{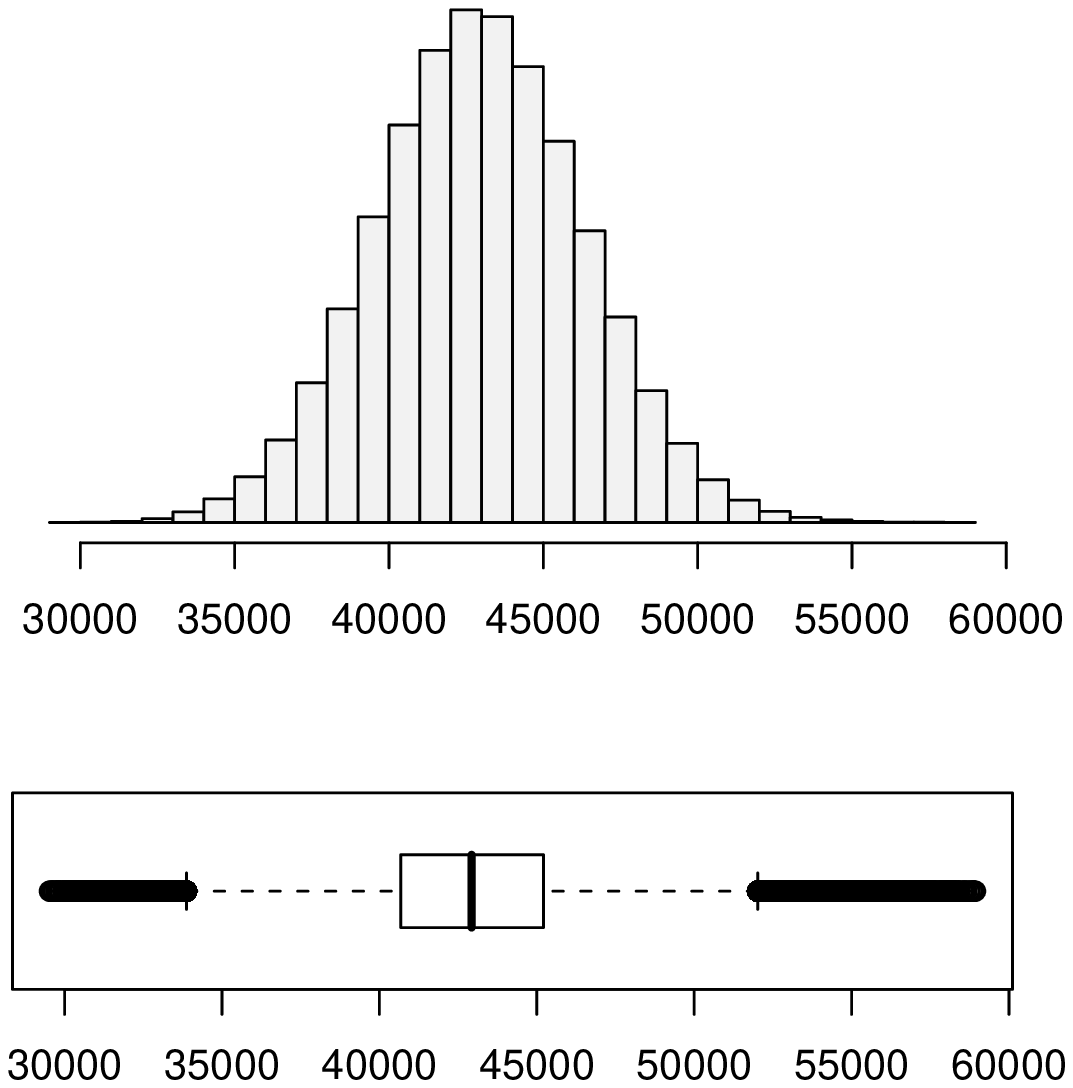}
  \caption{
    Histograms and boxplots of the total energy time series
    for $T_E=10,30,50$ (left, center, right, respectively).
  }
  \label{fig:histbox}
\end{figure}

\begin{figure}[p]
  \centering
  \includegraphics[width=0.3\textwidth]{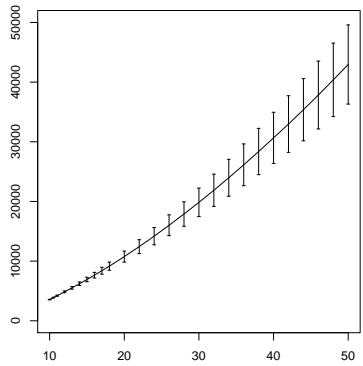}
  \caption{
    Time-averaged total energy (vertical axis)
    as a function of $T_E$ (horizontally),
    for each of the 21 selected values of $T_E$.
    Confidence bands (average plus or minus a 1.96 times sample
    standard deviation) are added.
    The full 6-hourly time-series of 1000 years have been
    used, see \secref{data}.
  }
  \label{fig:e32}
\end{figure}

The distribution of the total energy time series is visualized
by means of the histograms and boxplots in \figref{histbox},
for three values of $T_E$. Notice that, as $T_E$ increases,
\begin{itemize}
\item
  the upper tail of the distribution becomes heavier,
  whereas the lower tail shortens;
\item
  both the average value and the variability of the
  total energy time series increase.
\end{itemize}
The latter point is clearly visualized in \figref{e32}, where the
time-averaged total energy is displayed for each of the 21
stationary time series, together with confidence intervals.
Throughout the paper, confidence intervals are computed as
average plus/minus sample standard deviation multiplied by $1.96$.

In concluding this section a theoretical remark is in order here.
All examined strange attractors are implicitly \emph{assumed}
to possess a unique Sinai-Ruelle-Bowen (SRB) ergodic invariant
measure~\cite{ER}.
This is indeed a rather general and difficult problem in Dynamical
Systems and Physics: on the one hand,
existence of a unique SRB measure is \emph{necessary}
to rigorously associate a stationary stochastic process
with the dynamical evolution law.
On the other hand, existence of a unique SRB measure is a very strong
regularity assumption for a dynamical system: in general
it is even the question whether invariant measures exist at all and,
if so, whether a finite or infinite number of invariant measures
coexist for a given chaotic system.
Moreover, even if an SRB measure exists and is unique,
it is in general non-parametric: there is \emph{no explicit formula}
relating the statistical behavior to the system's
equation and parameters.

\subsection{Parameter Estimation and Model Assessment in GEV Inference}
\label{sec:GEVmethods}

As discussed in the previous section, the time series we work with
are characterized by fast decay of autocorrelations,
which implies weak (short time-range)
dependence of the observations, compare \figref{acf}.
Inference of threshold exceedance models~\cite{Col01,EKM97,LLR83}
is in this case complicated by the choices of suitable
threshold values and cluster size for \emph{declustering}
(see \emph{e.g.}~\cite[Chap. 5]{Col01}), which might be
somewhat arbitrary in the applications.
On the other hand, since the dependence is short-range,
maxima of the total energy time series, taken over sufficiently large
data blocks, are with good approximation independent.
This is why we have preferred the use of the GEV
with respect to threshold models.
Moreover, since we can generate time series of arbitrary length,
for simplicity we refrained from using the $r$-largest statistics,
which is often valid alternative to the GEV,
especially when data scarcity is an issue.
In this section, therefore, we recall the methods of GEV inference
as far as needed in the present work.
The exposition is largely based on~\cite{Col01}.
Also see~\cite{Cas88,Col01,EKM97,FT28,Gal78,Gne43,LLR83}
for methodology and terminology of extreme value theory.

Gnedenko's theorem~\cite{Gne43}, or the \emph{three types theorem},
first presented in a slightly less general form by
Fisher and Tippet~\cite{FT28}) states that, under fairly mild
assumptions, the distribution of the block-maxima of a sample of
independent identically distributed variables converges, in a
suitable limit, to one of three types of extreme value
distributions. The three types are in fact special cases of the
GEV distribution (also called von Mises type), having the
following expression:
\begin{equation}
  \label{eq:GEV}
  G(x)= \exp \left\{ - \left[
      1+ \xi\left( \frac{x-\mu}{\sigma} \right) \right]^{-1/ \xi}\right\}
\end{equation}
for $x$ in the set $\{ x:\, 1+\xi(x-\mu)/\sigma \,>\,0\}$ and
$G(x)=0$ otherwise, with $-\infty < \mu < +\infty$, $\sigma >0,$
and $-\infty < \xi < +\infty$. The quantities $(\mu,\sigma,\xi)$
are called location, scale and shape parameter, respectively. In
such a framework, statistical inference of extreme values amounts to
estimating the GEV distributional parameters $(\mu,\sigma,\xi)$ for
a given time series and assessing the quality of the fit. If
$\xi>0$ ($\xi<0$) the distribution is usually referred to as
Fr\'{e}chet (Weibull) distribution, if $\xi=0$ we have the Gumbel
distribution, which can be expressed as \eqref{gumbel}.
See \appref{theory} for further theoretical details
and~\cite{EKM97,Cas88,Col01,Gal78,LLR83} for examples and discussion.

In practical application of the extreme value theory the parent
distribution function of the data is typically unknown. Therefore,
both the type of limiting distribution and the parameter values
must be inferred from the available data and the quality of the
resulting estimates should always be assessed.
For GEV inference, a sequence of maxima is constructed by
subdividing the available data $\{x_i\}$ into blocks of equal
length and extracting the maximum from each block. The block
length is one of the choices playing the usual, critical role
between bias and variance in the parametric estimates. Too short
blocks increase the length of the maxima sequence but, at the same
time, they increase the risk of failure of the limit~\eqref{hypG}.
If the blocks are too long, the resulting scarcity of maxima induces
an enhanced uncertainty of the inferred values of the GEV parameters.
In many situations a reasonable (and sometimes compulsory)
choice is to consider the annual maxima (see~\cite{Col01}).

Assume that the observations in the time series are equispaced
in time and that none of them is missing (both conditions are often
violated in concrete cases, see \emph{e.g.}~\cite{PRT05}).
Let $n$ be the number of observations in a year and
denote by $M_{n,1},\ldots,M_{n,m}$ the sequence of the annual maxima,
\emph{i.e.}, the maxima over data blocks of length $n$.
Under the assumption of independence of the $X_n$, the variables
$M_{n,1},\ldots,M_{n,m}$ are independent as well.
In fact, approximate independence of the $M_{n,i}$ holds also
in the case of \emph{weak dependent} stationary sequences,
see~\cite{LLR83,Col01} for definitions and examples.

Among the numerous methods to infer the GEV parameters (graphical
or moment-based techniques, see~\cite{Cas88}), we adopt the
maximum likelihood estimator for its great adaptability to changes
of models. Denote by $\boldsymbol{\theta}=(\mu,\sigma,\xi)$ the
parameter vector for the GEV density $g(x;\boldsymbol{\theta})$,
the latter being the derivative of $G(x)=G(x;\boldsymbol{\theta})$
in~\eqref{GEV}.
In the stationary context, the \emph{block maxima} of the
observed data are assumed to be realizations of a stationary
stochastic process having density $g(x;\boldsymbol{\theta}^0)$,
where $\boldsymbol{\theta}^0$ is the unknown parameter vector. The
\emph{maximum likelihood estimator}
$\widehat{\boldsymbol{\theta}^0}$ of $\boldsymbol{\theta}^0$ is
defined as the value that maximizes the likelihood function
\begin{equation}
  \label{eq:likelihoodf}
  L(\boldsymbol{\theta})=\prod_{i=1}^n g(M_{n,i};\boldsymbol{\theta}).
\end{equation}
In loose words, maximizing $L(\boldsymbol{\theta})$ yields the
parameter values for which the probability of observing the
available data is the highest. It is often more advantageous to
maximize the \emph{log-likelihood function}
\begin{equation}
  \label{eq:logllGEVdef}
  l(\boldsymbol{\theta})  =  \log \; L(\boldsymbol{\theta}) =
  \sum_{i=1}^m \log \, g(M_{n,i};\boldsymbol{\theta})
\end{equation}
and, according to \eqref{GEV},we get
\begin{equation}
  \label{eq:logllGEV}
  \begin{split}
    &l(\mu,\sigma,\xi)=\\
    &=\begin{cases}
      -m\,\log \, \sigma -\left(1+ \frac{1}{\xi}\right)
      \sum_{i=1}^m \left \{
        \log \, \left[ 1+ \xi \left (\frac{M_{n,i}-\mu}{\sigma} \right) \right]
        - \left[  1+ \xi \left (\frac{M_{n,i}-\mu}{\sigma} \right)
        \right]^{-\frac{1}{\xi}}
      \right \},
      & \text{if\quad$\xi \neq 0$,}\\
      -m\,\log \, \sigma\, -\, \sum_{i=1}^m \left \{
        \left (\frac{M_{n,i} -\mu}{\sigma} \right )
        -\exp \left [-\left (\frac{M_{n,i} -\mu}{\sigma} \right ) \right ]
      \right \},
      & \text{if\quad$\xi = 0$,}
    \end{cases}
  \end{split}
\end{equation}
defined on the points $M_{n,i}$ that, in the case $\xi \neq 0$,
satisfy the condition
$1+ \xi \left(\frac{M_{n,i} - \mu}{\sigma} \right)\; > 0$
for all $i=1,\ldots,m$.
Indeed, since the logarithm is a monotonic increasing function,
the likelihood function reaches its maximum value at the same
point as the log-likelihood function.

Approximate confidence intervals for
$\widehat{\boldsymbol{\theta}^0}$ are constructed using the fact
that each component of $\widehat{\boldsymbol{\theta}^0}=
(\widehat{\theta^0}_1,\widehat{\theta^0}_2,\widehat{\theta^0}_3)=
(\widehat{\mu^0},\widehat{\sigma^0},\widehat{\xi^0})= $ is
asymptotically normal~\cite{Col01}:
\begin{equation}
  \label{eq:mnvDistribution}
  \widehat{\theta^0}_i \backsim
  N(\theta_i^0,\widehat{\psi}_{i,i}) \quad \forall\, i=1,\ldots,d,
\end{equation}
where $\widehat{\psi}_{i,j}$ is a generic element of the inverse
of the \emph{observed information matrix}
$I_0(\boldsymbol{\theta})$ defined by
\begin{equation}
  \label{eq:obinfm}
  I_0(\boldsymbol{\theta})=\left ( -\frac{\partial^2 \,
      l(\boldsymbol{\theta})}{\partial\, \theta_i \partial\,
      \theta_j}\right)_{i,j} \quad \forall\, i,j=1,\ldots,d
\end{equation}
and evaluated in
$\boldsymbol{\theta}=\widehat{\boldsymbol{\theta}^0}$. From
Eq.~\eqref{mnvDistribution} one obtains the $(1-\alpha)$-confidence
interval for $\widehat{\theta^0}_i$:
\begin{equation}
  \label{eq:observedinfo}
  \theta_i^0 \, \pm \,
  z_{\frac{\alpha}{2}}\sqrt{\widehat{\psi}_{i,i}}
\end{equation}
where $z_{\frac{\alpha}{2}}$ is the $(1-\alpha/2)$ quantile of the
standard normal distribution.
All confidence intervals in this paper are computed by
formula~\eqref{observedinfo}, except when a more detailed
analysis is presented. For example, in the assessment
of inference quality, confidence intervals are also computed by a
standard bootstrap procedure (applied to the sequence of annual maxima)
and by profile likelihood.
The latter technique consists in the following.
Consider the parameter $\xi$, to fix ideas.
The profile likelihood of $\xi$ is obtained by setting $\mu$ and $\sigma$
to their maximum likelihood estimates, $\widehat{\mu^0}$ and
$\widehat{\sigma^0}$, respectively, in the log-likelihood
function $l$~\eqref{logllGEV}.
The plot of $l(\widehat{\mu^0},\widehat{\mu^0},\xi)$
as a function of $\xi$ is a section of the likelihood surface
of~\eqref{logllGEV} as viewed from the $\xi$-axis.
Confidence intervals constructed from this graph are often more
accurate than those obtained by observed information matrix,
see~\cite{Col01} for examples.

One of the main goals of extreme value theory is estimating the
probability of occurrence of events that are \textit{more extreme}
than those that have been observed thus far.
Let $z_p$ be the value that has a probability $p$ to be exceeded
every year by the annual maximum: $P\{M_{n,i}>z_p\}=p$ with $0 < p
< 1$. In common terminology $z_p$ is called the \emph{return
level} associated with the \emph{return period} $1/p$. A maximum
likelihood estimator for $z_p$ is obtained by plugging the
estimates for $\widehat{\boldsymbol{\theta}^0}
=[\widehat{\mu},\widehat{\sigma},\widehat{\xi}]$ into the
quantiles of $G(x)$, obtained by inverting Eq.~\eqref{GEV}. This
yields the estimator
\begin{equation}
  \label{eq:returnlevel}
  \widehat{z}_p=
  \begin{cases}
    \widehat{\mu}-\frac{\widehat{\sigma}}{\widehat{\xi}}\left[ 1-
      \left
        \{-\log(1-p)\right\}^{-\widehat{\xi}} \right]&
    \text{for\ \ }\widehat{\xi} \neq 0, \\
    \widehat{\mu}-\widehat{\sigma}\, \log
    \left\{-\log(1-p)\right\}&
    \text{for\ \ }\widehat{\xi} = 0.
  \end{cases}
\end{equation}
The variance of the return level estimator $\widehat{z}_p$ is
approximated as
\begin{equation}
 Var(\widehat{z}_p)\approx \nabla z_p^T \, V \, \nabla z_p,
\end{equation}
where
$\nabla z_p^T= \left [ \frac{\partial \, z_p}{\partial \,
    \mu}, \frac{\partial \, z_p}{\partial \, \sigma}, \frac{\partial
    \, z_p}{\partial \, \xi} \right]$,
$V$ is the variance-covariance matrix:
\begin{equation}\label{eq:varcovarmatrix}
 V(\mu,\sigma,\xi)=\left (
   \begin{array}{ccccc}
     Var(\mu)& & Cov(\mu,\sigma) & & Cov(\mu,\xi)\\
     Cov(\sigma,\mu) & & Var(\sigma) & & Cov(\sigma,\xi) \\
     Cov(\xi,\mu) & & Cov(\xi,\sigma) & & Var(\xi)
   \end{array}
 \right),
\end{equation}
and both $\nabla z_p$ and $V$ are evaluated at the maximum likelihood
estimate
$\widehat{\boldsymbol{\theta}^0}=
[\widehat{\mu},\widehat{\sigma},\widehat{\xi}]$.
This allows the construction of confidence intervals for
$\widehat{z}_p$ and is referred to as the \emph{delta method}.
Again, profile likelihood and a boostrap technique are used
for goodness-of-fit assessment of the return level inferences.

Only for Weibull distributions ($\xi<0$) it is possible to have
$p=0$, corresponding to a return level with an infinite return
period. In this case,
\begin{equation}
  \label{eq:asymptote}
  z_0=\widehat{\mu}-\frac{\widehat{\sigma}}{\widehat{\xi}}.
\end{equation}
All information about the return levels is usually reported in the
\emph{return level plot}, where $\widehat{z}_p$ is plotted against
$\log \,y_p$, where $y_p=-\log(1-p)$ (compare
Eq.~\eqref{returnlevel}). The return level plot is linear for the
Gumbel distribution, concave for $\xi>0$ (Fr\'echet) and has the
horizontal asymptote~\eqref{asymptote} (Weibull). Notice that the
smallest vales of $p$ are usually those of interest, since they
correspond to very rare (particularly extreme) events. In the
return level plots, events with a short return period (large
probability $p$) are compressed near the origin of the axes, while
outliers and rare events (small $p$) are highlighted. For this
reason such plots are very useful tools for both model analysis
and diagnosis.

The above procedures for the estimates of return levels and GEV
parameters require assessment with reference to the available
data. Useful graphical checks are the \emph{probability plot}, the
\emph{quantile-quantile plot} (QQ plot) and the return level plot.
The first is the comparison between the estimated and the
empirical distribution function $\widetilde{G}(x)$. The latter is
a stepfunction defined by
\begin{equation}
  \label{eq:edf}
  \widetilde{G}(M_{(i)})=\frac{i}{m+1},
\end{equation}
where $M_{(i)}$ is the order statistics for the sequence
$M_{n,1},\ldots,M_{n,m}$ of $m$ block maxima. Notice that the
definition of the empirical d.f.~\eqref{edf} is not unique,
see~\cite{Cas88}.

The QQ-Plot, formed by the points
\begin{equation}
  \label{eq:QQ_plot}
  \left \{ \left (\widetilde{G}^{-1}\left (\frac{i}{m+1}\right ),
      m_{(i)} \right), \quad \forall \, i=1,\ldots,m \right \}
\end{equation}
highlights the behavior of the model tail, which is often the most
interesting part. Substantial departures of the above plots from
the diagonal indicate inadequacy of the GEV model or other
systematic errors.
Another diagnostic plot is constructed by adding confidence
intervals for $\widehat{z}_p$ and return levels of the empirical
d.f., according to Eq.~\eqref{QQ_plot}, to the return level plot
(see above). Agreement of the empirical d.f. with the return level
curve suggests goodness of fit and adequacy of the GEV model.

All computations and plots in this paper have been made with the
software \textbf{R}~\cite{IG96}, available under the GNU license
at \texttt{www.r-project.org}. The library \texttt{ismev}
(\texttt{www.cran.r-project.org}),
which is an \textbf{R}-port of the routines written by Stuart Coles
as complement to~\cite{Col01}, has been used with minor
modifications.

\section{GEV Inferences for 1000 Annual Maxima}
\label{sec:GEVinference}

\begin{figure}[p]
  \centering
  \includegraphics[width=0.32\textwidth]{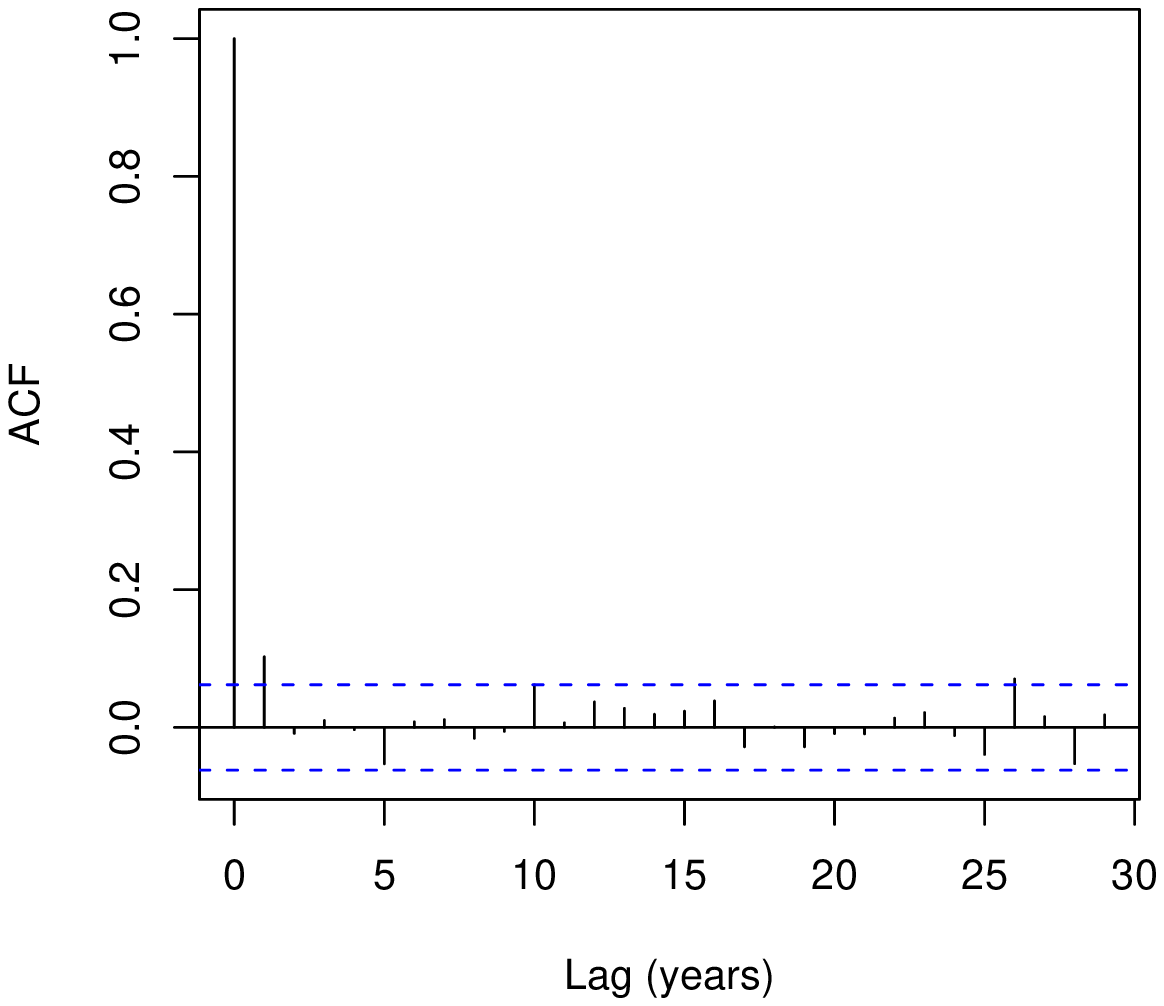}
  \includegraphics[width=0.32\textwidth]{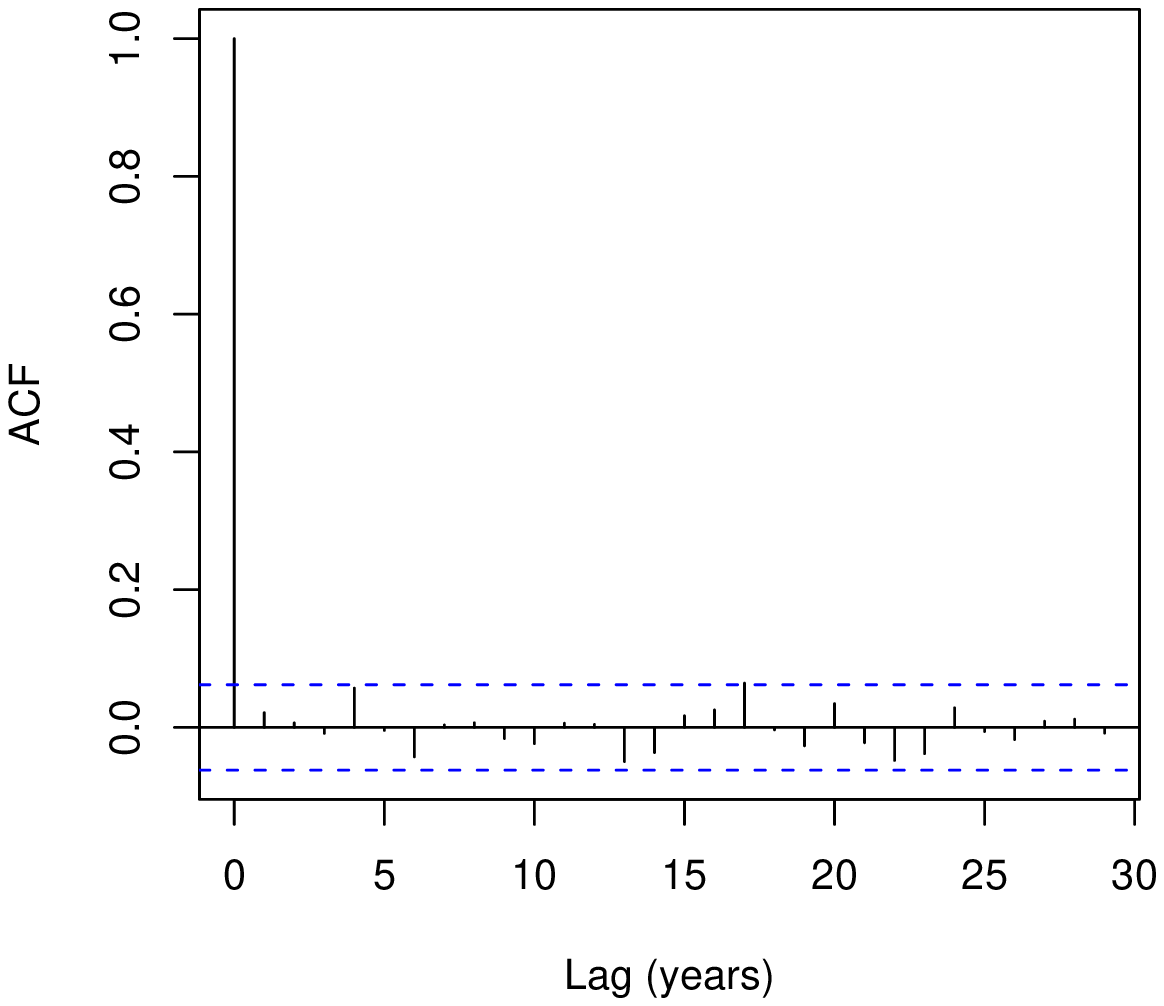}
  \includegraphics[width=0.32\textwidth]{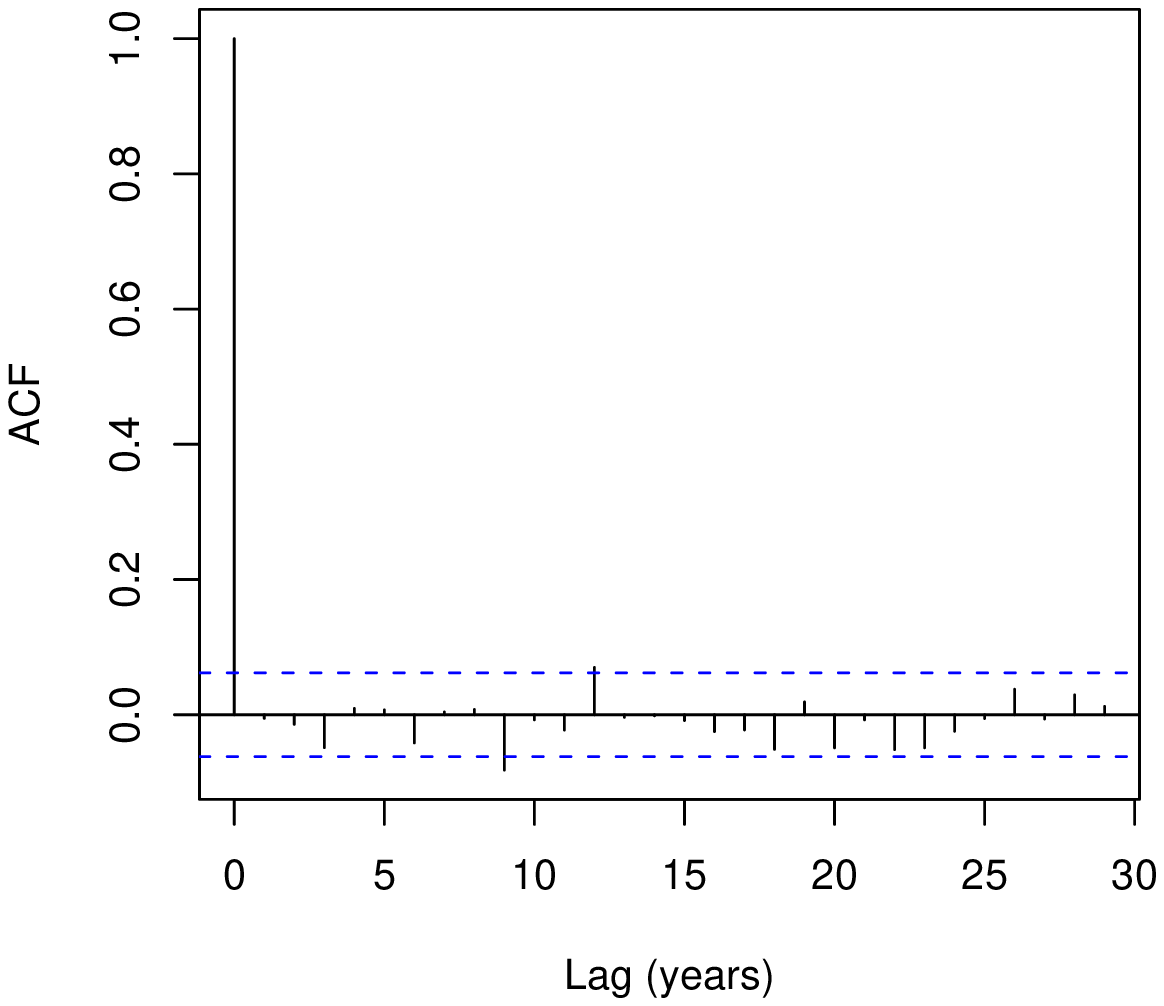}
  \caption{
    Autocorrelations of the sequences of 1000 yearly maxima
    of the total energy time series for $T_E=10,30,50$
    (from left to right, respectively).
  }
  \label{fig:acf-max}
\end{figure}

The annual maxima are extracted from the 6-hourly time series of
the energy described in \secref{setup}.
Each series contains $4 \times 365 \times 1000=1460000$ data.
We then set $n=1460$ in~\eqref{maxima}, thereby obtaining sequences
of 1000 annual extremes of the total energy.
The yearly maxima are linearly uncorrelated (\figref{acf-max}),
suggesting that it is safe and reasonable to assume independence.
Also compare with the autocorrelation decay time in \figref{acf}

On theoretical grounds we can at least deduce one constraint on
the distribution of extremes for the energy time series. Indeed,
since the attractor is contained within a bounded domain of the
phase space and since the energy observable $E(t)$ defined
in~\eqref{energydensity} is a continuous function of the phase space
variables, it turns out that the total energy is bounded on any
orbit lying on (or converging to) the attractor. Therefore, the
energy extremes are \emph{necessarily} Weibull distributed ($\xi$
is negative). This provides a theoretically founded criterion for
quality assessment of the obtained GEV inferences.

\begin{figure}[p]
  \centering
  \includegraphics[width=0.32\textwidth]{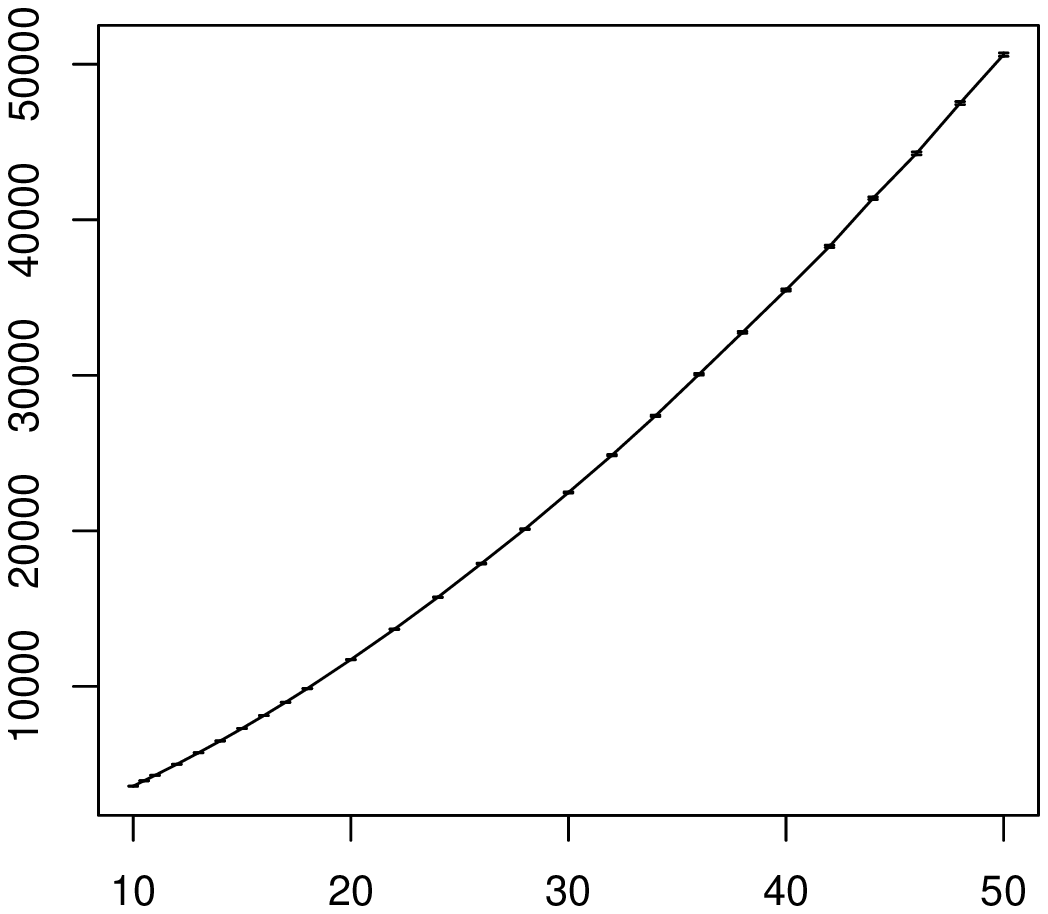}
  \includegraphics[width=0.32\textwidth]{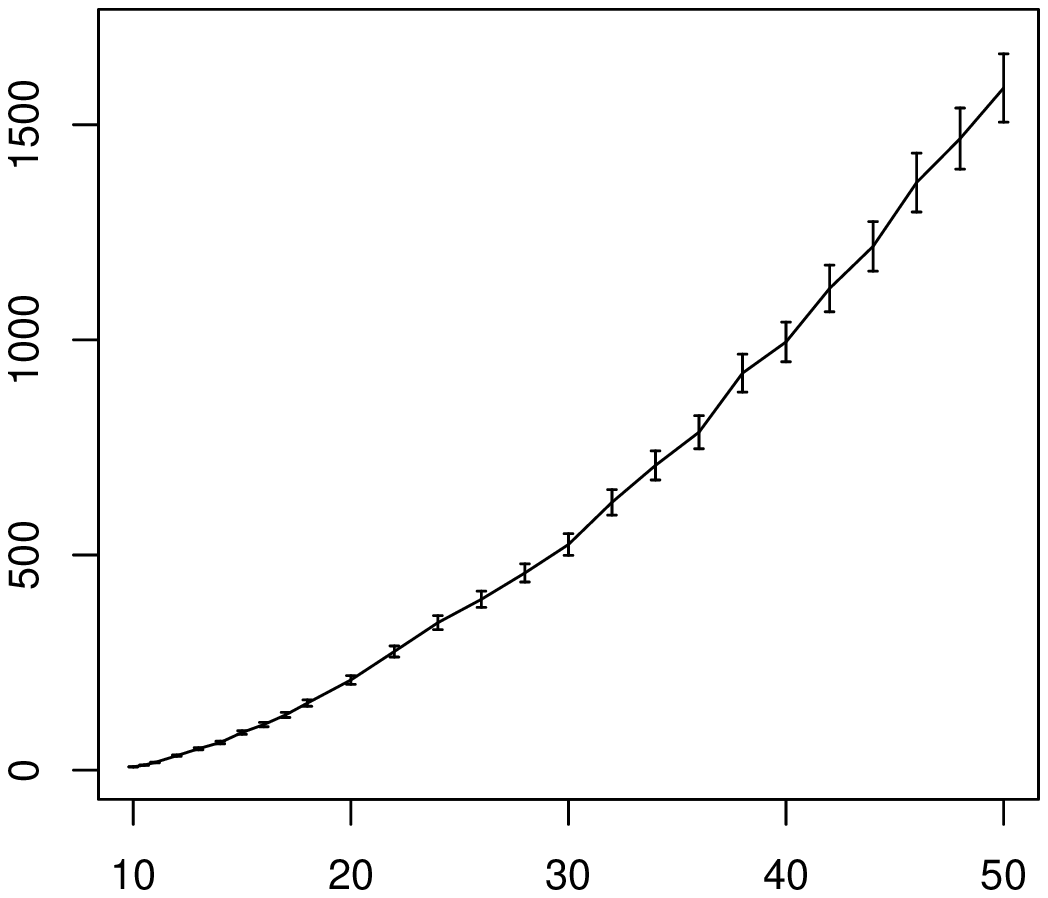}
  \includegraphics[width=0.32\textwidth]{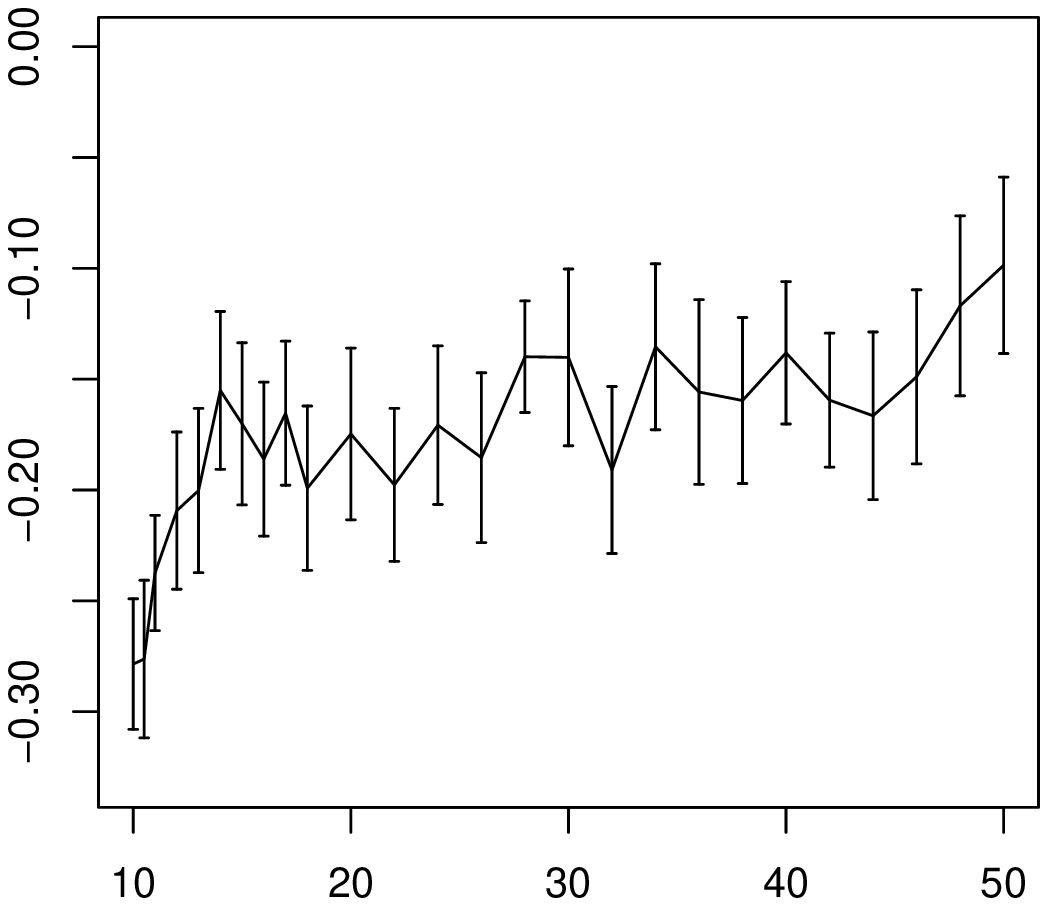}
  \caption{
    From left to right:
    maximum likelihood estimates of $\mu$, $\sigma$, and $\xi$,
    respectively (vertical axis), for each of the 21 sequences
    of 1000 maxima of the total energy,
    against the corresponding values of $T_E$ (horizontal axis).
    Confidence intervals are added with errorbars but are hardly visible
    for $\mu$ (leftmost panel) at the selected scale.
  }
  \label{fig:stationaryGEV}
\end{figure}

The GEV parameters $(\mu,\sigma,\xi)$ are estimated by the maximum
likelihood method (see \secref{GEVmethods}) from the sequences
of yearly maxima. The fitted values of $(\mu,\sigma,\xi)$,
together with confidence bands (computed by the observed
information matrix, formula~\eqref{observedinfo})
are plotted as functions of $T_E$ in \figref{stationaryGEV}.
The inferred parameters $\mu$ and $\sigma$ increase
monotonically with $T_E$. Estimates of $\xi$ are in each case
negative and the related confidence intervals are markedly bounded
away from zero: observed information matrix, profile likelihood
and bootstrap yield similar estimates.
The latter result matches quite well the theoretical expectation
discussed in the previous section.
Also notice that the uncertainty in $\xi$ may reach up to $21\%$
of its value, whereas the parameters $\mu$ and $\sigma$ are quite
accurately estimated: the maximal uncertainties in $\mu$ and
$\sigma$ are $0.1\%$ and $2.5\%$ of the corresponding value,
respectively.

\begin{figure}[p]
  \centering
  \includegraphics[width=0.49\textwidth]{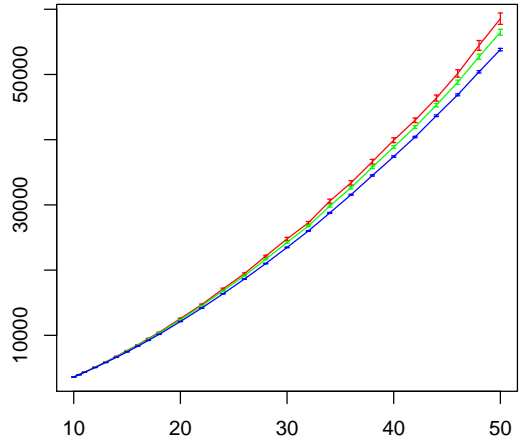}
  \caption{
    Maximum likelihood estimates of the 10-, 100-, and 1000-year
    return levels of the total energy
    (red, green, and blue, respectively)
    for each of the 21 stationary series of 1000 maxima
    of the total energy, against the corresponding values of $T_E$
    (horizontal axis).
    Confidence intervals are added with errorbars but are hardly
    visible (at the selected scale).
    }
  \label{fig:retfits}
\end{figure}

Information on the tails of the energy distribution is
straightforwardly expressed by the return level plots, where
$z_p=G^{-1}(1-1/p)$ is the return level associated to the
$p$-year return period and $G$ is the GEV distribution~\eqref{GEV}.
In \figref{retfits}, return levels with
return periods of 10, 100 and 1000 years are plotted as functions
of $T_E$. Each graph is monotonically increasing with $T_E$ and,
for $T_E$ fixed, the return levels increase with the return period.

\begin{figure}[p]
  \centering
  \includegraphics[width=0.45\textwidth]{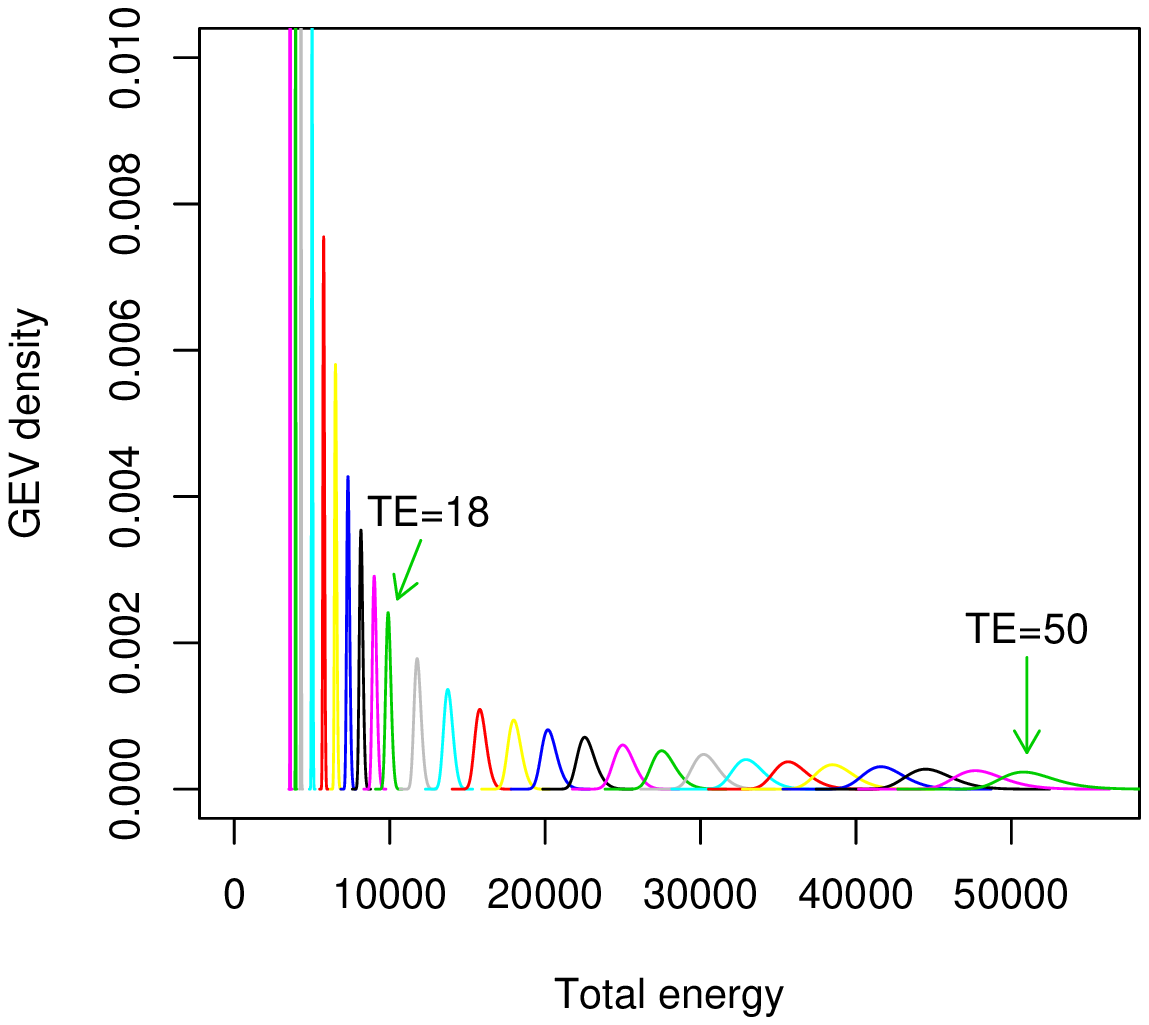}
  \includegraphics[width=0.45\textwidth]{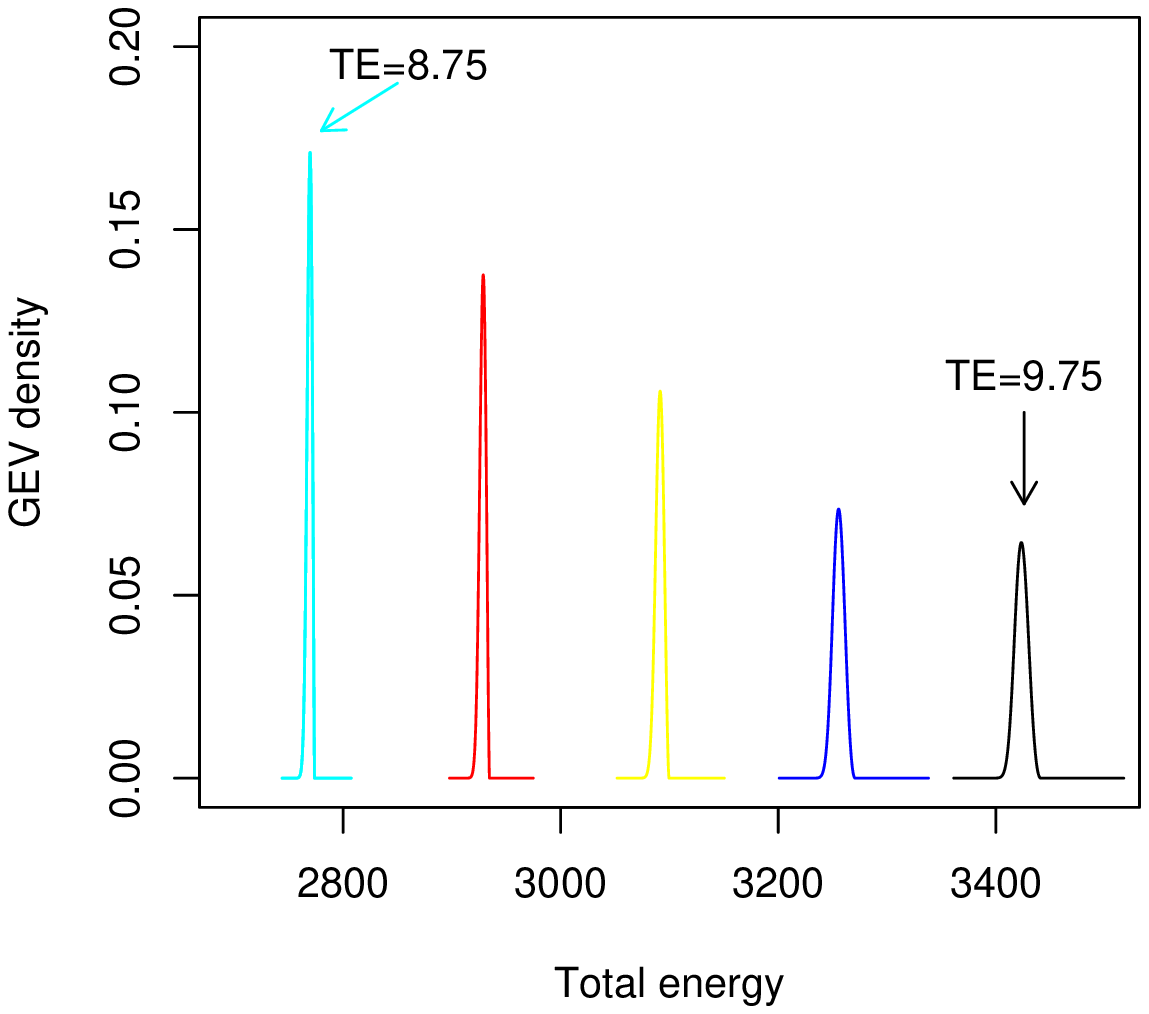}
  \caption{
    Left:
    Probability density functions of the GEV for the 21
    values of $T_E$ in the considered range $[10,50]$.
    Right: same as left for $8.75\le T_E\le9.75$.
  }
  \label{fig:densTOT}
\end{figure}

The dependence of the GEV probability density with respect to
$T_E$ is illustrated in \figref{densTOT}. The increase of scale
and location parameters with $T_E$ induces a rightward shift and a
spread of the probability density. In particular, from the
geophysical point of view, both the range and severity of possible
extreme values of the total energy increase with $T_E$. In fact,
this behavior sets in for $T_E$ right after the creation of the
chaotic attractor, see \figref{densTOT}~right.

\subsection{Smoothness of GEV Inferences with Respect to
  System Parameters}
\label{sec:smooth}
The dependence from $T_E$ of the time-averaged total energy
and of the inferred GEV parameters (including the return levels)
is rather smooth, see \figref{e32} and \figref{stationaryGEV}.
This strongly suggests the existence of functional relations of the form
\begin{equation}
  \label{eq:powermusigma}
  \mu=\alpha T_E^\gamma \quad\text{and}\quad \sigma=\alpha' T_E^{\gamma'}.
\end{equation}
Such power laws are fitted to the graphs of $\mu$ and $\sigma$ as
follows.

To set ideas, we consider $\mu$ and denote by $\hat{\mu}(T_E^j)$
and $\sigma_{\hat{\mu}}(T_E^j)$ the maximum likelihood estimate of
$\mu$ and the related standard deviation (calculated by the
observed information matrix), respectively, where $T_E^j$ is one
of the 21 chosen values in the interval $[10,50]$. A bootstrap
procedure is performed where iterated realizations of a sequence
of 21 independent Gaussian variables with mean $\hat{\mu}(T_E^j)$
and standard deviation $\sigma_{\hat{\mu}}(T_E^j)$ are simulated.
For each realization, a power law fit as in~\eqref{powermusigma}
is performed. The sample average and standard deviation of the so
obtained fits, constructed independently for $\mu$ and $\sigma$,
are reported in \tabref{powermu} and \tabref{powersigma}.

\begin{figure}[p]
  \centering
  \includegraphics[width=0.3\textwidth]{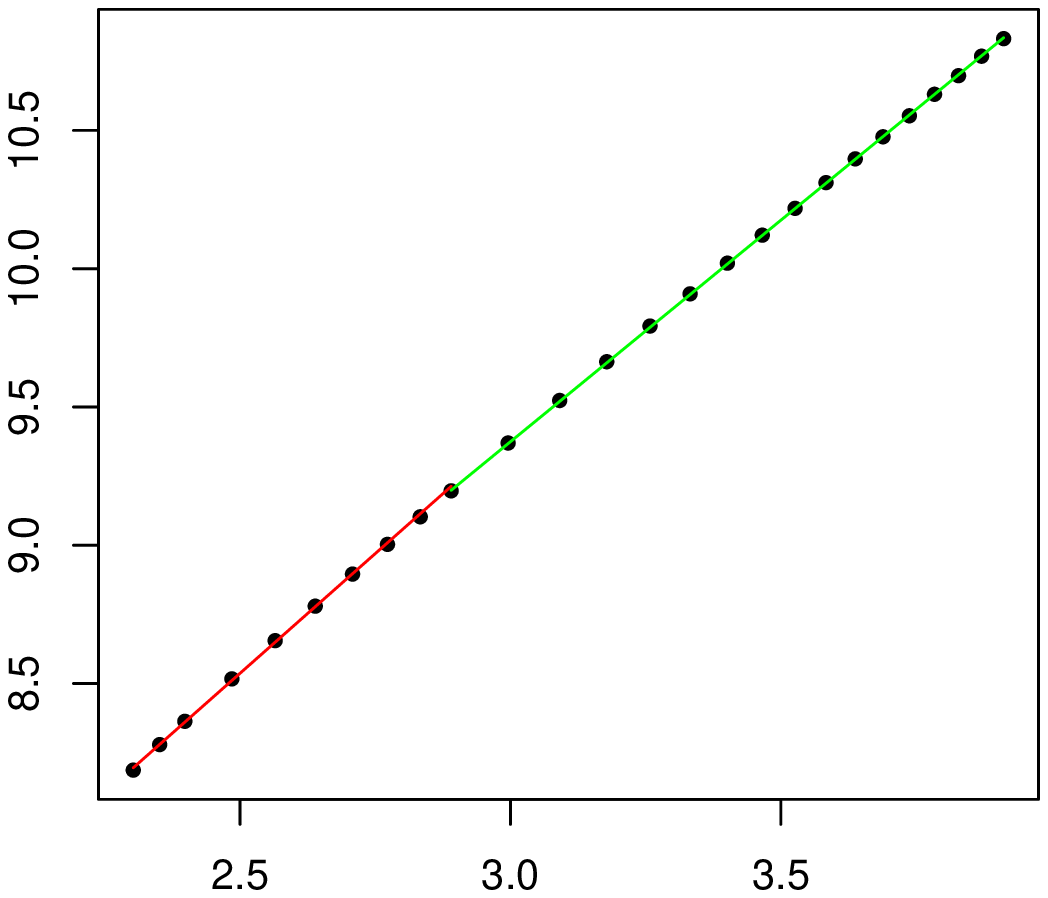}
  \includegraphics[width=0.3\textwidth]{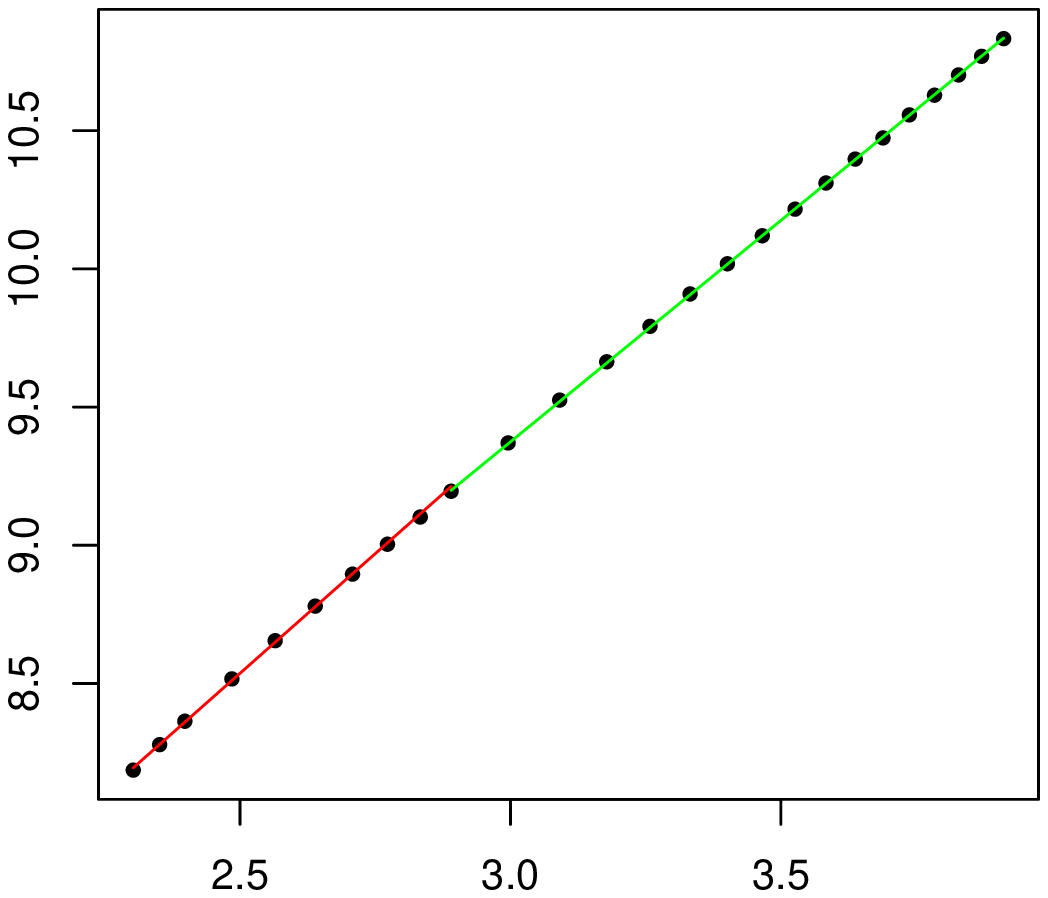}
  \includegraphics[width=0.3\textwidth]{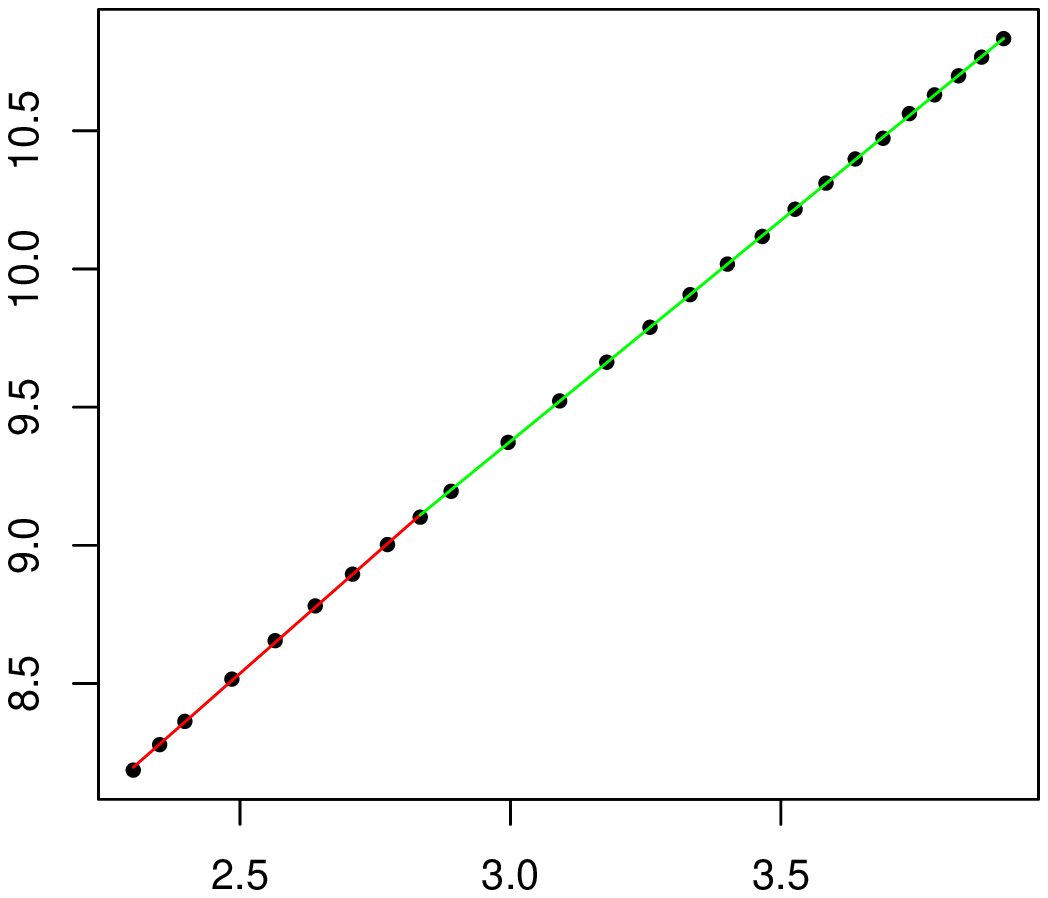}
  \caption{
    Power law fits of the inferred values of $\log(\mu)$ (vertical axis)
    as a function of $\log(T_E)$ (horizontal axis).
    From left to right: 1000, 300, and 100 yearly maxima have been used.
    In each case, there are two intervals of $T_E$, separated by a point
    $T_E^b$, characterized by a different scaling exponent,
    compare \tabref{powermu}.
  }
  \label{fig:powermu}
\end{figure}

\begin{figure}[p]
  \centering
  \includegraphics[width=0.3\textwidth]{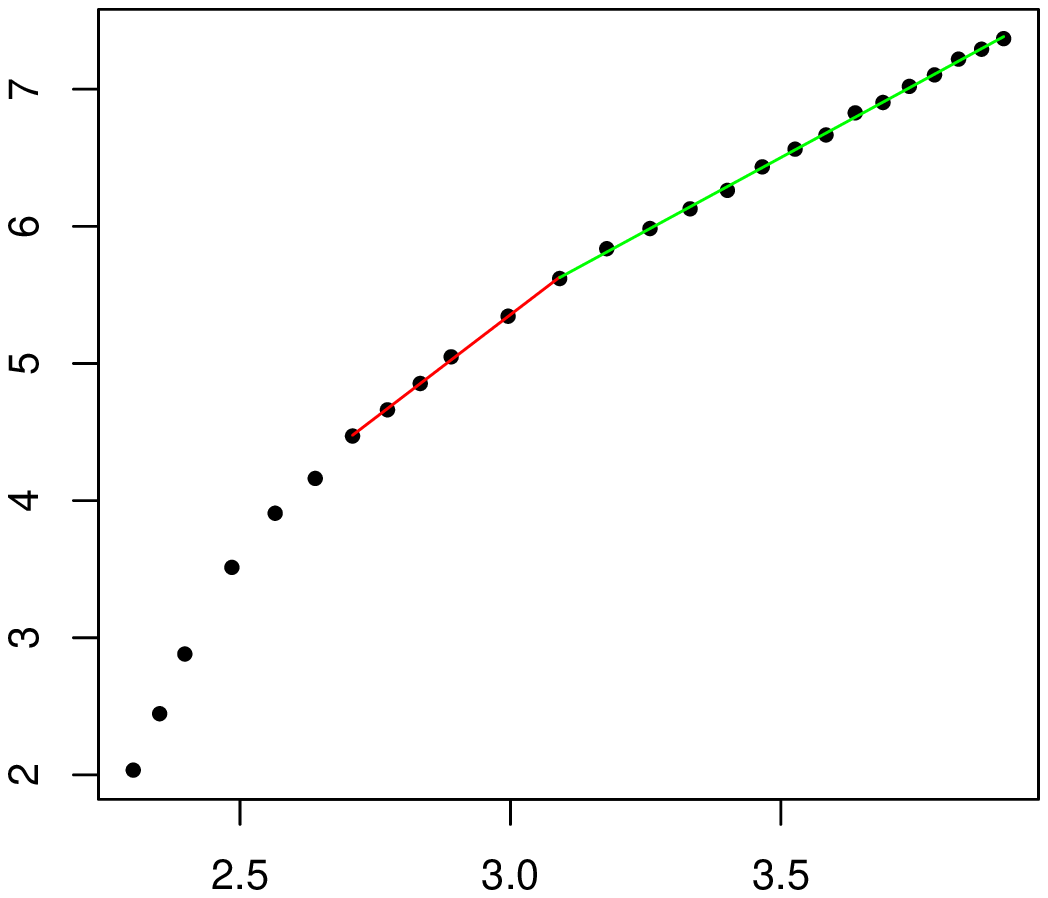}
  \includegraphics[width=0.3\textwidth]{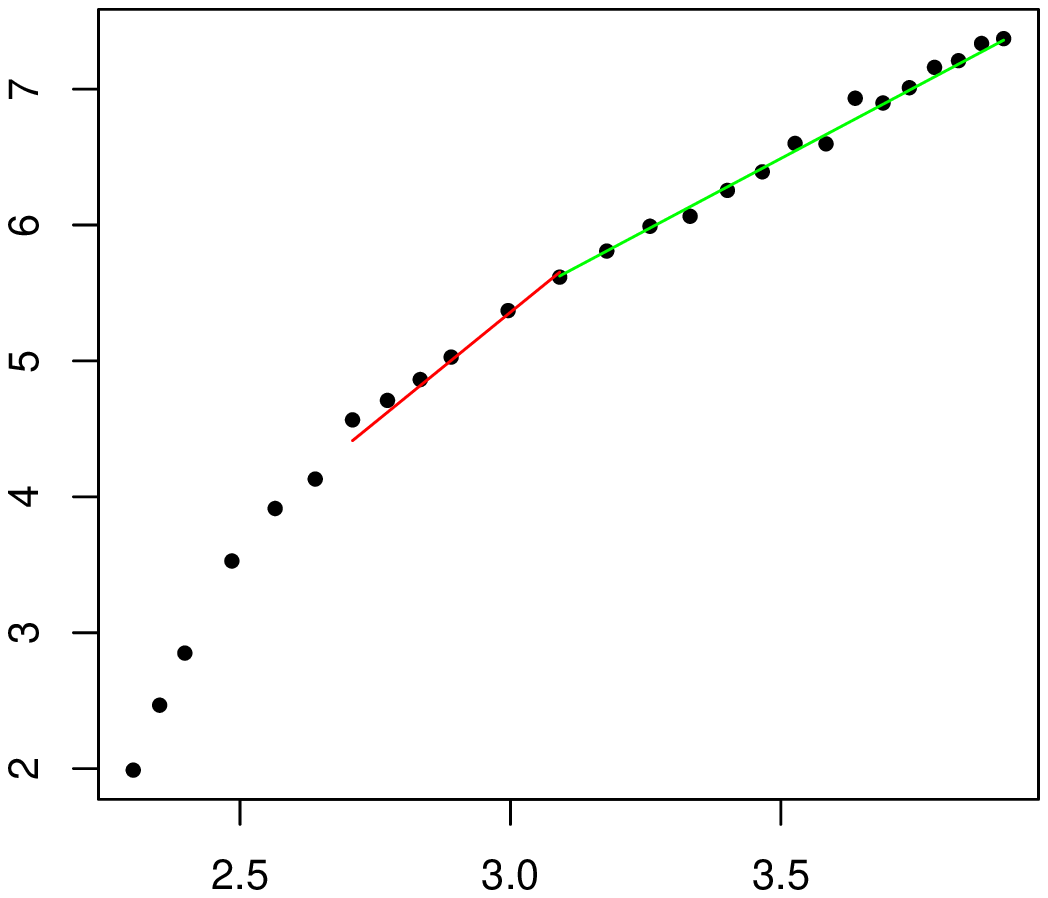}
  \includegraphics[width=0.3\textwidth]{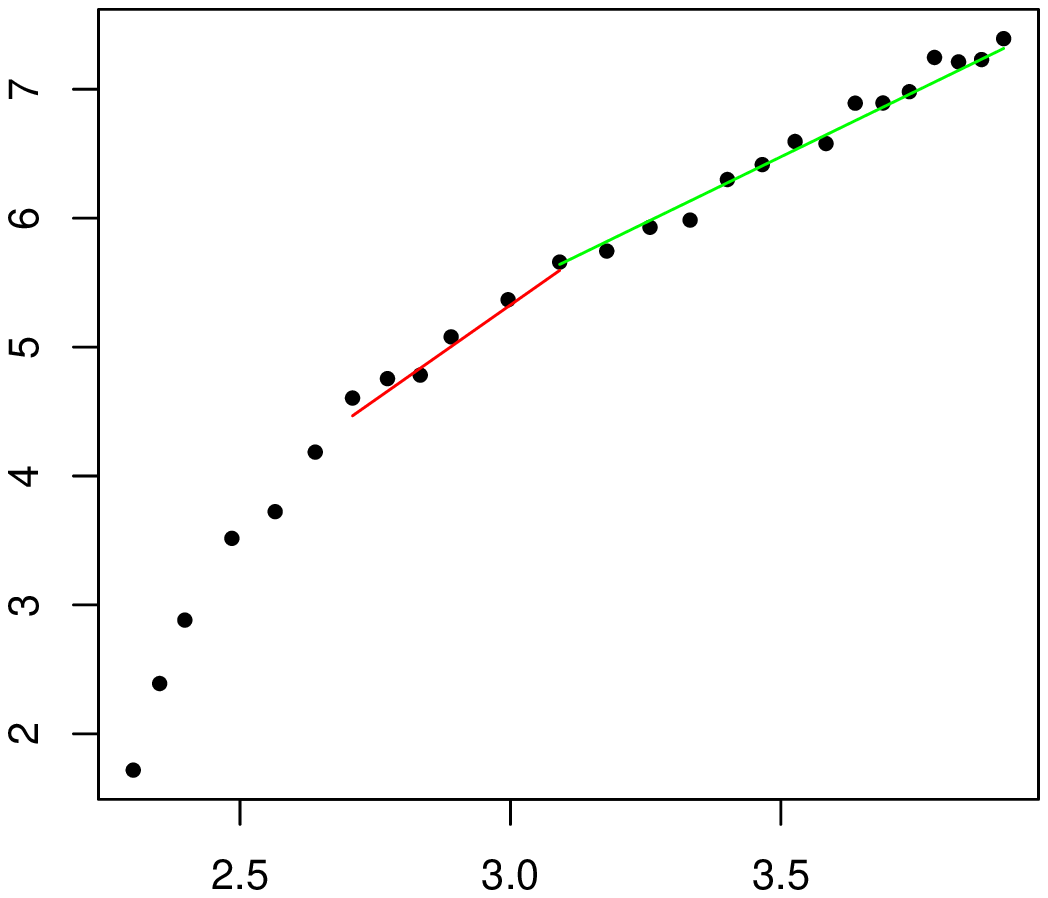}
  \caption{
    Same as \figref{powermu} for $\log(\sigma)$,
    see \tabref{powersigma}.}
  \label{fig:powersigma}
\end{figure}

Two distinct ranges of $T_E$ are identified, where $\mu$ scales by
a different exponent, also see \figref{powermu}~left. For
$T_E\lesssim18$ $\gamma_\mu$ is $\sim 1.73$ while it decreases to
$\sim 1.6$ for $T_E\gtrsim 18$.  The time-mean total energy of the
system has a rather similar power-law dependence on $T_E$
\cite{LSV05}. In the upper $T_E$-range the exponent of the power
law of the extremes is larger than that of the time-mean total
energy ($\sim 1.52$), which implies that asymptotically the
extremes tend to become relatively \textit{more extreme}. When
considering $\sigma$, there is an initial interval of $T_E$ where
no power law is obeyed, see \figref{powersigma}~left. For
$22\gtrsim T_E\gtrsim 15$ $\gamma_\sigma$ is $\sim 3.0$ while it
decreases to $\sim 2.1$ for $T_E\gtrsim 22$. Since
$\gamma_\sigma>\gamma_\mu$ for high values of $T_E$, we have that
asymptotically with $T_E$ the spread of the maxima tends to become
consistent with respect to their average location, thus suggesting
a larger variability in the maxima.
Shorter yearly maxima sequences, of length 300 and 100, lead to
nearly identical estimates for both $\gamma_\mu$ and
$\gamma_\sigma$ and their confidence intervals, thus implying that
this is a rather robust property of the system.

Apart for the total energy, it turns out that analogous power law
dependence with respect to $T_E$ is detected in the considered
model for several dynamical and physical observables, such as
Lyapunov dimension, maximal Lyapunov exponent and average zonal
wind \cite{LSV05}. This suggests that the whole attractor of the
model (more precisely, its SRB measure) has some scaling laws with
respect to $T_E$. The qualitative features described above for
$T_E$ sufficiently large, such as the form of $(\mu,\sigma)$ as
functions of $T_E$ and the fact that $\xi$ seems to approach a
constant negative value, are most probably related to this scaling
behavior. An important question we address elsewhere is whether
this is a peculiarity of the baroclinic model used here
or if analogous smoothness properties are common
(\emph{generic} or \emph{robust} in some sense)
for models of atmospheric dynamics, including General
Circulation Models.

\section{Sensitivity of the GEV inferences}
\label{sec:GEVsensitivity}
The length of 1000 for the sequences of yearly maxima turns out to
yield good accuracy for the GEV inferences. The sensitivity of
such results has been tested by relaxing the experimental
conditions considered in the previous section. This has been done
in several ways:
\begin{itemize}
\item
  by varying the \textit{number of extreme events}
  (length of the sequences of yearly maxima);
\item
  by using \textit{soft extremes}
  (maxima are computed over data blocks corresponding
  to time spans shorter than one year).
  \item
  by varying the resolution of the model.
\end{itemize}
The best estimates and related uncertainties of the GEV parameters
obtained under modified experimental conditions have been first
compared at phase value to what obtained in the reference case, in
order to detect mismatch due to biases and changing precision.
Moreover, the resulting differences in the GEV distributions have
been inspected also with by adopting the standard graphical
diagnostics, such as quantile-quantile and return level plots, and
by computing bootstrap confidence intervals and profile likelihood
both for the critical parameter $\xi$ and for the return levels.

\subsection{Sensitivity with Respect to the Extreme Events
Sample Size}

\begin{figure}[p]
  \centering
  \includegraphics[width=0.24\textwidth]{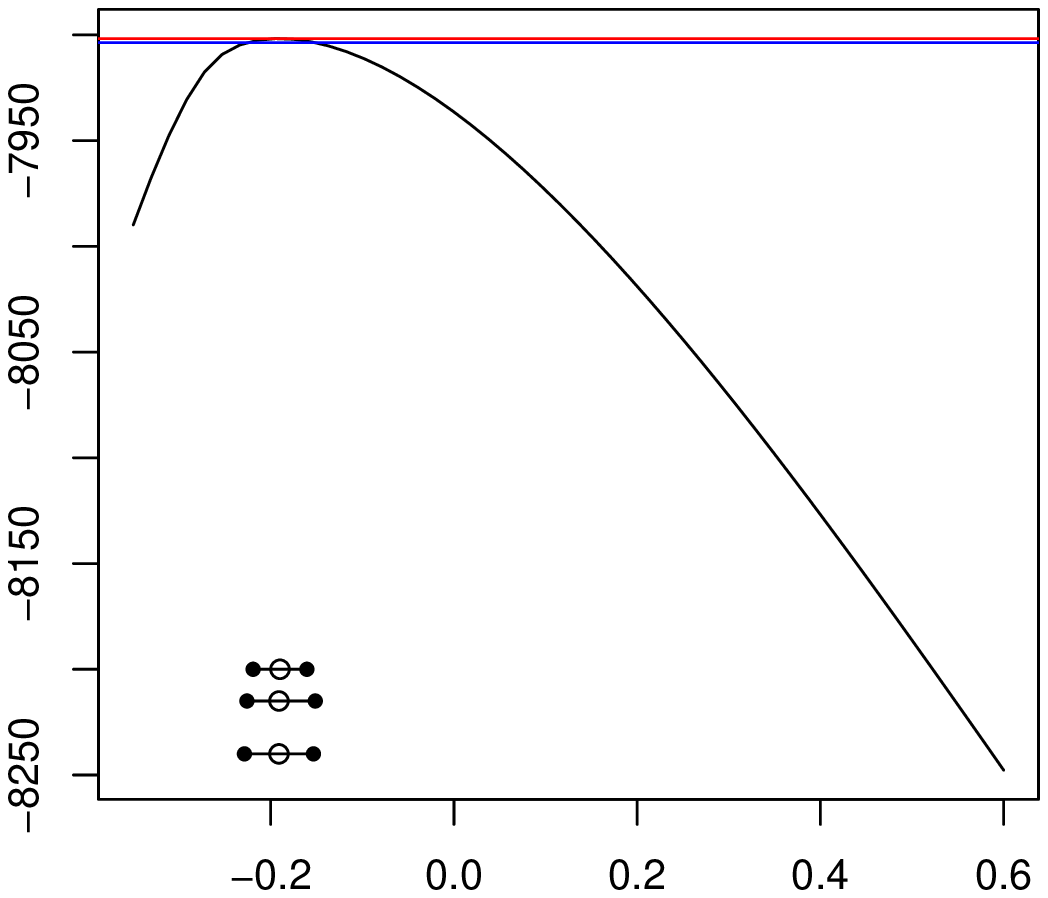}
  \includegraphics[width=0.24\textwidth]{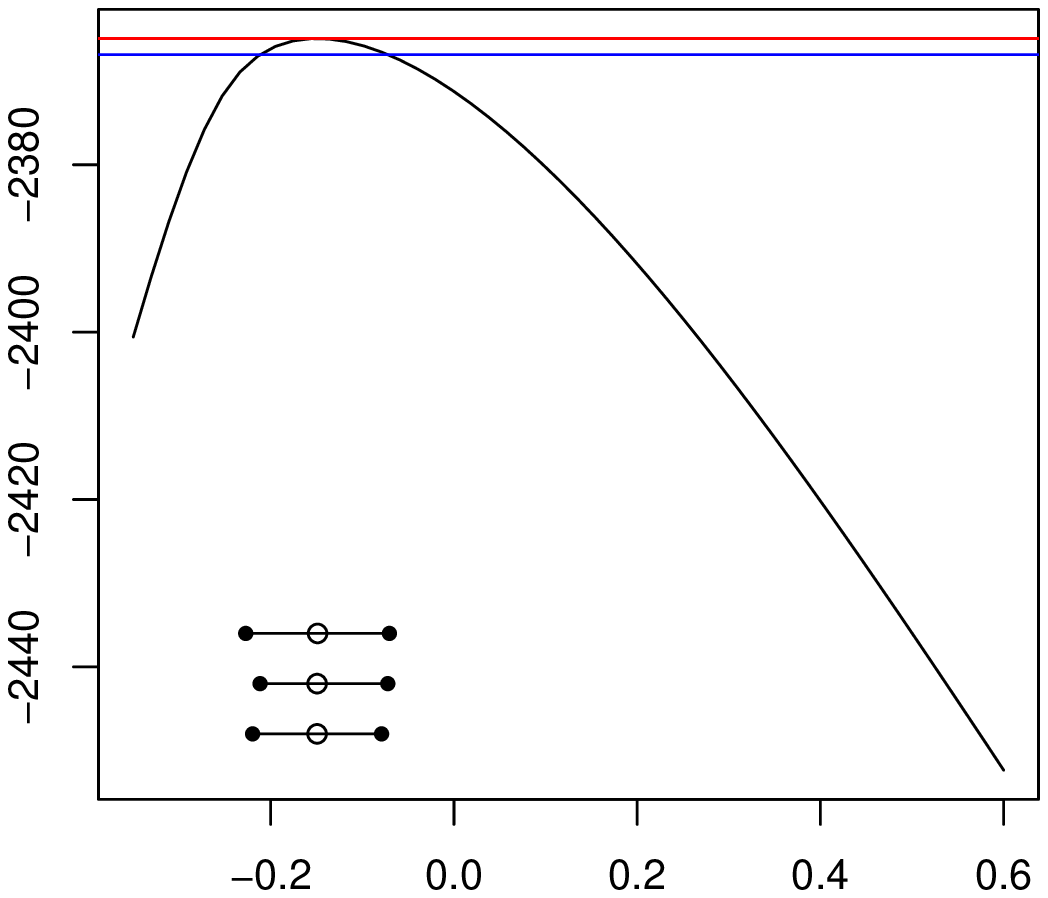}
  \includegraphics[width=0.24\textwidth]{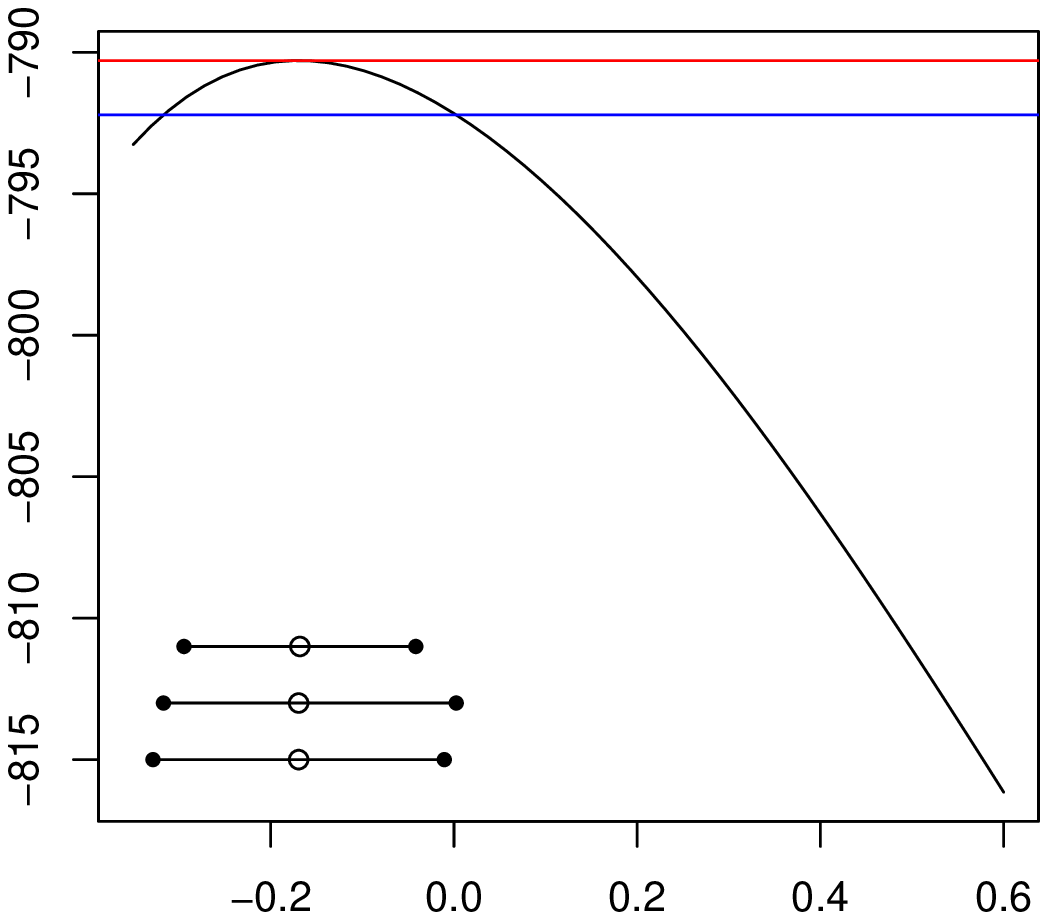}
  \includegraphics[width=0.24\textwidth]{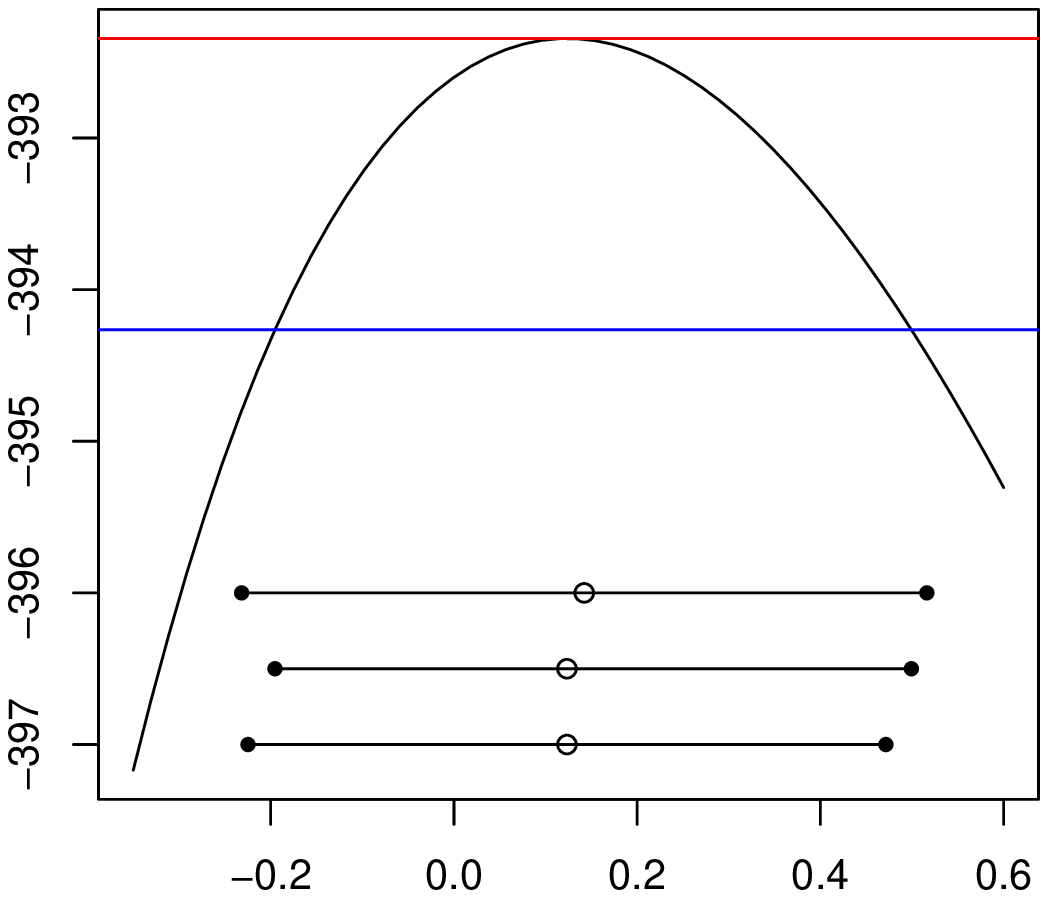}
  \caption{
    Profile likelihood plots of $\xi$ for $T_E=32$.
    From left to right, 1000, 300, 100, and 50 yearly maxima
    have been used, respectively.
    In the latter case, the estimate of $\xi$ is positive.
    Confidence intervals are computed by bootstrap, profile likelihood,
    and observed information matrix (the three stacked lines
    at the bottom part of the plots, from top to bottom, respectively).
    Notice the increasing width of the confidence intervals
    (quite large already for length 100) and the
    agreement between confidence intervals computed by the three
    methods, also for the \emph{wrong} estimate obtained with 50 maxima.
  }
  \label{fig:xi-len}
\end{figure}

We describe what is found when reducing the number of yearly
maxima used for GEV inference. The, particularly unfortunate, case
occurring for $T_E=32$ is first analyzed by means of profile
likelihood for the GEV parameter $\xi$. Sequences of 1000, 300,
100, and 50 yearly maxima of the total energy are used to produce
the plots in \figref{xi-len}. The cases 1000, 300, and 100 yield
coherent estimates for $\xi$. For detailed diagnostics,
confidence intervals are computed by the observed information matrix
(formula~\eqref{observedinfo}) and compared by those obtained by
profile likelihood and by a standard bootstrap procedure.
The three methods yield similar results in all cases,
both for the estimates and for the confidence intervals.
However, for 50 maxima the confidence intervals become very wide
and a positive value for $\xi$ is inferred, which is unphysical.

\begin{figure}[p]
  \centering
  \includegraphics[width=0.24\textwidth]{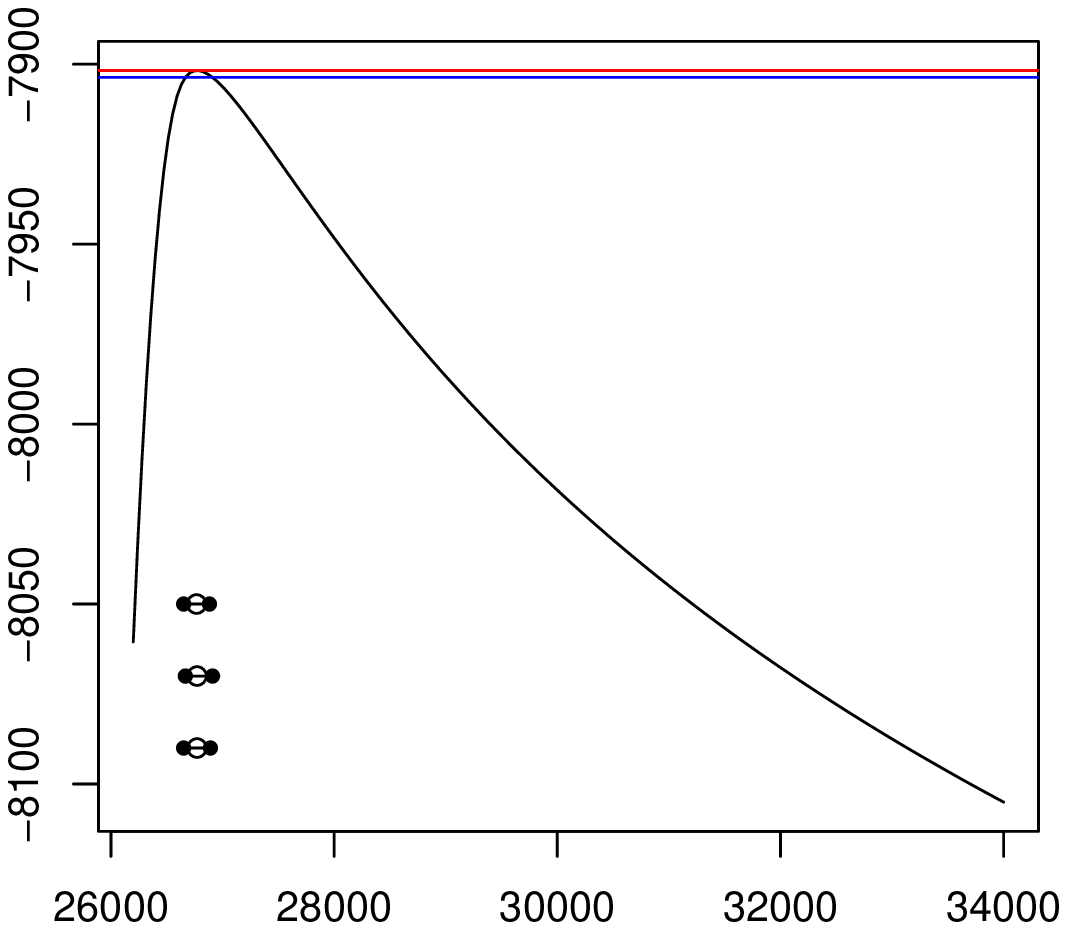}
  \includegraphics[width=0.24\textwidth]{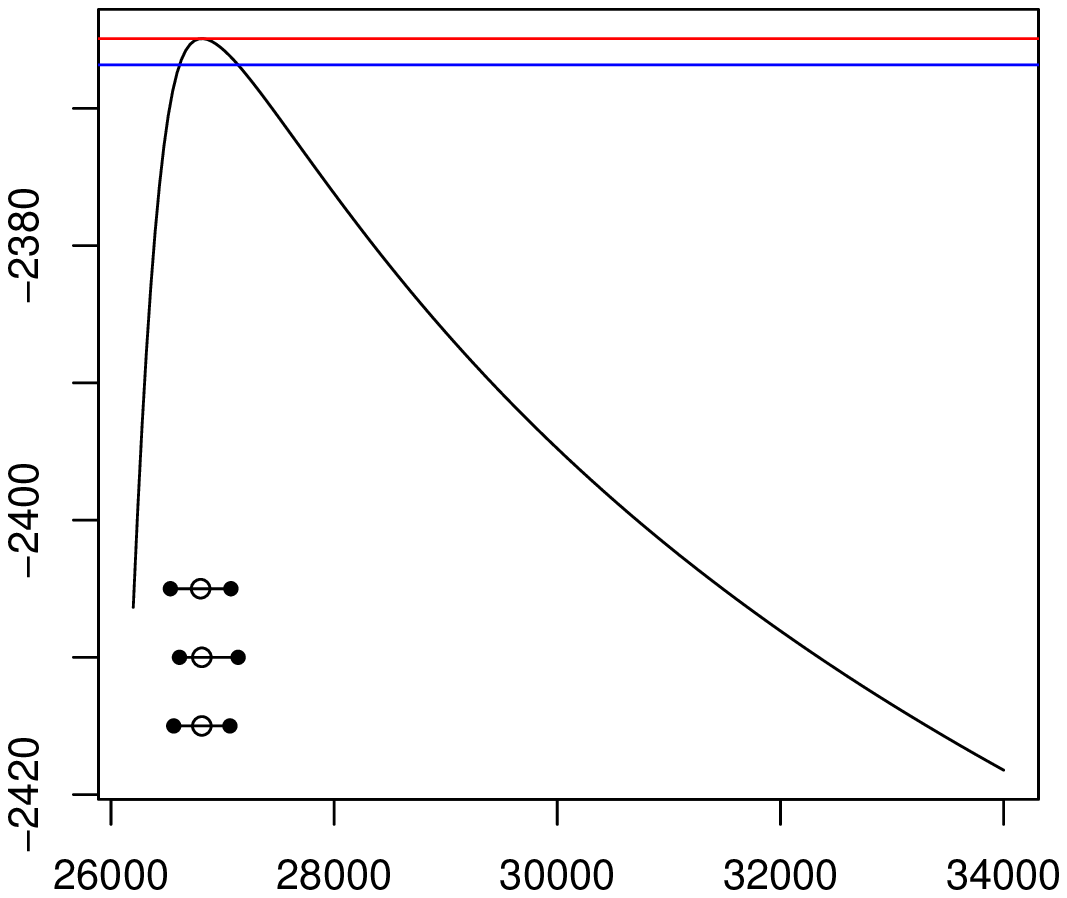}
  \includegraphics[width=0.24\textwidth]{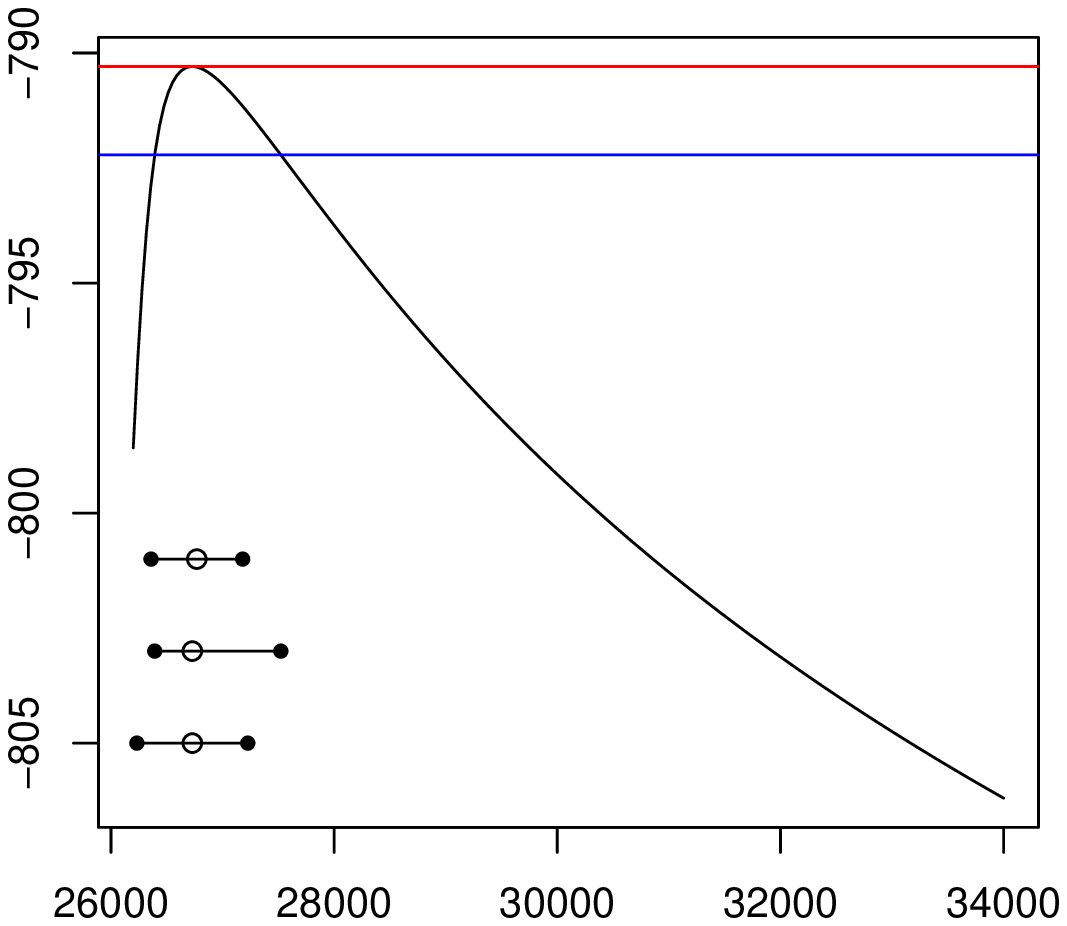}
  \includegraphics[width=0.24\textwidth]{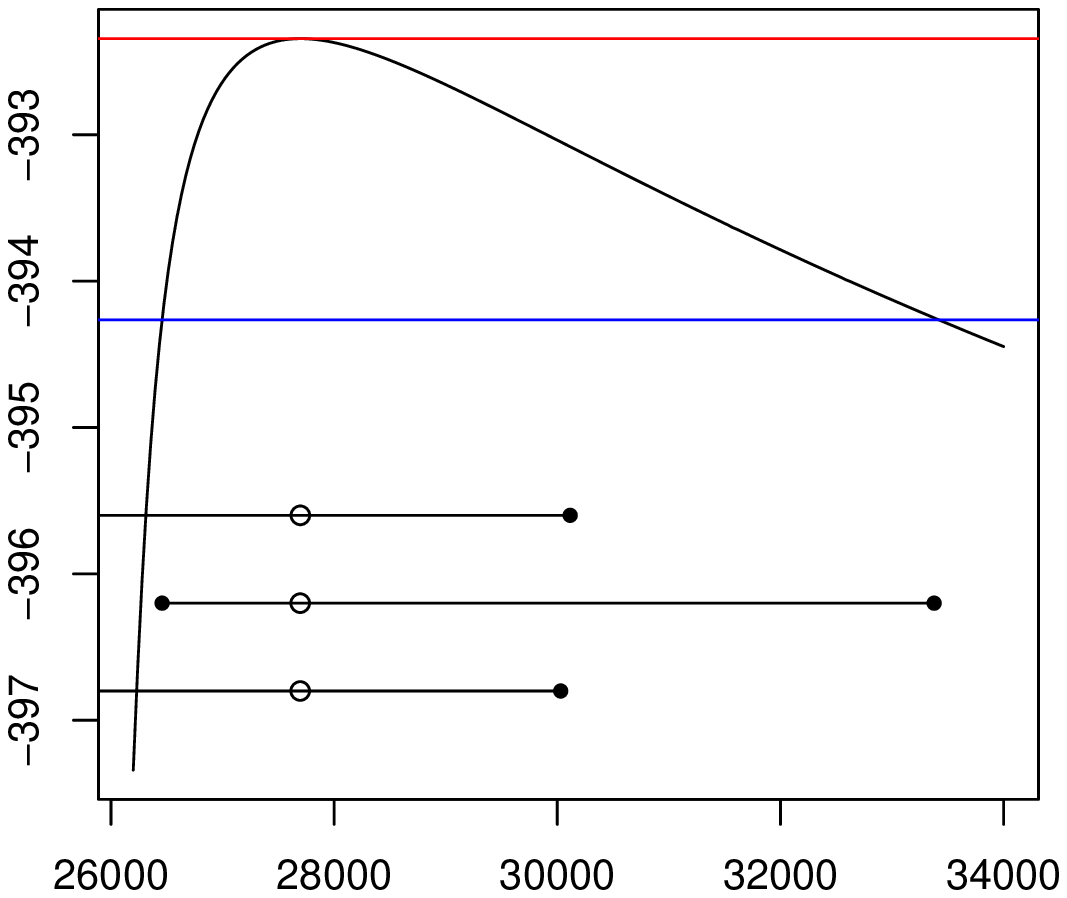}
  \caption{
    Profile likelihood plots of the 100-year return level
    for $T_E=32$ for different lengths of the sample of yearly maxima:
    1000, 300, 100, and 50 yearly maxima have been used from left
    to right, respectively.
    Confidence intervals are computed by bootstrap, profile likelihood,
    and observed information matrix (the three stacked lines
    at the bottom part of the plots, from top to bottom, respectively).
    Notice the increasing width of the confidence intervals
    and increasing skewness of those obtained by profile likelihood.
  }
  \label{fig:retlev-len}
\end{figure}

The decay of the inference quality is revealed in a different way
by the profile likelihood plots for the 100-year return levels
(\figref{retlev-len}). It is, in general, not quite safe to infer,
from a series of $n$ annual extremes, return levels with return
periods larger than $n$ years. Extrapolation to larger return
periods may produce incorrect values and is likely to yield
significant uncertainties. For the considered case, the estimates
are coherent for 1000, 300, and 100 yearly maxima. As expected,
the confidence intervals (computed by the $\delta$ method, see
\secref{GEVmethods}) become larger as shorter sequences of
maxima are used. This also holds for bootstrap and profile
likelihood. However, for 50 maxima the profile likelihood
confidence intervals become very skewed as opposed to bootstrap or
$\delta$ method. This clearly indicates poor approximation of
normality for the GEV estimators~\cite{Col01}, revealing the
intrinsic unreliability of the estimates.

\begin{figure}[p]
  \centering
  \includegraphics[width=0.24\textwidth]{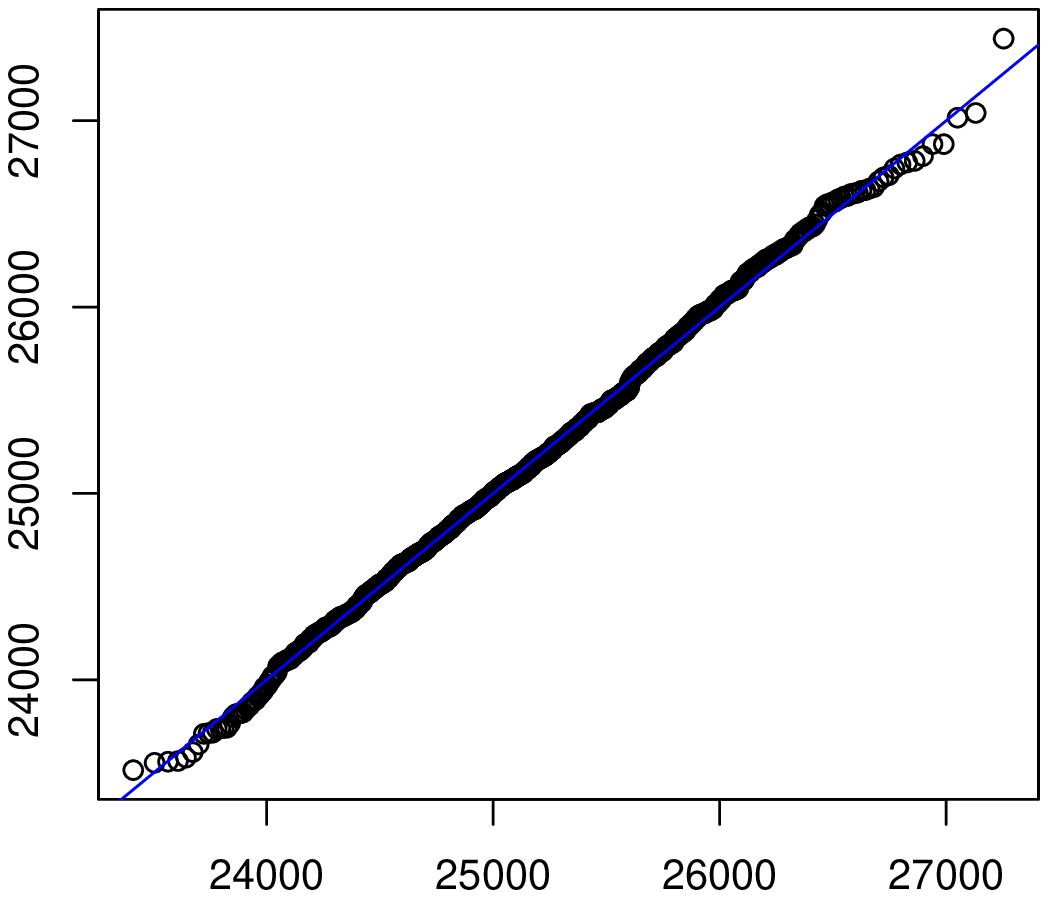}
  \includegraphics[width=0.24\textwidth]{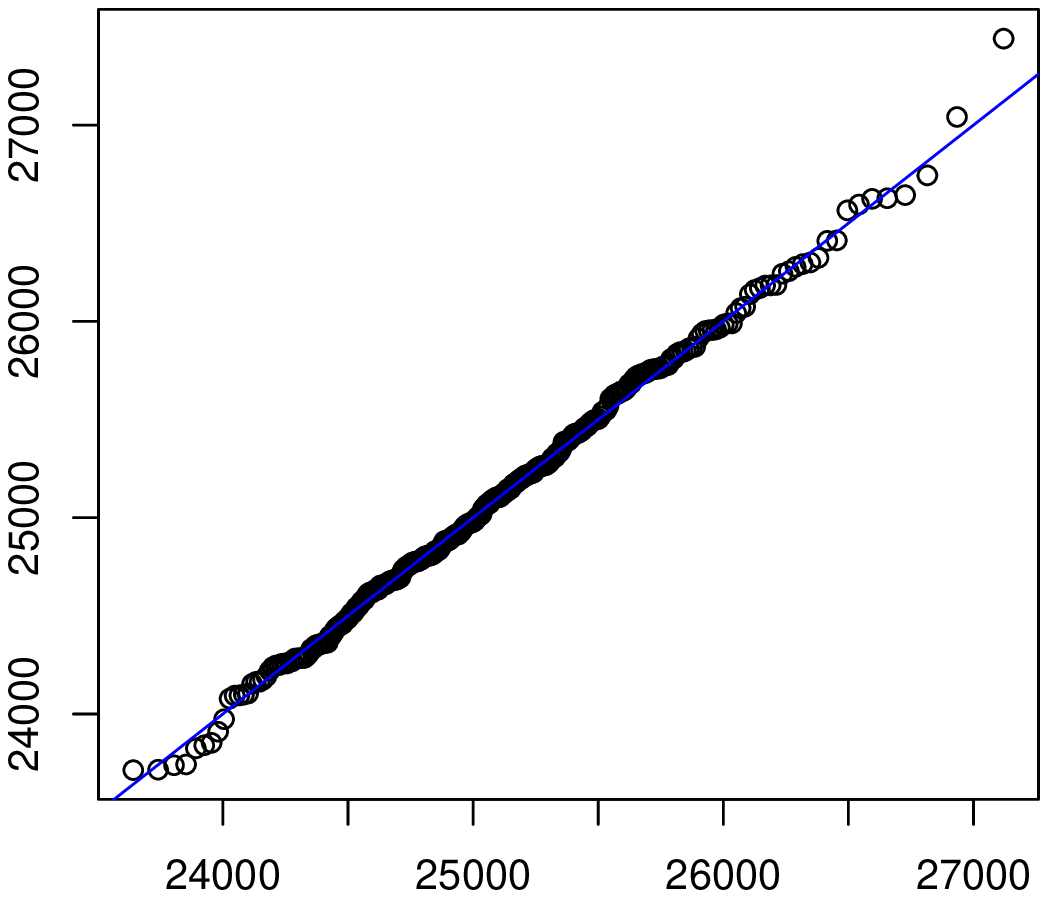}
  \includegraphics[width=0.24\textwidth]{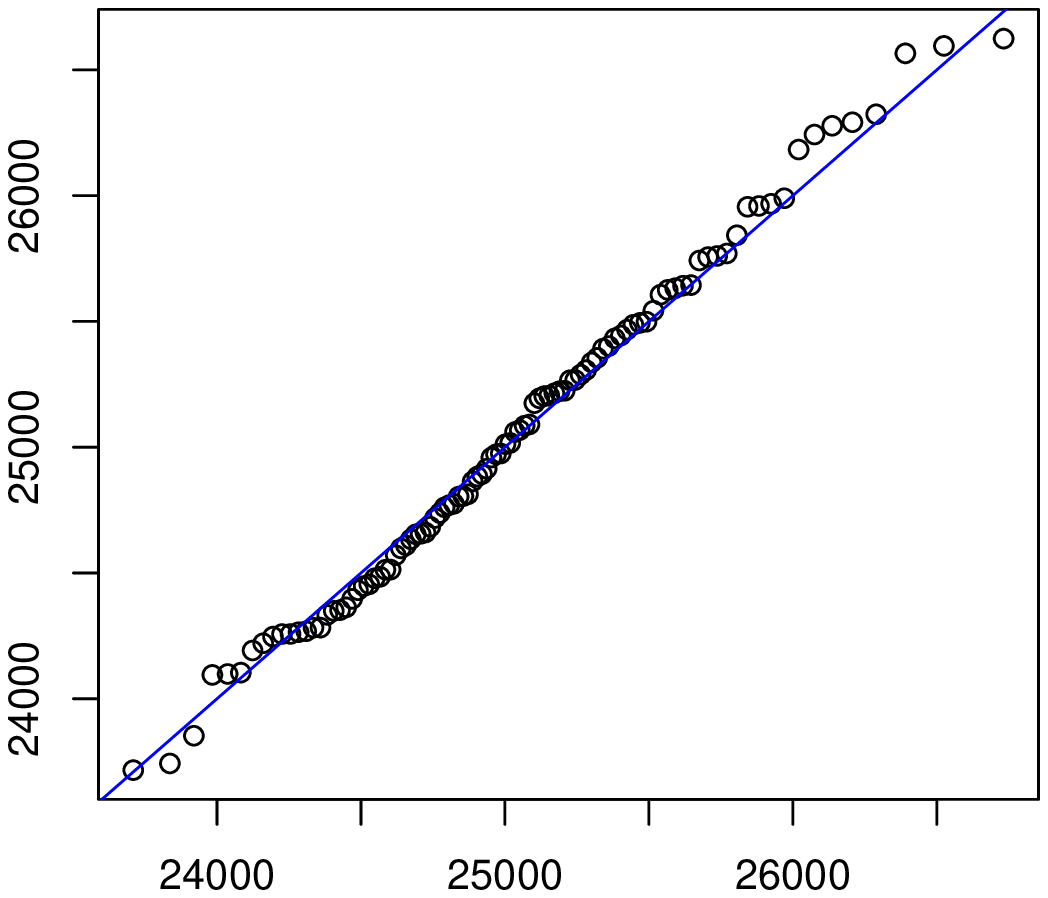}
  \includegraphics[width=0.24\textwidth]{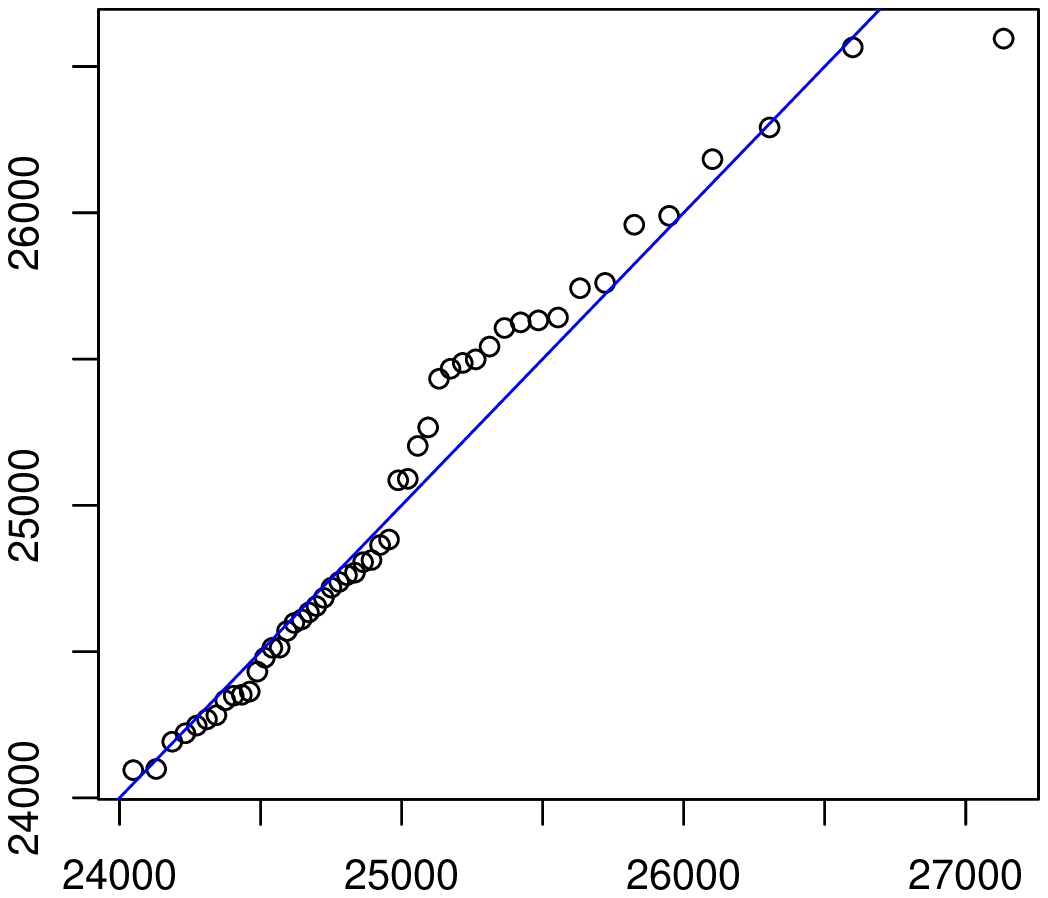}\\

  \includegraphics[width=0.24\textwidth]{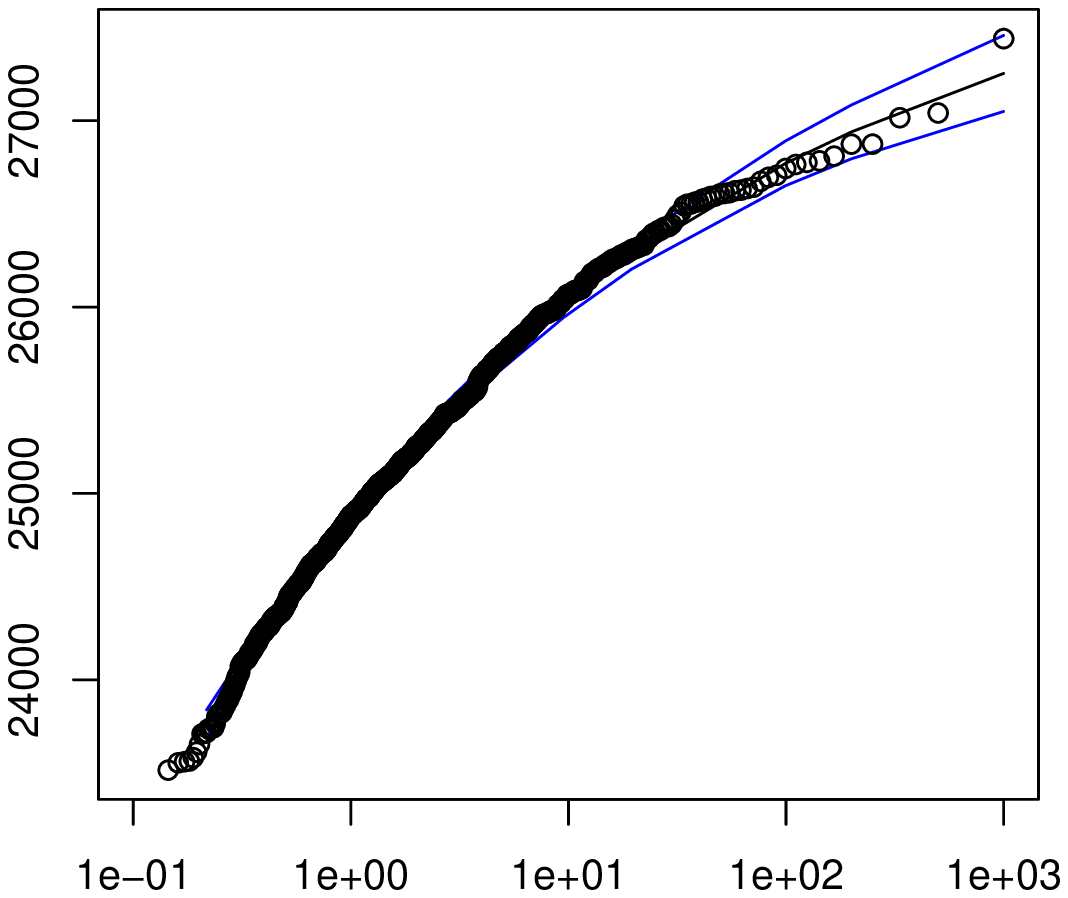}
  \includegraphics[width=0.24\textwidth]{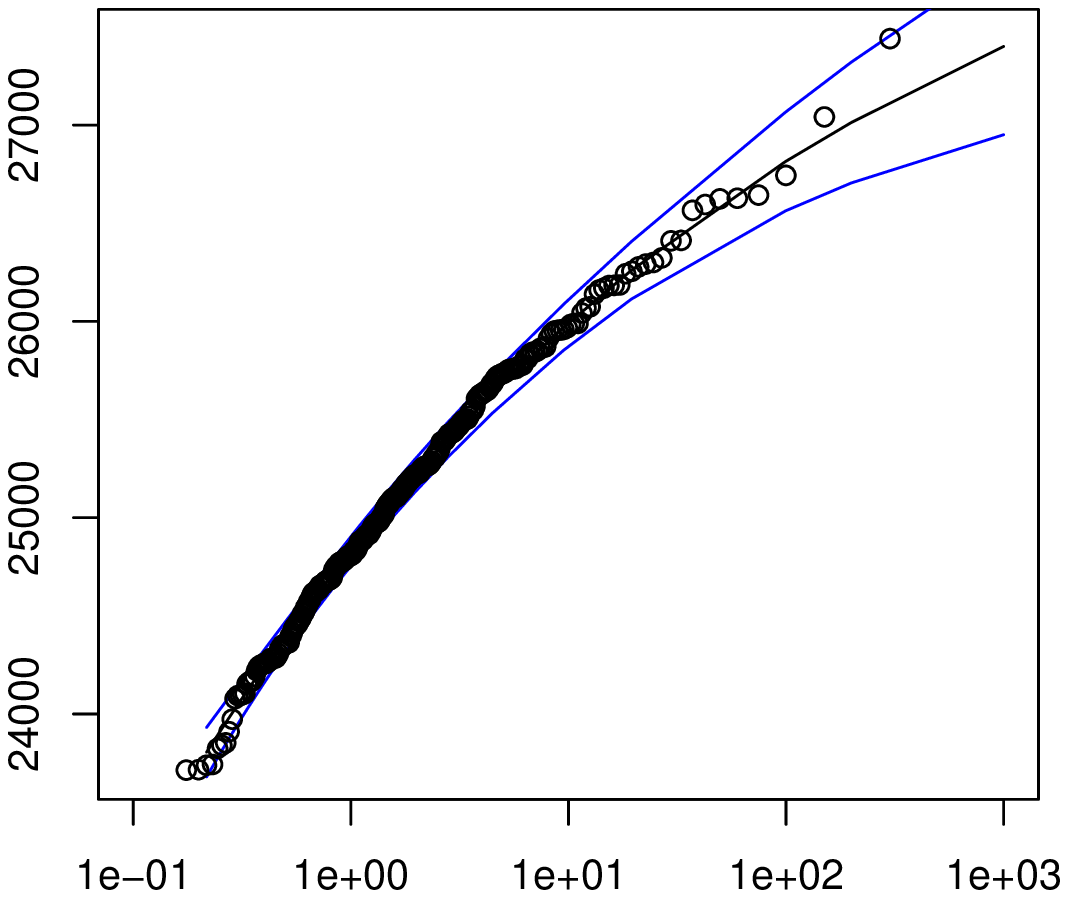}
  \includegraphics[width=0.24\textwidth]{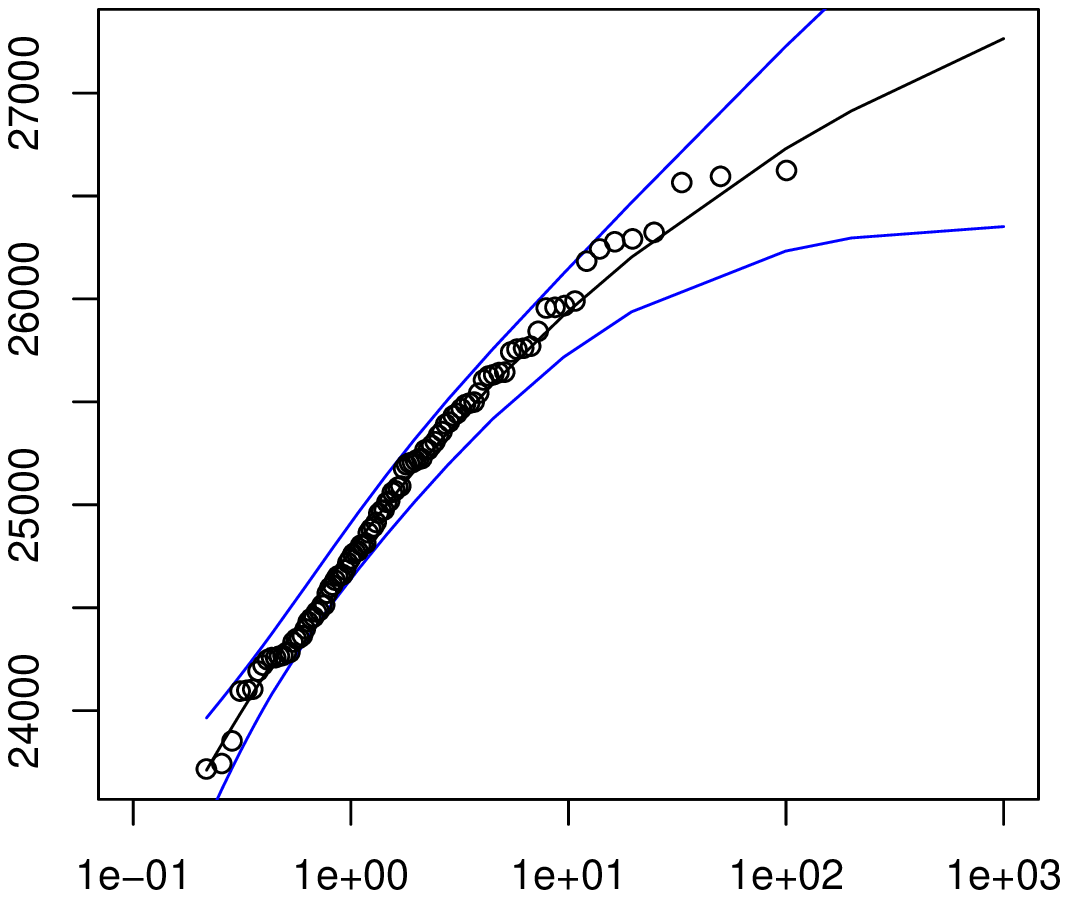}
  \includegraphics[width=0.24\textwidth]{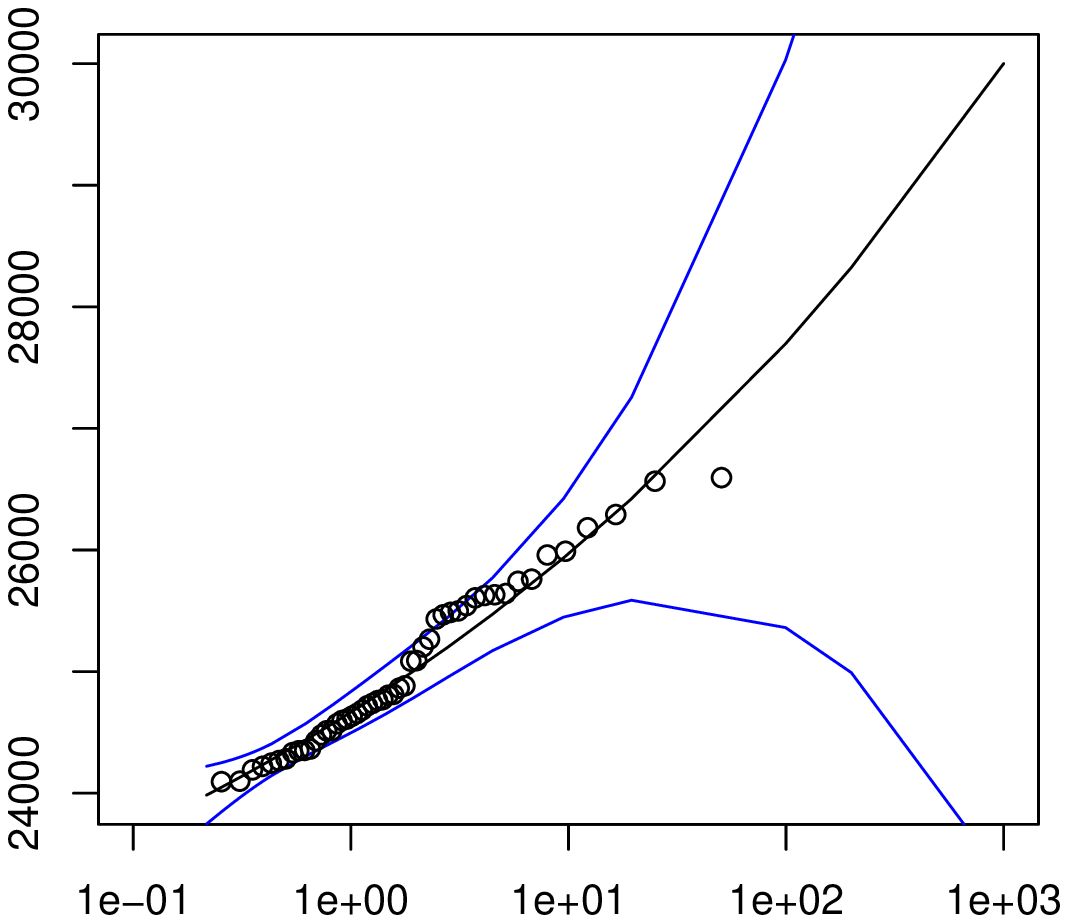}
  \caption{
    Diagnostic plots of the GEV inferences for $T_E=32$.
    Top and bottom row: quantile-quantile and return level plots,
    respectively (see \secref{GEVmethods} for definitions).
    From left to right column: sequences of yearly maxima
    of the total energy are used, having lengths
    1000, 300, 100, and 50, respectively.
    Notice the different scale of the vertical axis in
    the rightmost return level plot.
  }
  \label{fig:diagnostat}
\end{figure}

Quantile-quantile and return level plots for the above inferences
are reported in \figref{diagnostat}. These confirm excellent
quality for 1000, 300, and 100 maxima, whereas they reveal that
something must be wrong for 50. In the quantile-quantile plots
(top row of \figref{diagnostat}), from left to right there appear
increasing departures from the diagonal in the tails, especially
the upper tail, whereas the central part of the distribution does
not suffer from sample reduction, except in the case of 50 maxima.
Analogous effects occur in the upper tail of the return level
plots. The main point here is that the most delicate part of an
extreme value inference is the behavior of the tails. Usually,
this is also the aspect one is most interested in.
Notice how the black line in the middle of the return level plot
for 50 maxima erroneously suggests unboundedness of the return
levels (which is only possible for $\xi\ge0$, see
\secref{GEVmethods}). Therefore, extrapolations to high levels
should be avoided in this case.

We emphasize that the value of $T_E$ just examined corresponds to
a particularly \emph{bad} inference for 50 years. An overview,
throughout the considered range of $T_E$, of GEV inference
sensitivity to length reduction is summarized in
\figref{comparelen}, where the cases of 300, 100, 50 yearly maxima
are plotted against 1000. The quality of the fits, of course,
generally decreases when using shorter series of maxima. Inference
of $\xi$ is particularly sensitive to the length of the series of
maxima: the maximal value of the ratios between uncertainty in $\xi$
and value of the corresponding maximum likelihood estimate of $\xi$
is $600\%$, $1387\%$, $45\%$, and $21\%$,
for 50, 100, 300, and 1000 maxima, respectively.
The median of those ratios is $48\%$, $30\%$, $19\%$, $10\%$,
respectively.
Taking only 50 maxima yields two positive estimates of $\xi$ (for
$T_E=32$ and $50$), which is an unphysical result, and overall
very large uncertainties: for many values of $T_E$, confidence
bands for $\xi$ include part of the positive axis. The bias in the
estimates of $\xi$ induces a significant alteration in those of
$\sigma$, although the inferred values of $\mu$ remain quite
stable. Also notice the big uncertainties in the return levels for
the two cases $T_E=32$ and $50$, corresponding to positive
estimates of $\xi$.

If we further consider additional sources of difficulty present in
nature (for example the annual seasonal modulation), skepticism with
respect to several inferences on extremes proposed in meteo-climatic
literature seems justified.

\subsection{Sensitivity with Respect to the Extreme Events Selection Procedure: Soft Extremes}

We now turn to the second type of inference sensitivity mentioned
above, obtained by using so-called \emph{soft extremes}~\cite{KtK03}
instead of~\emph{genuine} extremes. In the present setting,
we simulate the usage of \emph{soft extremes} by
considering sequences of maxima over data blocks that correspond
to time spans shorter than one year, in particular 0.6, 1.2, and 3
months. In the first two cases, and especially in the first, we
are not even sure that the considered maxima are effectively
uncorrelated, which is the typical situation in real systems. In
each case, the number of considered extremes is kept fixed to
1000, so that the difference is only determined by block length.

\begin{figure}[p]
  \centering
  \includegraphics[width=0.24\textwidth]{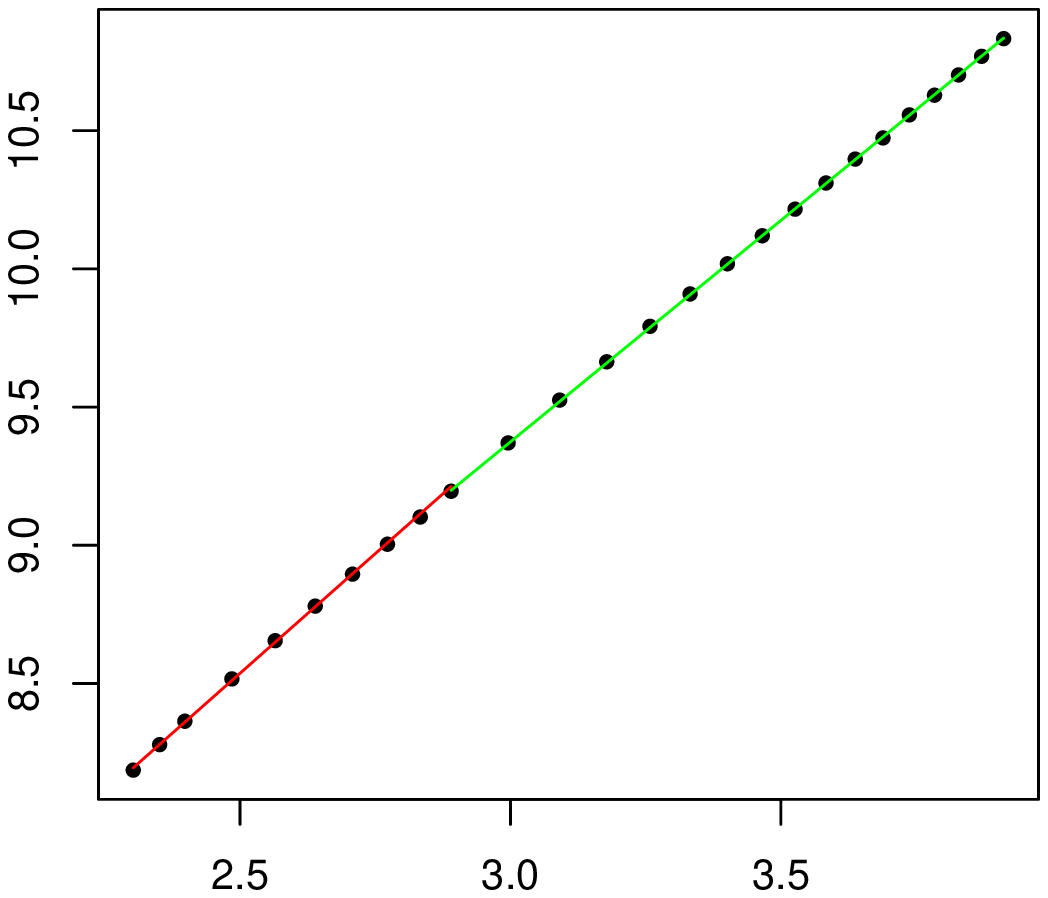}
  \includegraphics[width=0.24\textwidth]{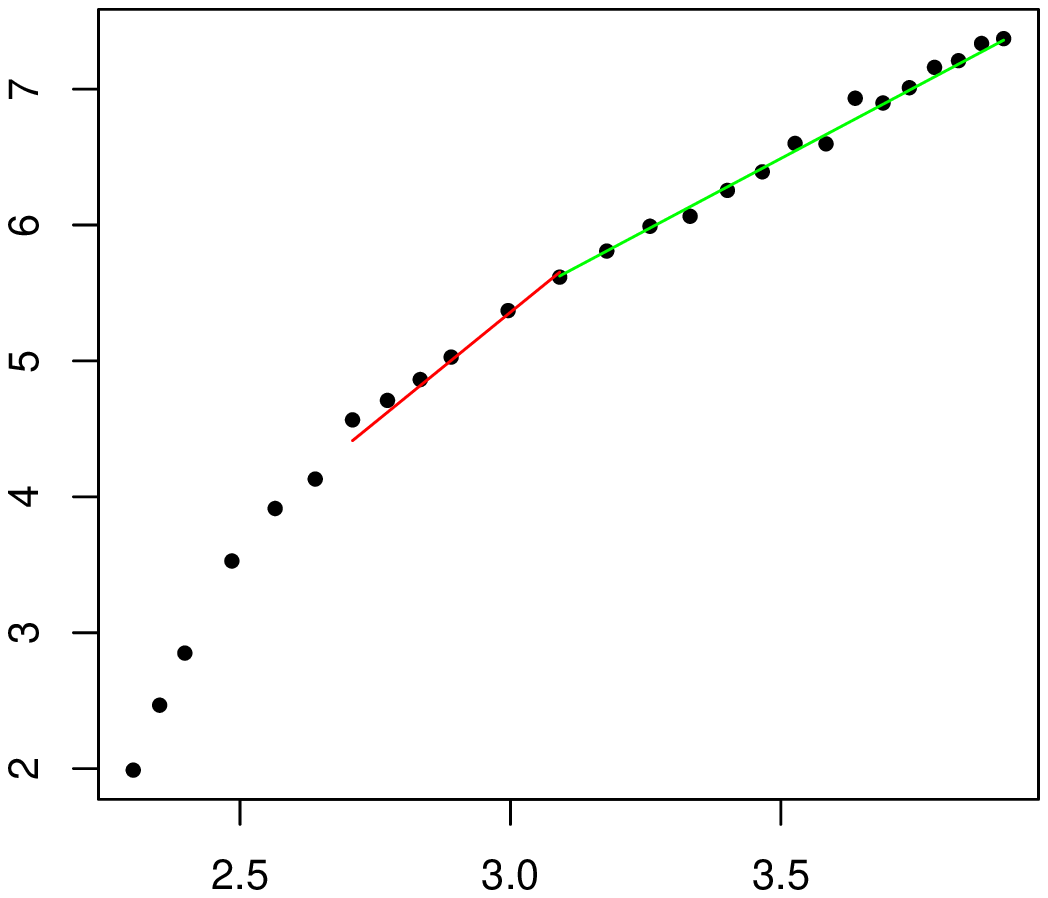}
  \includegraphics[width=0.24\textwidth]{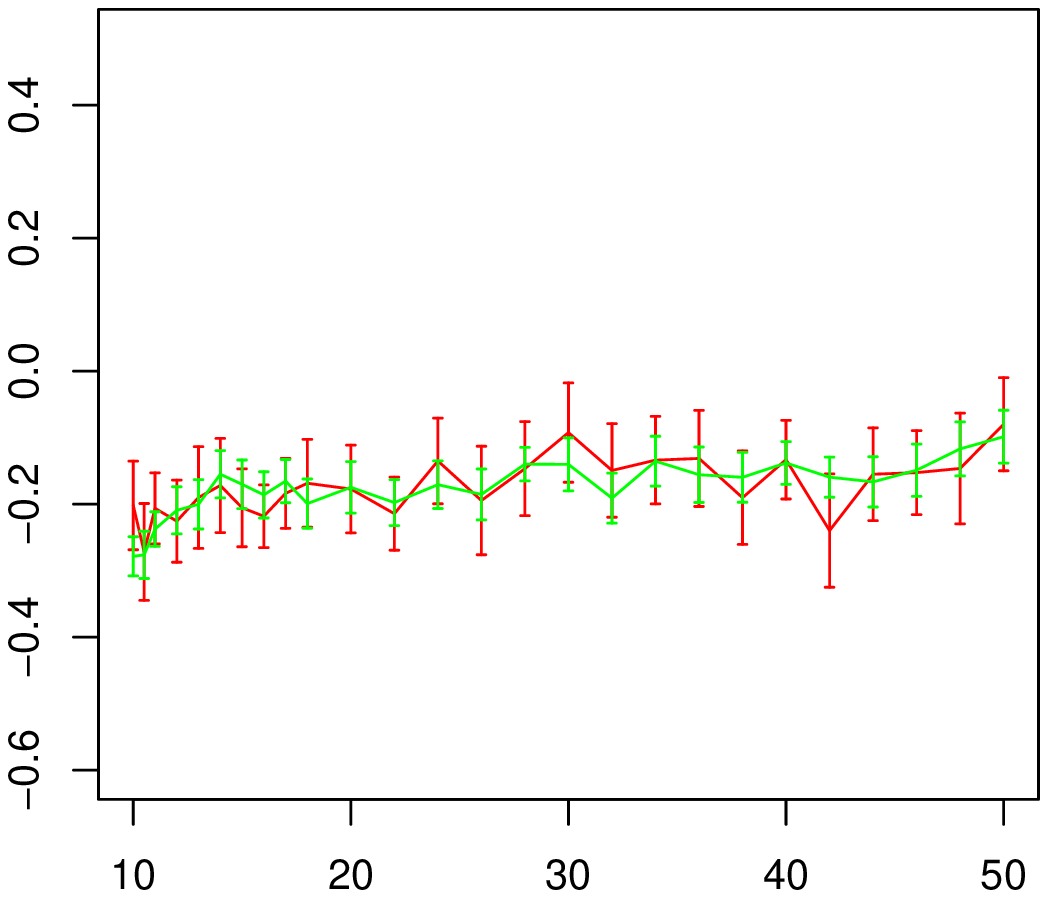}
  \includegraphics[width=0.24\textwidth]{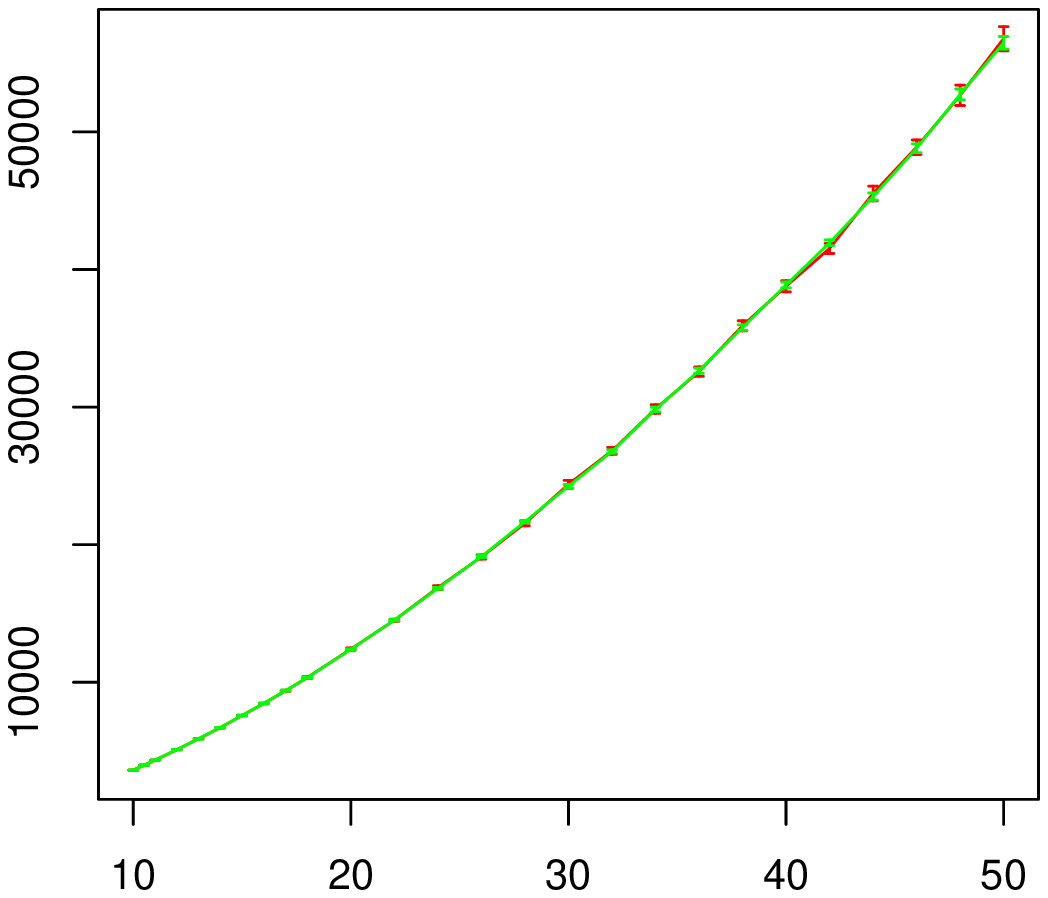}\\

  \includegraphics[width=0.24\textwidth]{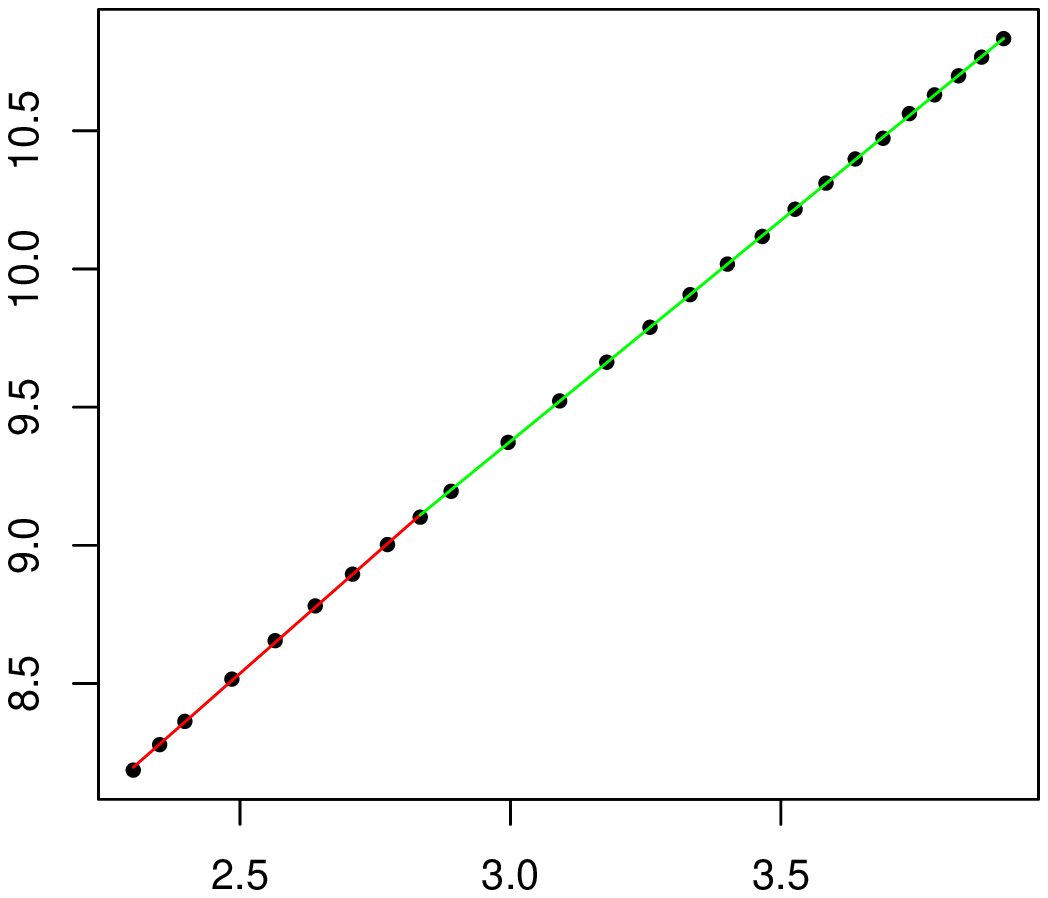}
  \includegraphics[width=0.24\textwidth]{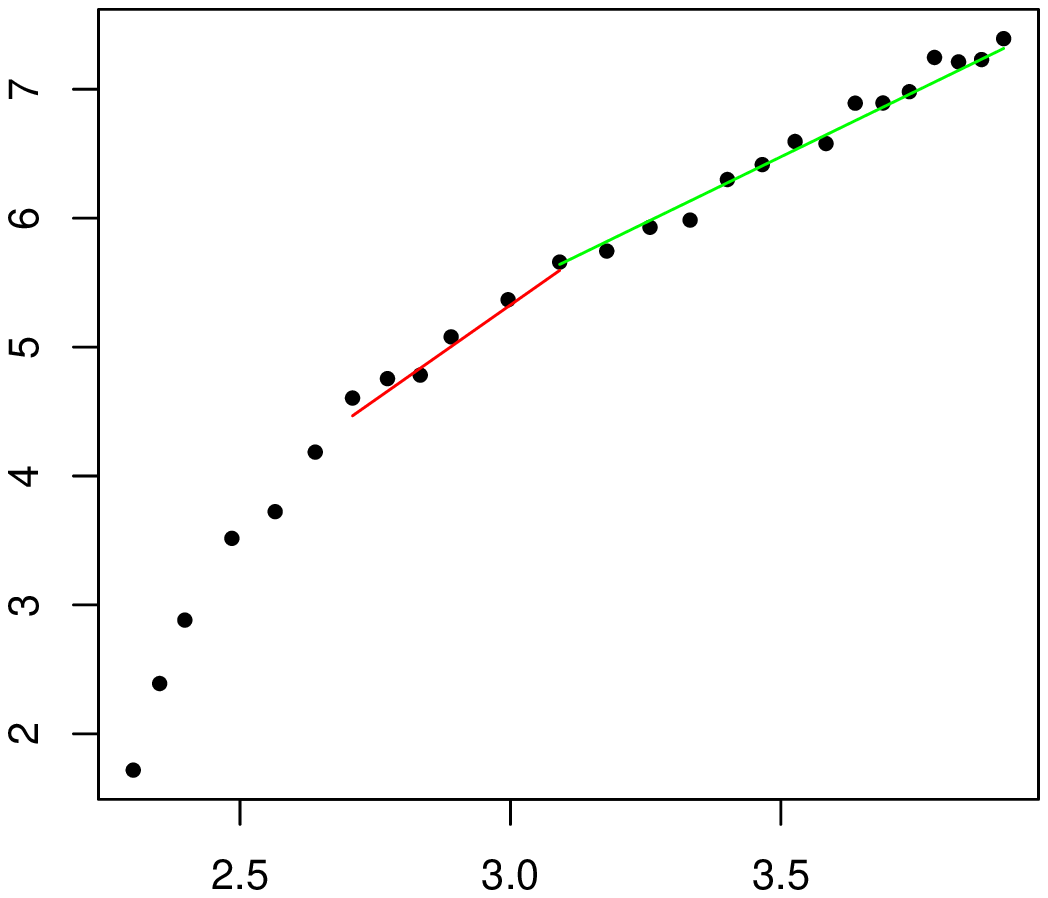}
  \includegraphics[width=0.24\textwidth]{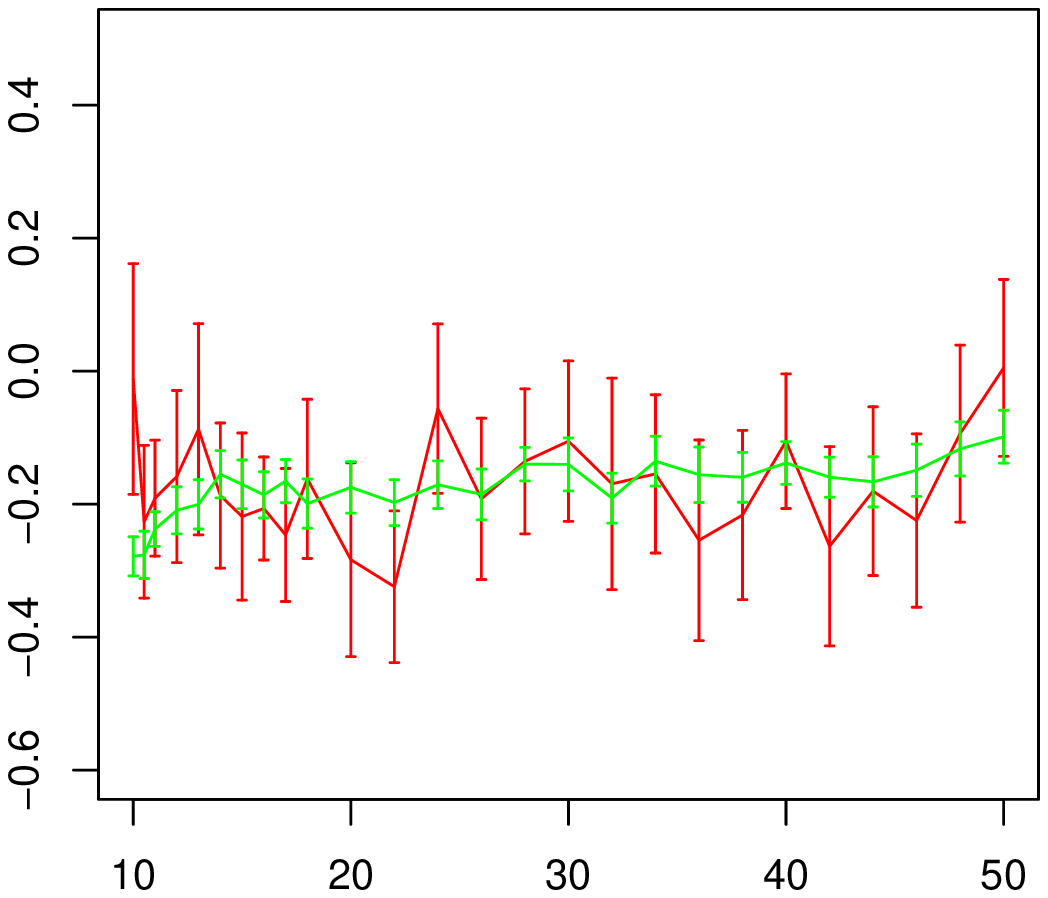}
  \includegraphics[width=0.24\textwidth]{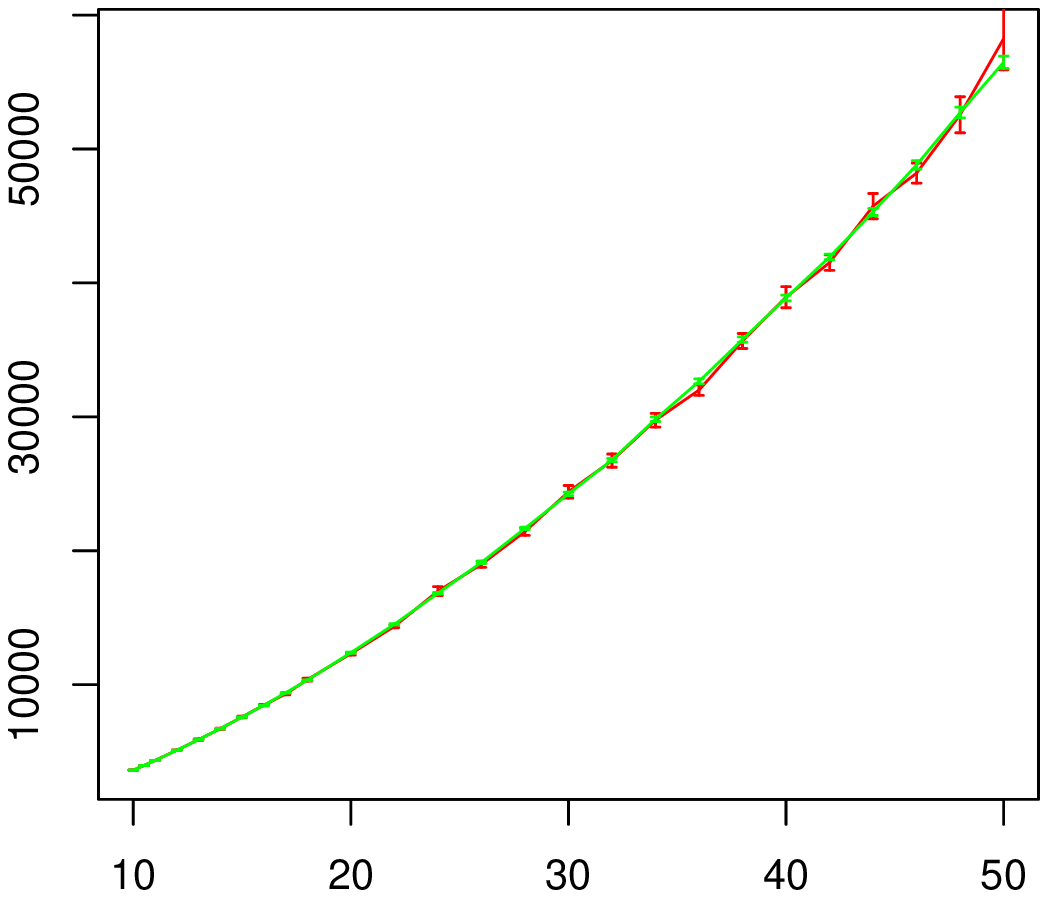}\\

  \includegraphics[width=0.24\textwidth]{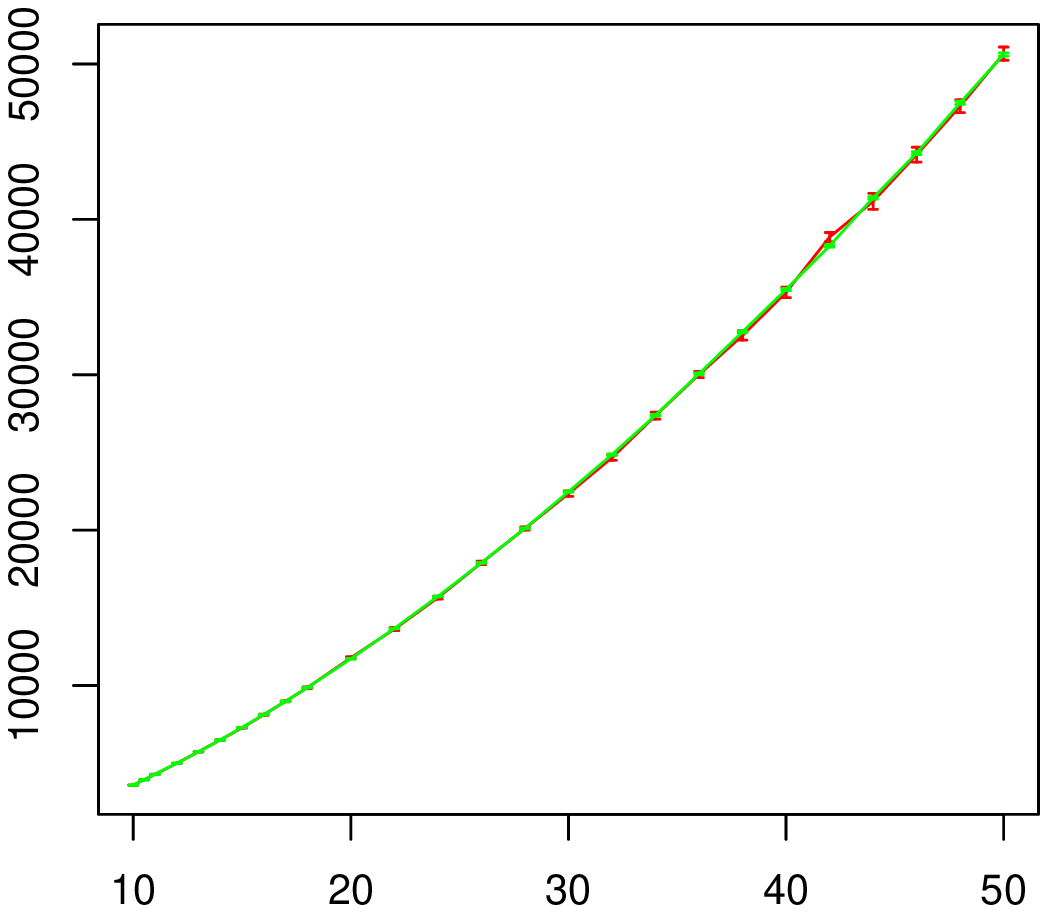}
  \includegraphics[width=0.24\textwidth]{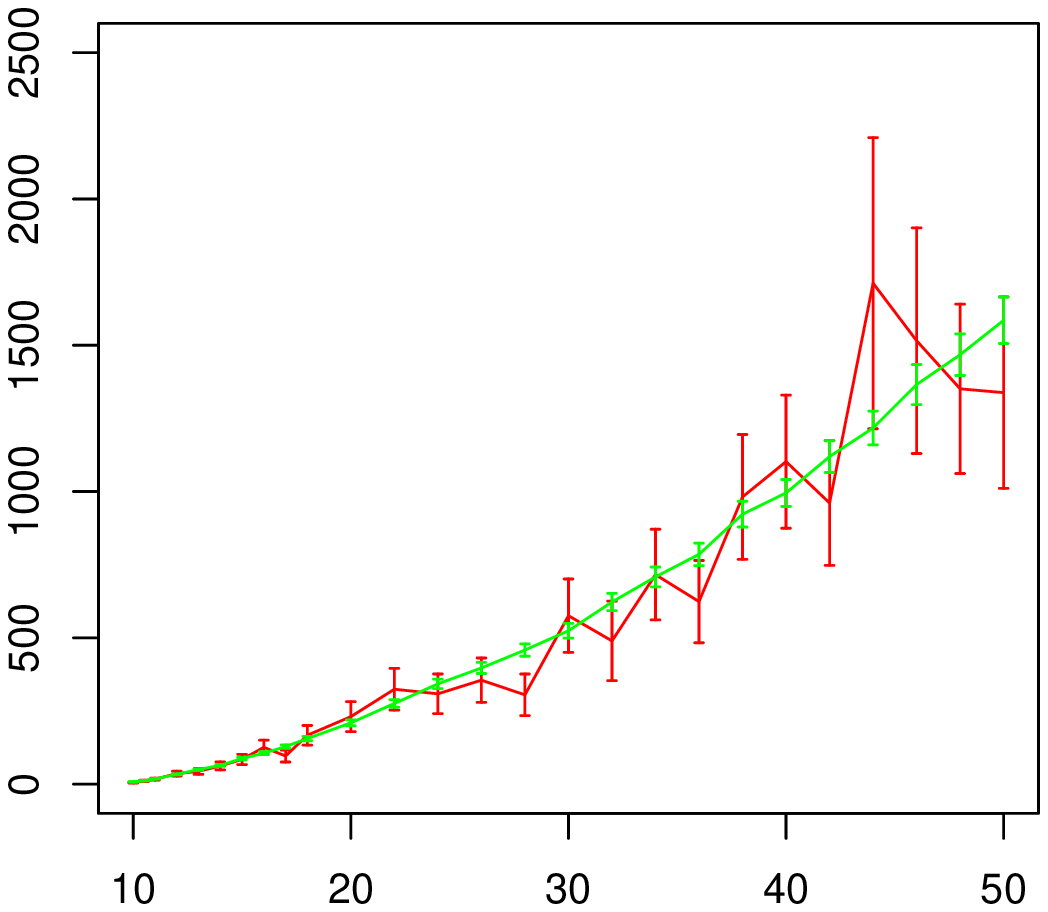}
  \includegraphics[width=0.24\textwidth]{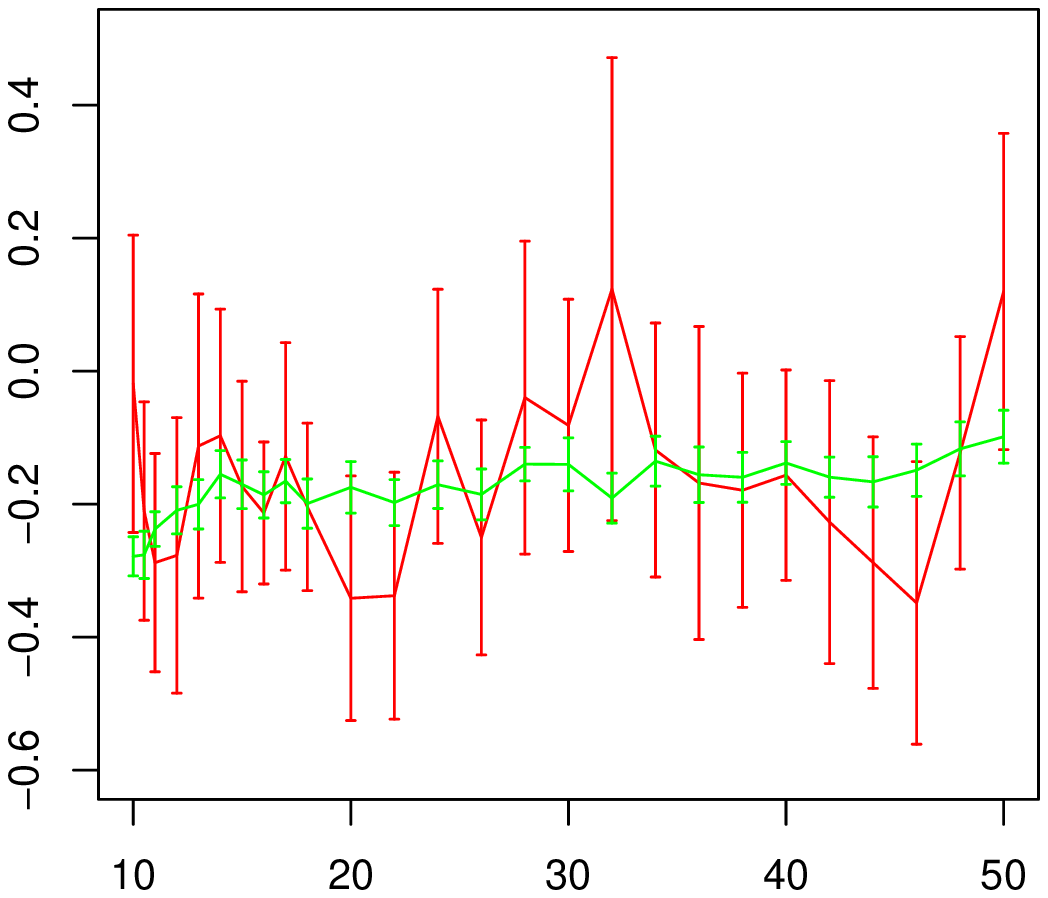}
  \includegraphics[width=0.24\textwidth]{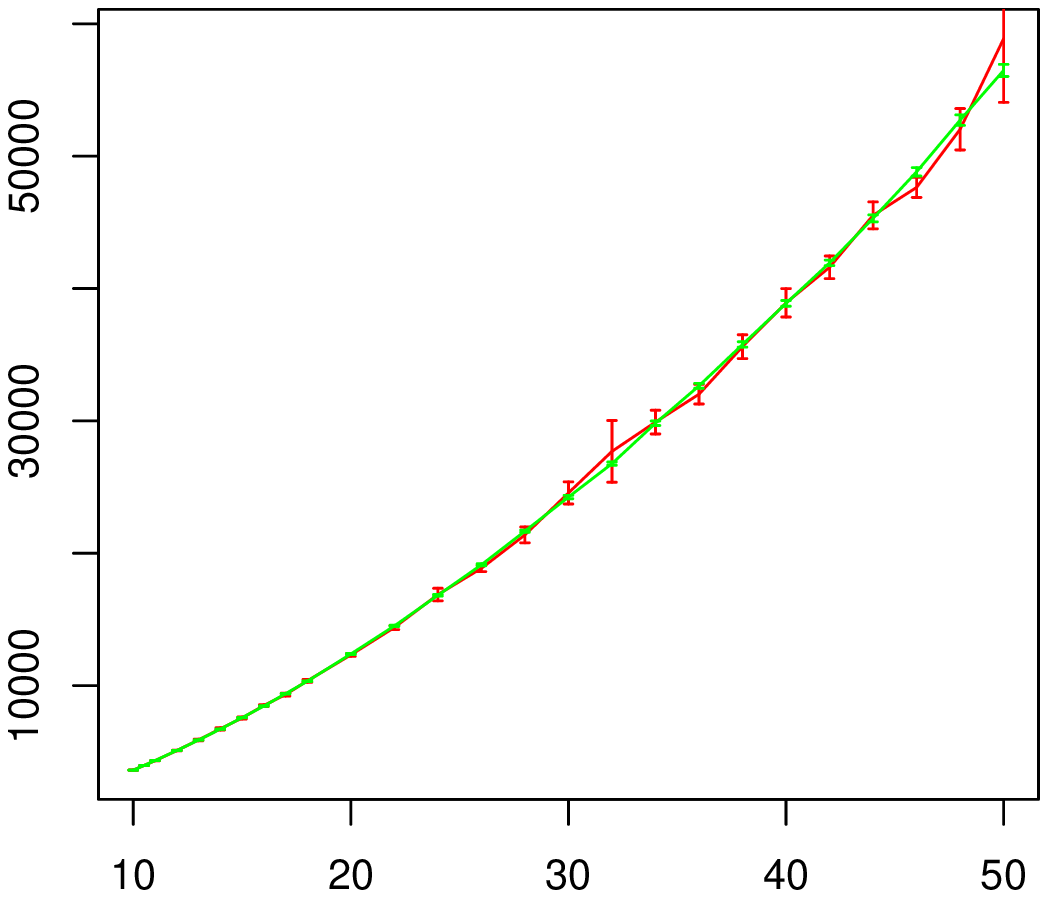}\\
  \caption{
    Top row:
    maximum likelihood estimates of $\mu$, $\sigma$, and $\xi$
    (from left to right, respectively) for 1000 and 300 yearly maxima
    (green and red respectively), with confidence intervals
    computed by the observed information matrix~\eqref{observedinfo}.
    Center, bottom row:
    same as top, for 100 and 50 yearly maxima, respectively,
    instead of 300.
    In the case of 50 maxima, for $T_E=32$ and $50$
    the inferred values of $\xi$ are positive
    (thus completely wrong according to the theoretical
    expectation, see text) and the uncertainties are very large
    for $\sigma$ and $\xi$.
  }
  \label{fig:comparelen}
\end{figure}

\begin{figure}[p]
  \centering
  \includegraphics[width=0.24\textwidth]{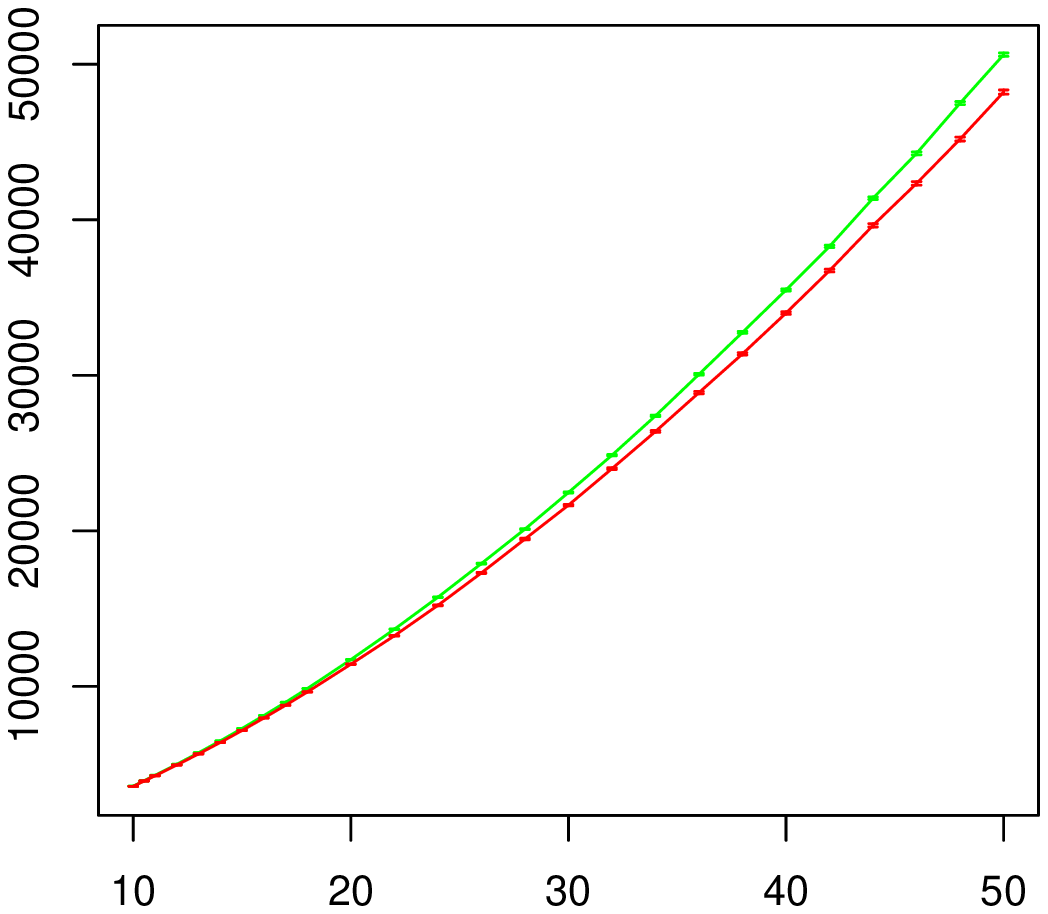}
  \includegraphics[width=0.24\textwidth]{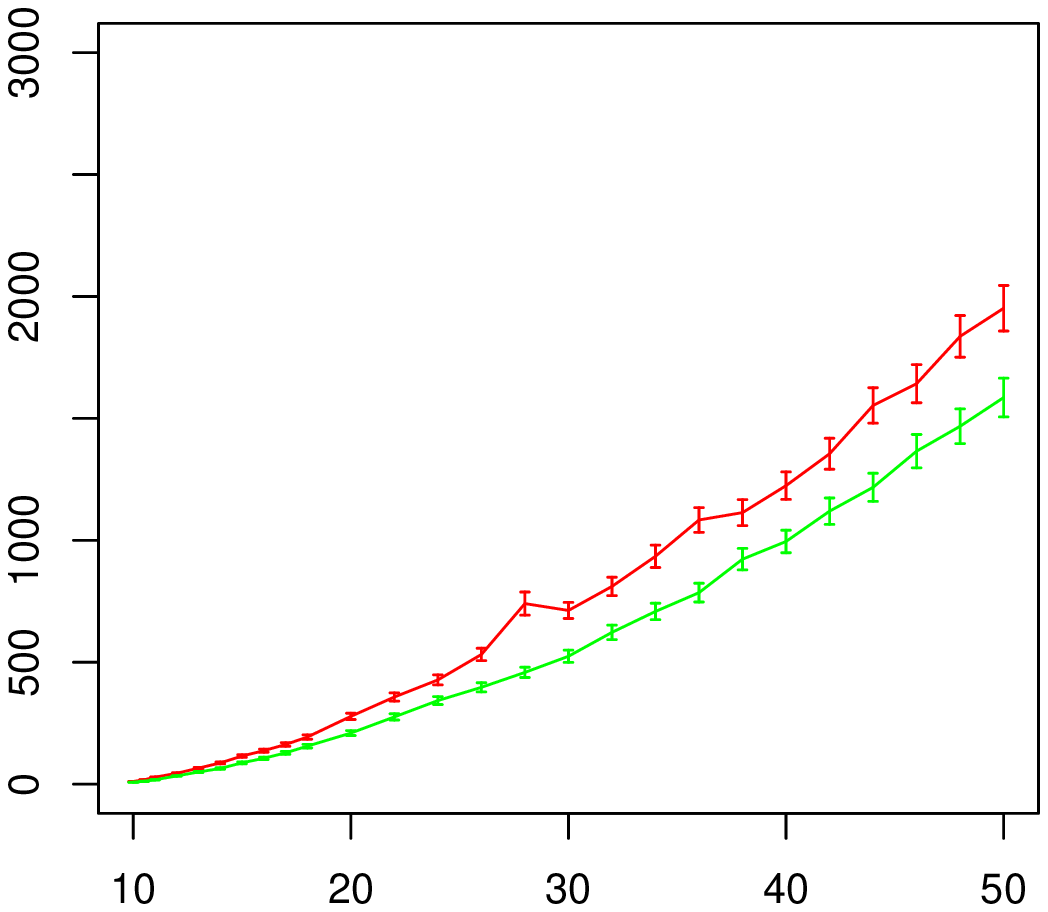}
  \includegraphics[width=0.24\textwidth]{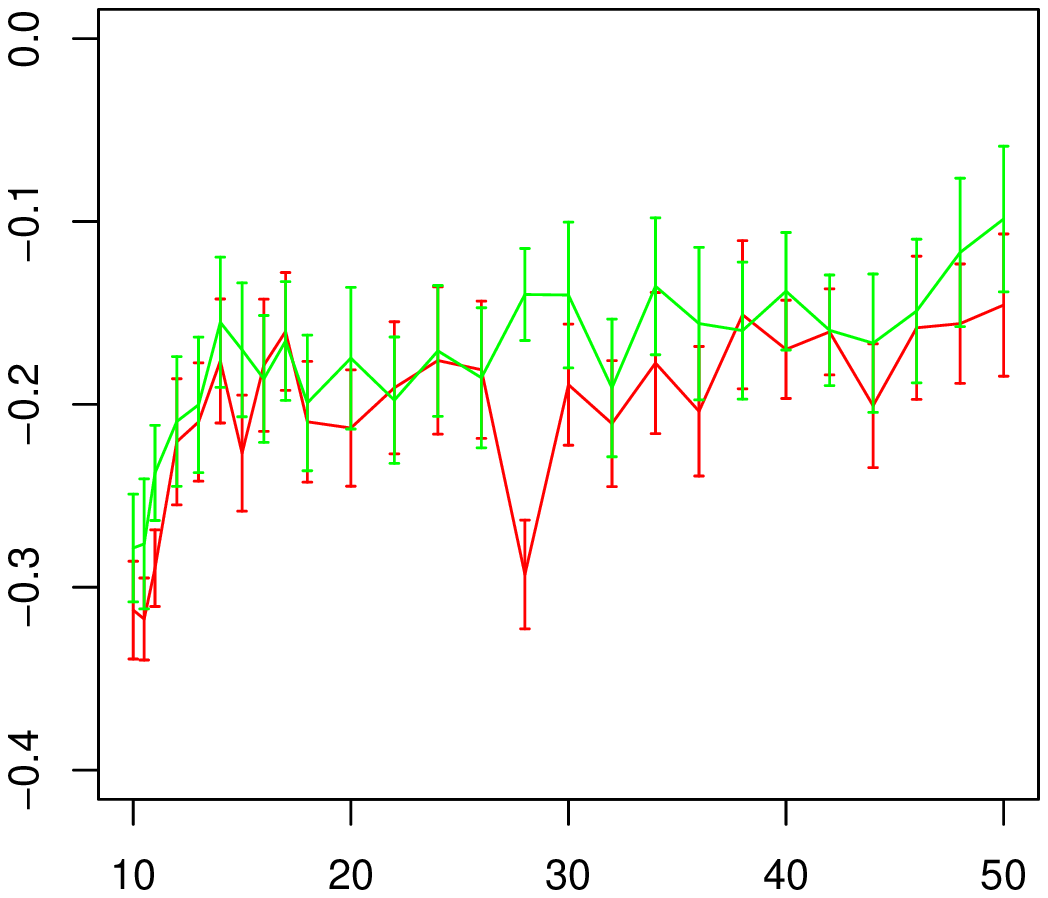}
  \includegraphics[width=0.24\textwidth]{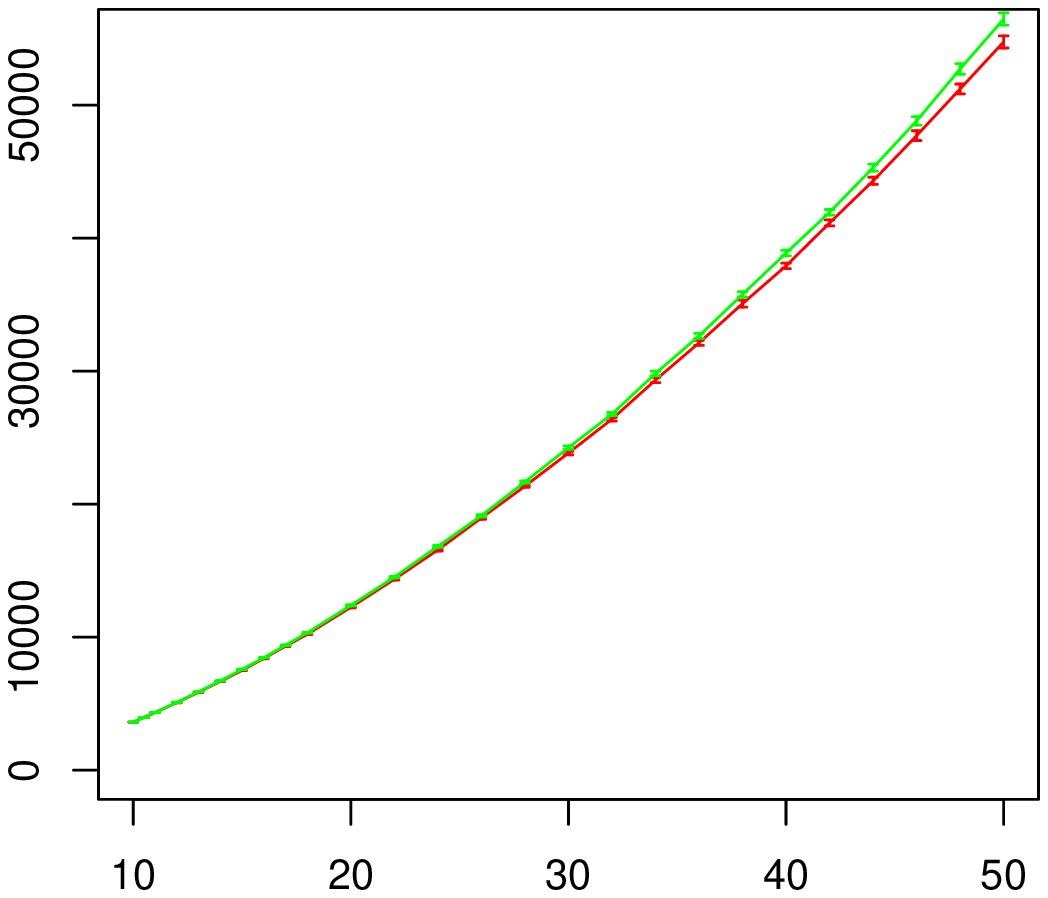}\\

  \includegraphics[width=0.24\textwidth]{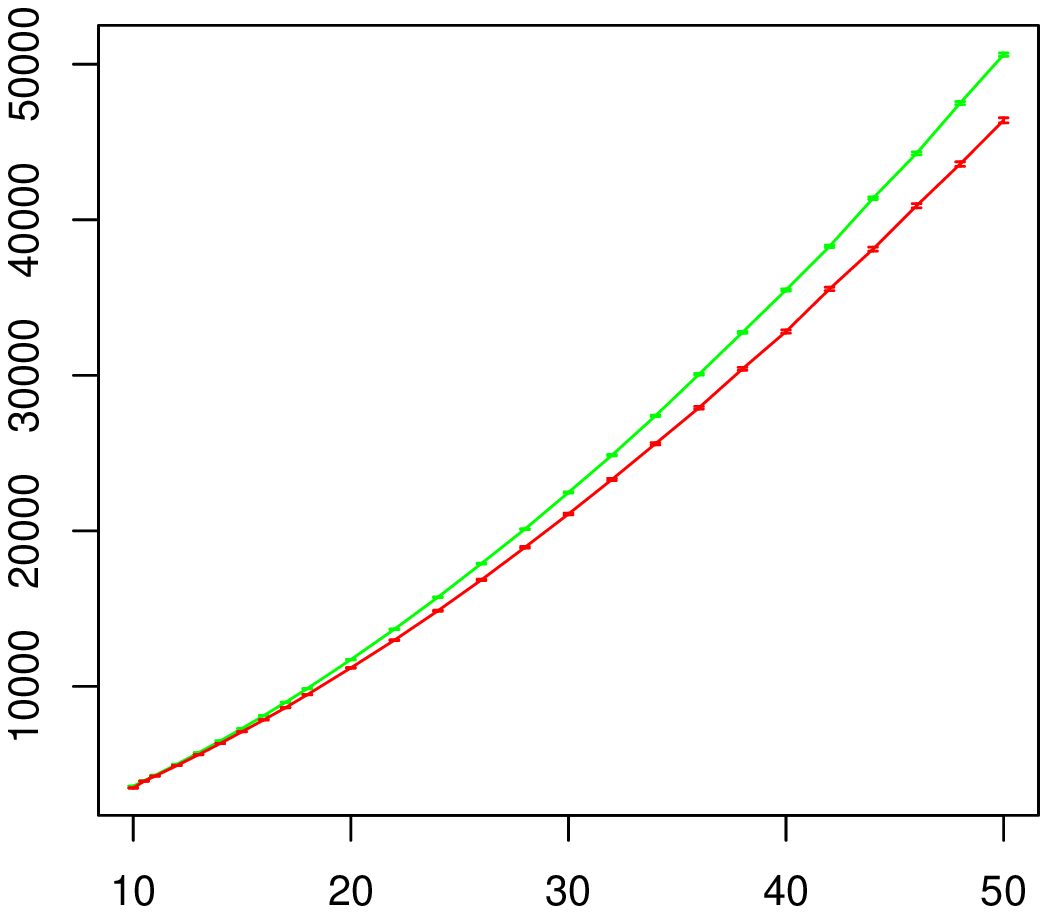}
  \includegraphics[width=0.24\textwidth]{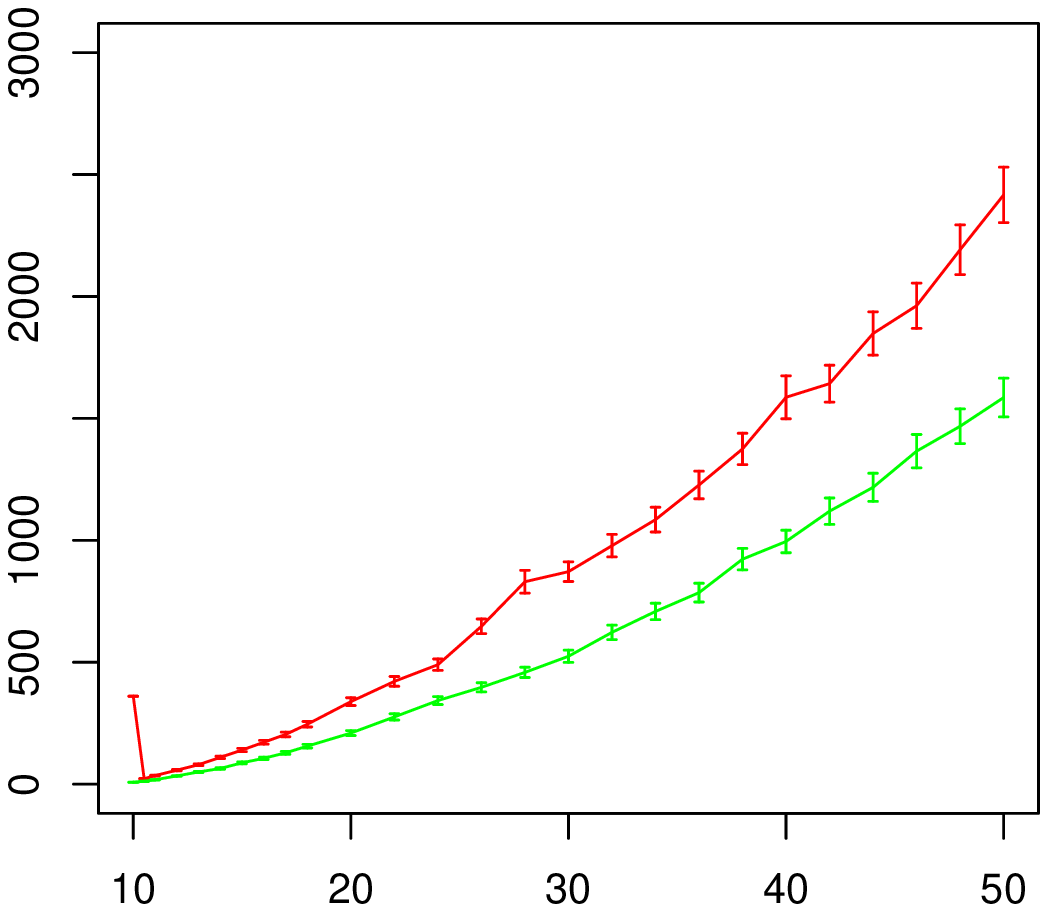}
  \includegraphics[width=0.24\textwidth]{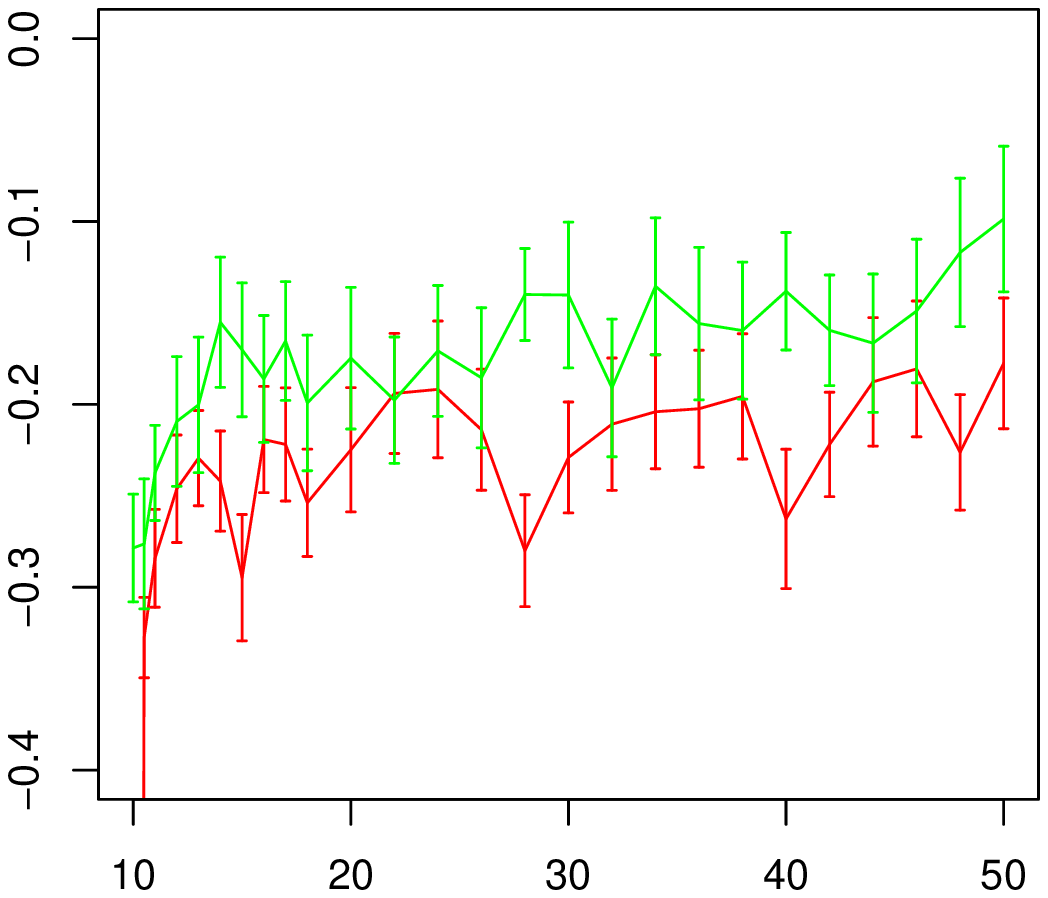}
  \includegraphics[width=0.24\textwidth]{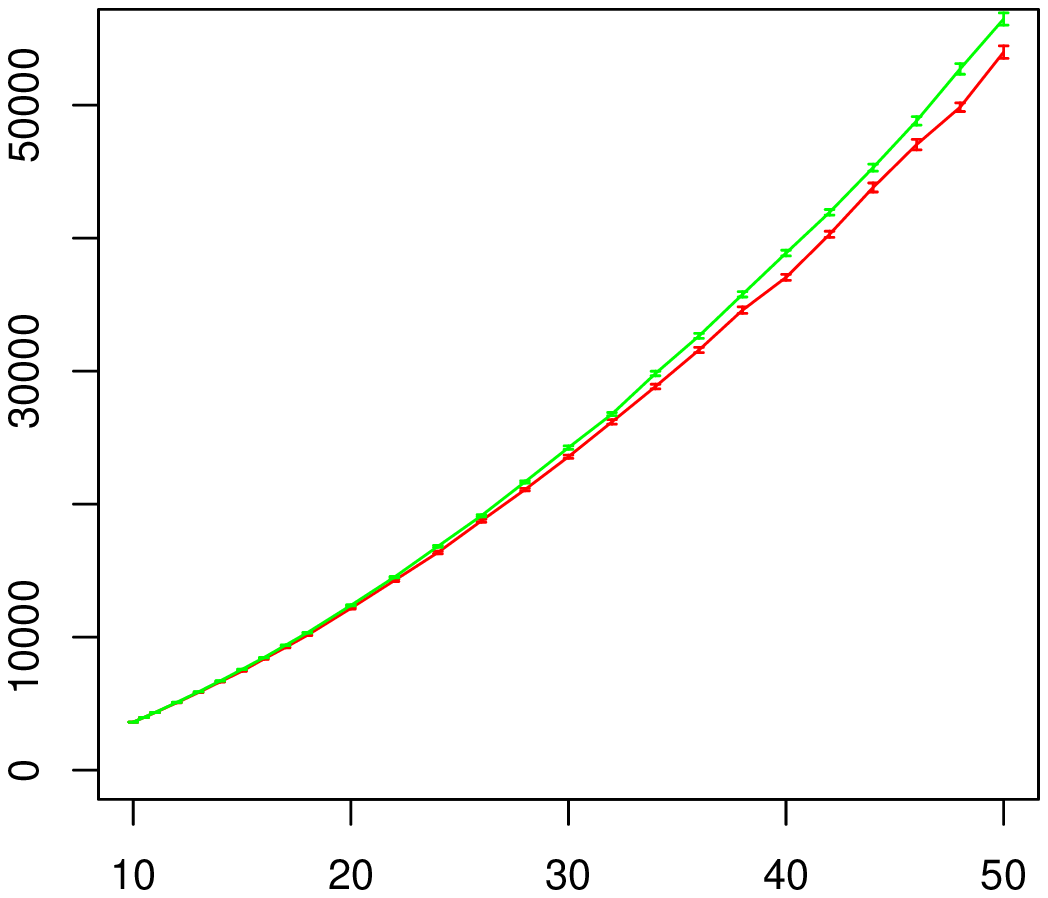}\\

  \includegraphics[width=0.24\textwidth]{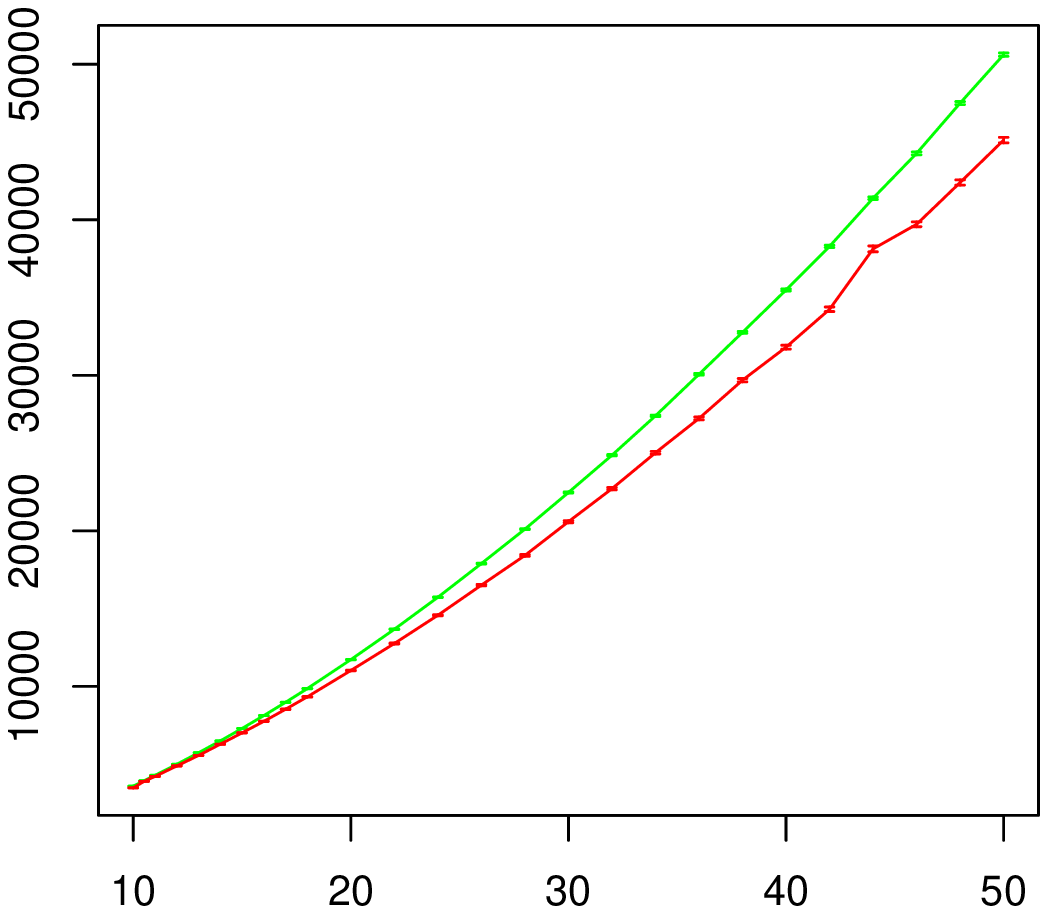}
  \includegraphics[width=0.24\textwidth]{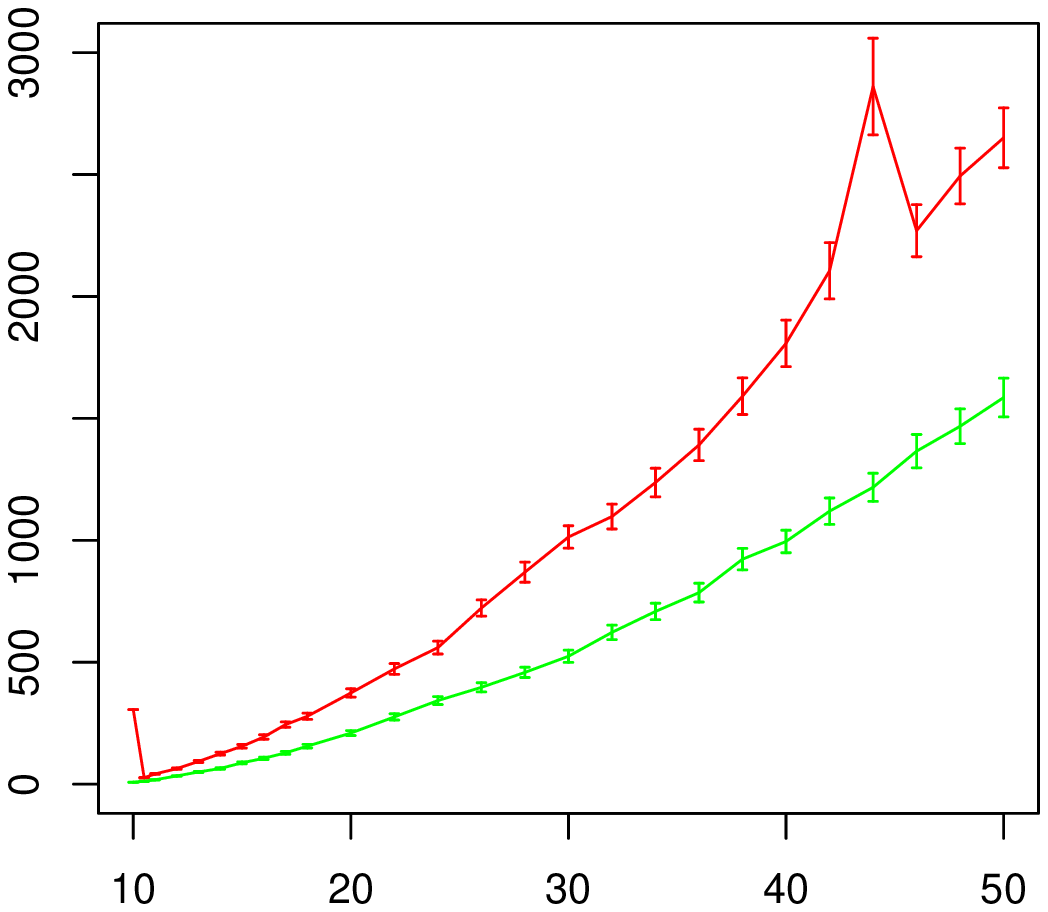}
  \includegraphics[width=0.24\textwidth]{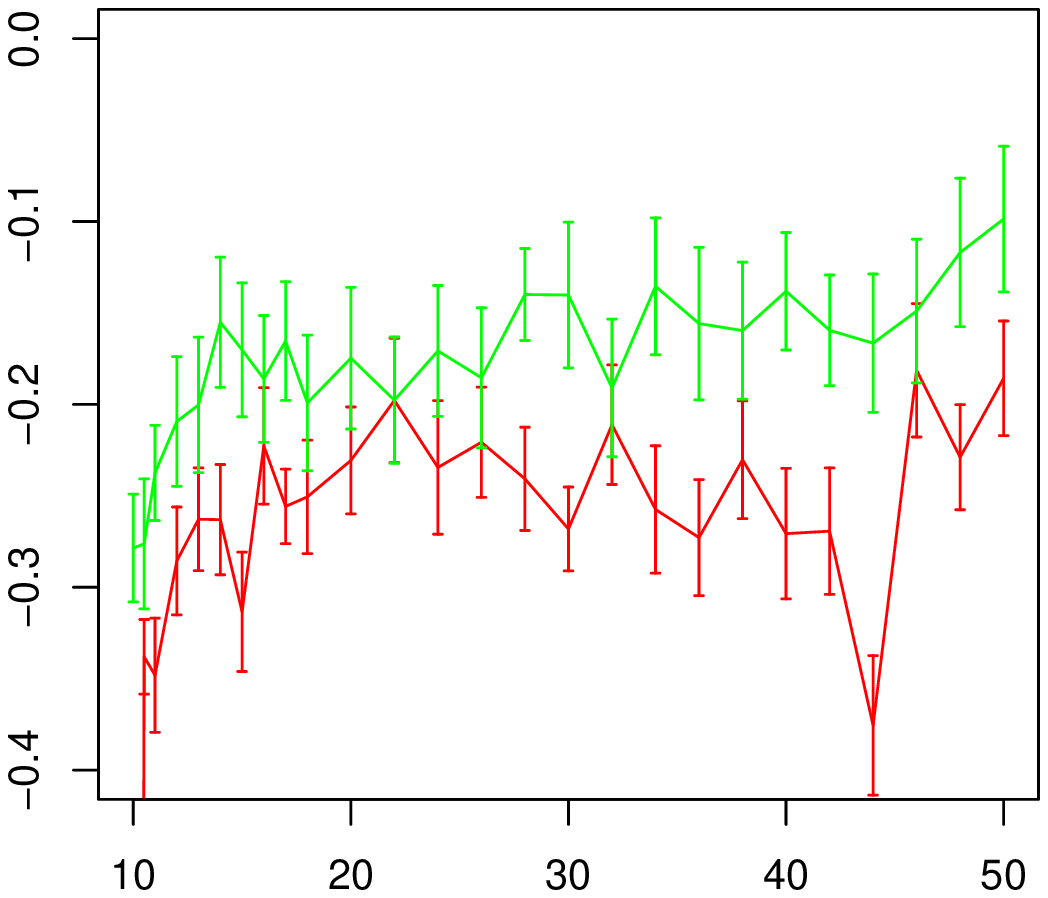}
  \includegraphics[width=0.24\textwidth]{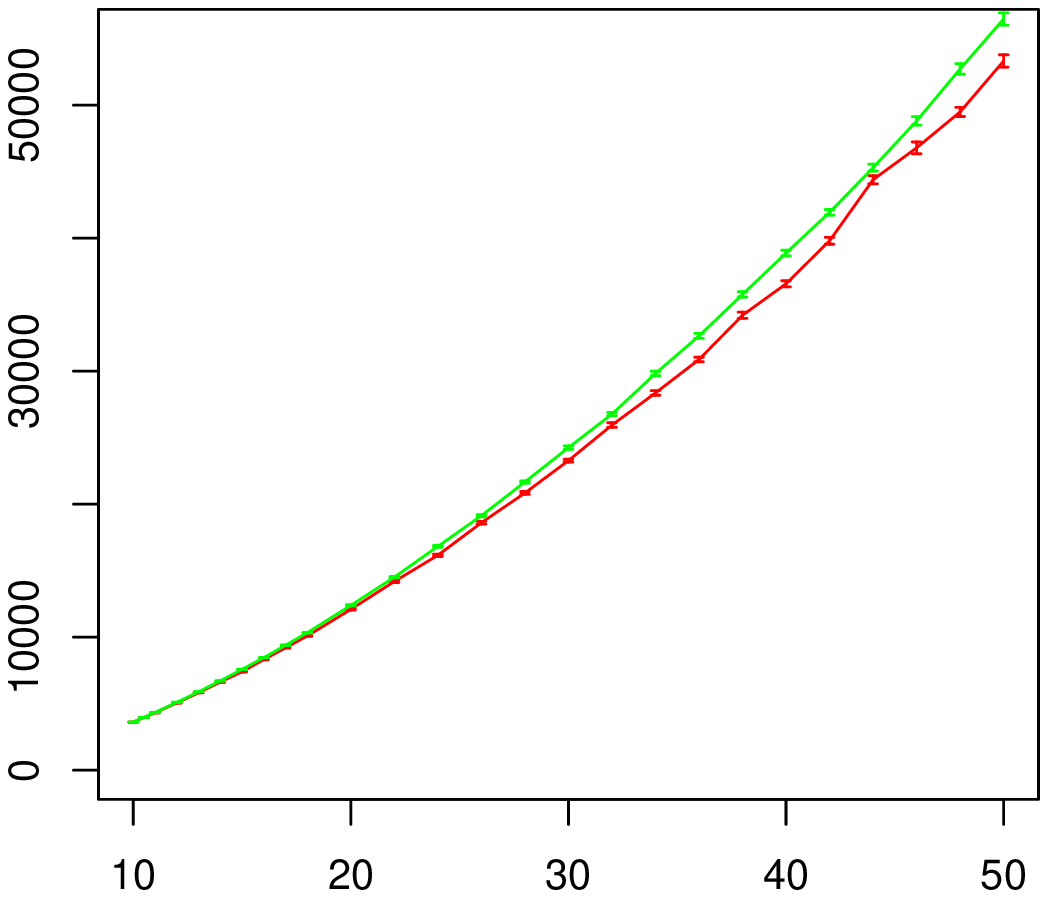}\\
  \caption{
    Inferred values of GEV parameters as a function of $T_E$
    (horizontal axis) in a \textit{soft extremes} experiment:
    from top to bottom row, sequences of 1000 maxima of the
    total energy time series are used,
    where the maxima are determined over data blocks corresponding
    to 3, 1.2, and 0.6 months.
    From left to right column, $\mu$, $\sigma$, $\xi$,
    and 100-year return levels are plotted.
    In green the estimates obtained for the yearly maxima
    (as in \figref{stationaryGEV}) are displayed for reference.
    Notice how the magnitude of the uncertainties shows little
    dependence on the temporal block length.
  }
  \label{fig:softextremes}
\end{figure}

The net result of using shorter time spans is the introduction
of a progressively larger bias in the GEV inferences.
The location and shape parameters are systematically underestimated.
For the location parameter $\mu$
(leftmost column in \figref{softextremes}) the underestimation
increases when taking maxima over shorter time-spans, but it also
increases with $T_E$. Notice that this is quite different
from the effect of reduction of the number of maxima,
compare \figref{comparelen} (leftmost column).
The sample medians of the relative differences between the estimates
of $\mu$ for 12 months and those for 3, 1.2, and 0.6 months
(where the sample is indexed by the values of $T_E$
for which the estimates are computed) are $3.2\%$, $5.7\%$, and $7.5\%$
for 3, 1.2, and 0.6 months, respectively.
Due to the definition of $z_p$, see~\eqref{returnlevel},
the underestimation of the 100-year return levels
is a consequence of that of $\mu$. 
Also notice that the variations in the return levels connected to
increase in $T_E$ are much larger than those induced by usage of
either soft extremes or shorter data sets, also compare with
\figref{comparelen} (rightmost column).
Conversely, the scale parameter $\sigma$ (second column from left
in \figref{softextremes}) is \emph{largely} overestimated:
the sample median of the relative differences between the estimates
of $\sigma$ are $31\%$, $59\%$, and $82\%$ for 3, 1.2, and 0.6 months,
respectively.
So in our case, taking soft extremes mistakingly suggests
an enhanced variability in the extreme values.

Qualitatively, the response of the GEV estimates to the usage of
\emph{soft extremes} is explained by the introduction of much more
data in the central part and in the lower tail of the distribution
of the selected extreme values.
From this fact, the underestimation of $\mu$ follows directly.
Moreover, since the range of the extreme events distribution gets wider,
a larger variability is \emph{artificially} introduced and this is
indicated by an overestimated scale parameter $\sigma$.
Lastly, the upper tail of the so obtained distribution of extremes
looks more \emph{squeezed}, given the wider extension at lower values.
This corresponds to a more negative value of $\xi$,
compare the third column from left in \figref{softextremes}.

\subsection{Sensitivity with Respect to the Model Resolution}
\label{sec:modelsens}

In this section we analyze the response of the GEV inferences to
variations in the model. In fact, this question is a further
aspect of the smoothness and robustness issues discussed
in \secref{smooth}, which is of great practical importance:
we would not like our estimates to drastically change
if the model is slightly altered.
Different choices are possible, such as introducing an orography
in the bottom layer or changing the lateral boundary conditions. In the
present setting, however, we confine ourselves to compare
simulations of the baroclinic model computed at different
resolutions (\emph{i.e.}, with different spectral discretization
order $JT$, see~\hbox{\eqref{weight1}-\eqref{weight4}}).

\begin{figure}[htb]
  \centering
  \includegraphics[width=0.32\textwidth]{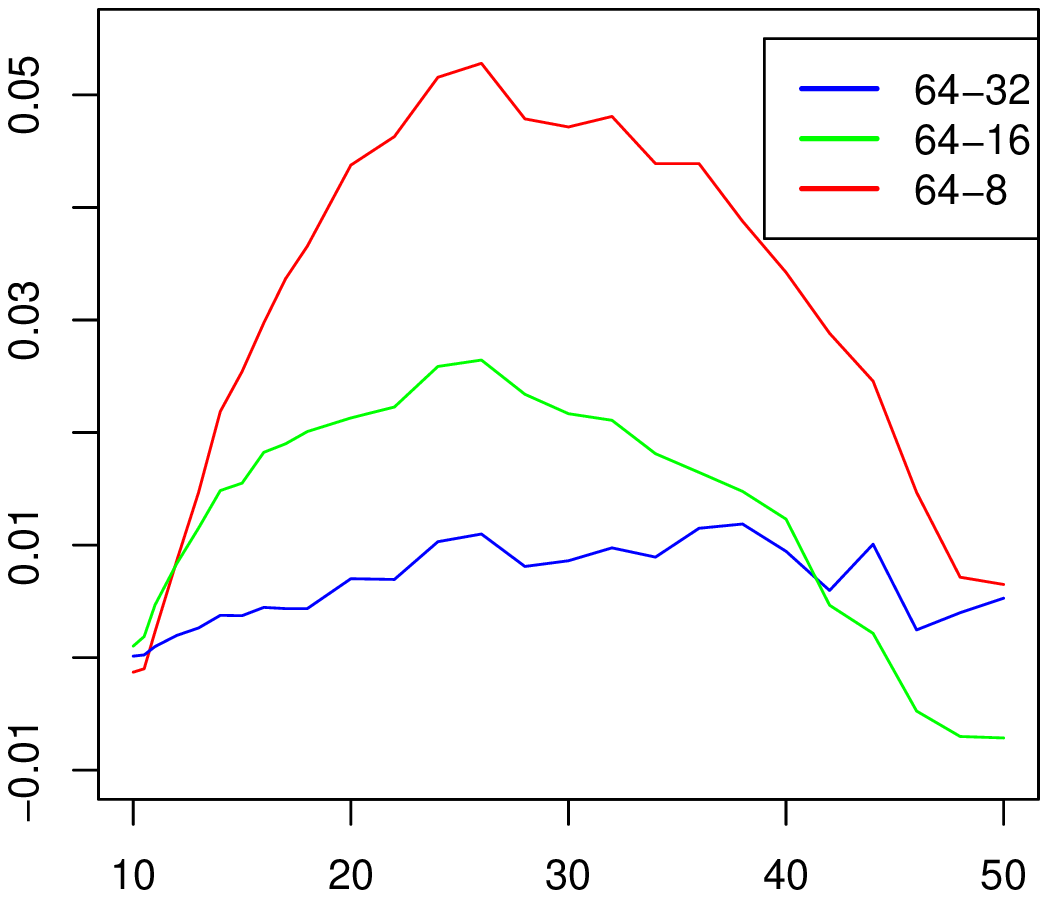}
  \includegraphics[width=0.32\textwidth]{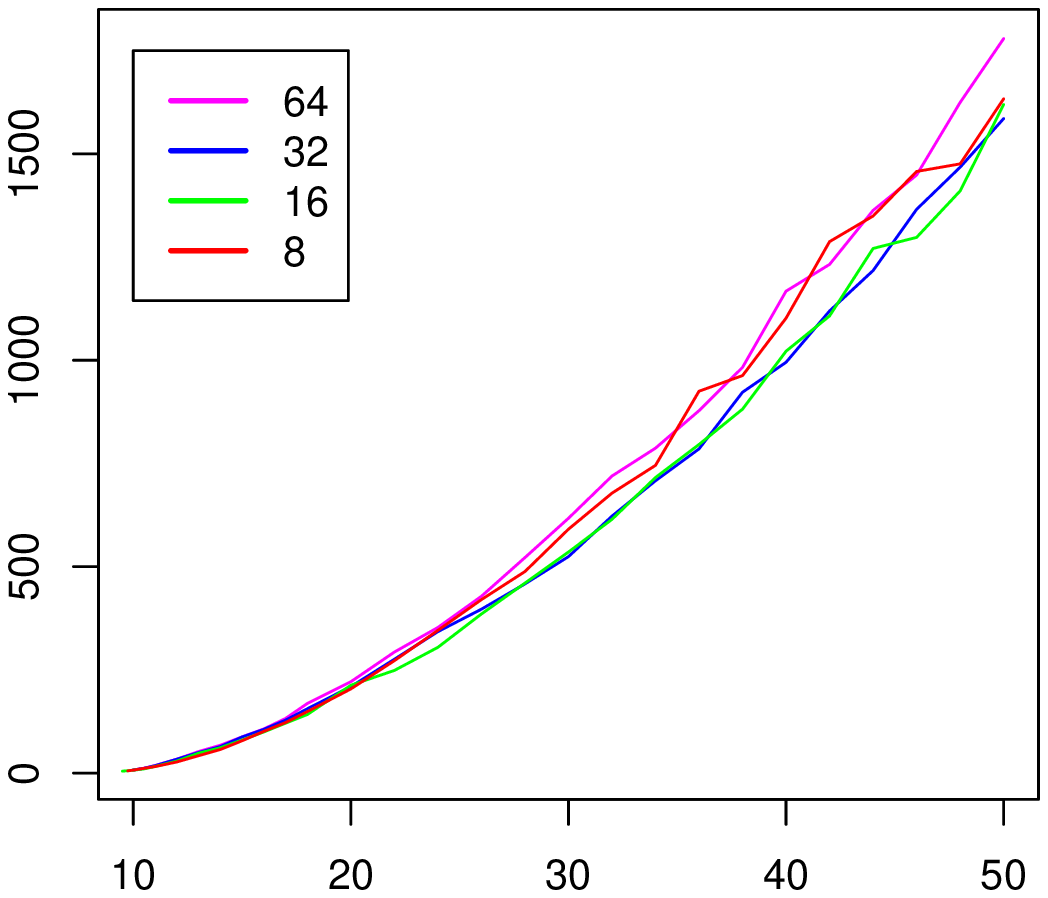}
  \includegraphics[width=0.32\textwidth]{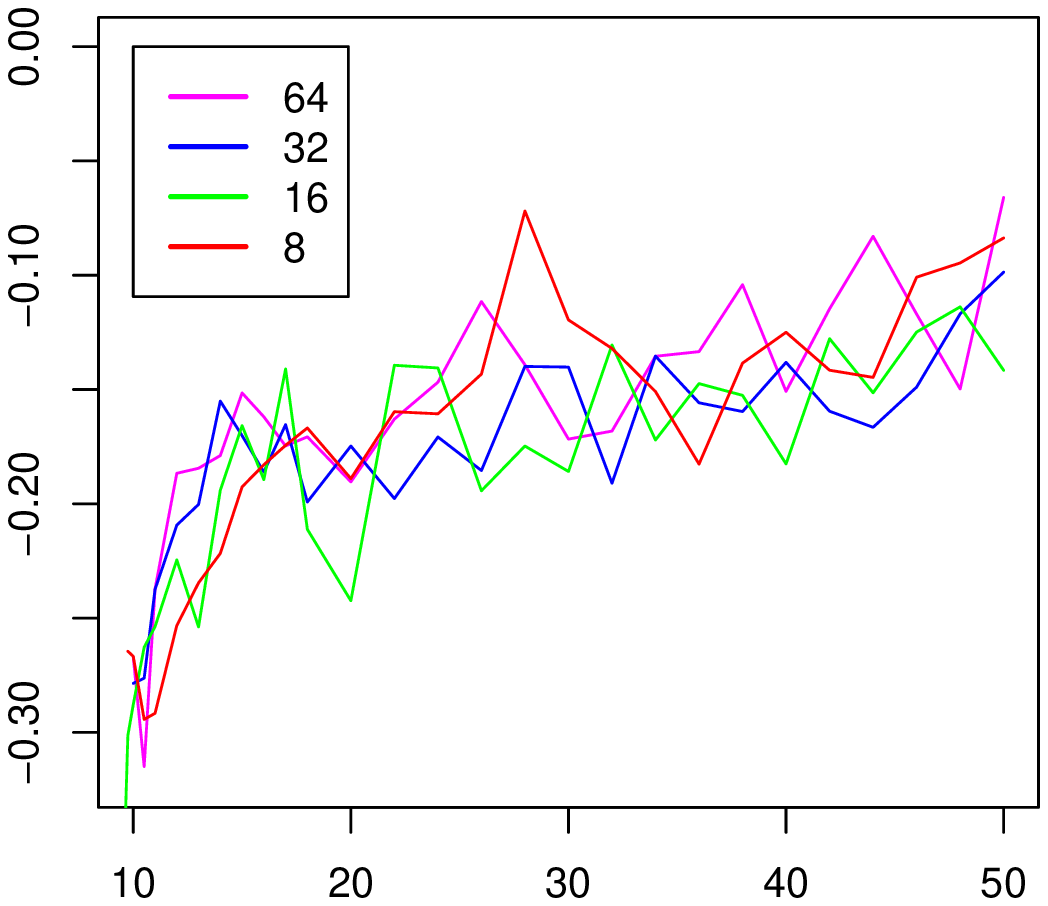}
  \caption{
    Left: relative difference $(\mu_{JT}-\mu_{64})/\mu_{64}$
    of the maximum likelihood estimates of the GEV parameter $\mu$
    (vertical axis) for resolutions $JT=8,16,32$
    (red, green, and blue, respectively)
    with respect to the reference case $JT=64$.
    Middle, right: estimates of $\sigma$ and $\xi$, respectively
    (vertical axis), for the cases $JT=8,16,32,64$
    (red, green, blue, magenta, respectively),
    where sequences of 1000 maxima are used.
    On the horizontal axis, the value of $T_E$ is given for which
    the simulations are performed.
  }
  \label{fig:gevcontrol}
\end{figure}

In particular, time series of the total energy, of length 1000
years, are computed with the baroclinic model using four different
resolutions: $JT=8,16,32,64$ (resolution $JT=32$ is used
throughout the rest of this paper). In each case the GEV
parameters are estimated from sequences of 1000 yearly maxima. The
results are compared with each other in \figref{gevcontrol}. The
relative differences of the estimated values of $\mu$ between the
case $JT=32$ and each of the other three cases (panel~(A)) remain
rather small: they are less than $1.5\%$ for $JT=16$ and $64$ and
grow up to about $4\%$ for $JT=8$. Also the estimates of $\xi$ in
general agree quite well for all the considered resolutions.

More pronounced differences appear in the inferred values of the
scale parameter $\sigma$: for $T_E\ge26$, the estimates obtained
with resolutions $JT=8$ and $64$ are larger than those for $JT=16$
and $32$. The estimates for $\mu$ closely reflect the behavior of
the time-averaged the total energy (computed on the same time
series from which the yearly maxima are extracted). Considering,
to fix ideas, the range $T_E\in[26,36]$, for each fixed $T_E$ both
the inferred values of $\mu$ and the time-averaged total energy
(not shown) \emph{decrease} as $JT$ increases. Conversely, there
is no simple relation between the sample standard deviation
$\sigma_E$ of the total energy time series and the GEV scale
parameter $\sigma$: for the mentioned values of $T_E$, the sample
standard deviation $\sigma_E$ decreases for larger $JT$ (not
shown), whereas this is not so for the scale parameter, see above.

Power law fits of $\mu$ and $\sigma$ as functions of $T_E$ are
performed for 1000 yearly maxima of the total energy, where the
baroclinic model is run with four different resolutions: $JT=8$,
$16$, $32$, $64$. As in \secref{smooth}, the range of $T_E$ is
divided into two intervals for the fits of $\mu$ and into three
for $\sigma$ (in the latter case, no power law is found in the
leftmost interval). Remarkable accuracy and coherence of the laws
for $\mu$ is observed. There is more variability in the power laws
for $\sigma$, although again a striking coherence is observed for
large $T_E$.

Summarizing, we have observed no dramatic model sensitivity for
the GEV estimates. However, it is to be emphasized that a
particularly \textit{stable} observable has been examined here
(the total energy) and only one type of model alteration has been
considered, namely a change in the spectral resolution.
We believe, though, our results are quite ``generic'' for the class
of models considered in this paper.

\newpage
\section{Summary and Conclusions}
\label{sec:conclusions}

In this paper we have performed statistical inference of extreme values
on time series obtained by means of a dynamically minimal two-level
quasi-geostrophic model of the atmosphere at mid-latitudes. The
physical observable used to generate the time series is the total
energy of the system and the statistical model for the extremes is
the Generalized Extreme Value distribution (GEV). Several
physically realistic values of the parameter $T_E$, descriptive of
the forced equator-to-pole temperature gradient and responsible
for setting the average baroclinicity in the atmospheric model,
are examined. In the standard setting, the maxima of the total
energy are computed over one year long data blocks, and 1000
maxima are used as basis for the inference.

A result of the present investigation, having potential relevance
in atmospheric dynamics, is the detection of a piecewise smooth
dependence of the location and scale GEV parameters $(\mu,\sigma)$
on the model parameter $T_E$ controlling average baroclinicity.
Two distinct power-laws, holding in different intervals of $T_E$,
are obtained both for $\mu$ and for $\sigma$ as functions of $T_E$,
where the fit for $\mu$ is quite accurate.
This regularity is put in relation with the results in~\cite{LSV05},
where analogous scaling laws are found for other dynamical indicators,
such as Lyapunov exponents and dimension, and physical observables,
such the time-space average of total energy and zonal wind.
The shape parameter $\xi$ also increases with $T_E$ but is always negative,
as \textit{a priori} required by the boundedness of the total energy of the
system. We conjecture that also the dependence of $\xi$ on $T_E$
becomes smooth when much longer time series are considered.
All these problems wiil be further explored in connected work.

After the assessment of the goodness-of-fit by means of standard
statistical diagnostics, such as return level and
quantile-quantile plots and computation of confidence intervals by
different procedures, we have consistently verified that:
\begin{itemize}
\item
  the selected block length of one year guarantees that the extremes
  are uncorrelated and genuinely extreme; guaranteeing this property
  may result more problematic when dealing with real observations
  because of seasonal modulations, etc.;
\item
  the considered length of the series of maxima (1000 data)
  yields reliable parameter estimates;
\item
  the GEV inferences are not dramatically affected by structural
  changes in the atmospheric model adopted in the present work.
\end{itemize}
The sensitivity of the statistical inference process is first
studied with respect to the selection procedure of the maxima:
we analyze the effects of reducing either the length of maxima sequences
or the length of data blocks over which the maxima are computed.

The first point is checked by repeating the GEV inferences with
sequences of maxima having lengths 300, 100, and 50 years.
The estimates are coherent for 1000, 300, and 100 yearly maxima,
but the confidence intervals of the best estimates, not surprisingly,
widen up as shorter sequences of maxima are used.
Moreover, markedly unreliable estimates are obtained when only 50
yearly maxima are considered:
the estimated long-term return levels are patently wrong,
the uncertainty of the inferred shape parameter $\xi$ is very large,
and the best estimate of $\xi$ is positive (that is, unrealistic)
for a few values of $T_E$.

In order to address the second point, we have taken maxima over data
blocks corresponding to shorter time spans, to explore the effects of
using \emph{soft extremes}~\cite{KtK03}. Specifically, the
sensitivity of the GEV inferences is analyzed with respect to
shortening the length of the data blocks to 3, 1.2, and 0.6
months. The obtained statistics is ``polluted'': a bias is introduced
which is unacceptable for the cases of 1.2 and 0.6 months and still
significant (at least for the GEV parameter $\sigma$) for 3
months. Moreover, the parameter $\xi$ tends to be underestimated.
Taking shorter maxima sequences results in even larger uncertainties,
very large for the case of 50 yearly maxima. Physically
unrealistic values of $\xi$ may also be obtained.

Lastly, issues related to model sensitivity are also explored by
varying the (spectral) resolution of the system. It turns out that
the GEV estimates are in general rather robust under this sort of
perturbation.
Summarizing, to get a good inference \emph{many} maxima are required
and they must be \emph{genuinely extreme}, that is, taken over
sufficiently large data blocks. Failing to fulfill these requirements
may result in affecting the GEV inferences much more critically than
adopting a baroclinic model with lower resolutions.

We conclude by highlighting that the parameterization of physical
observables with respect to an external forcing is indeed a rather
general and difficult problem in the dynamical analysis of the
physical system. Existence of a unique Sinai-Ruelle-Bowen measure
is \emph{required} to rigorously associate a stationary stochastic
process to the dynamical evolution law. However, even if an SRB
measure exists and is unique, typically there is no explicit expression
in terms of the system's equations and parameters~\cite{ER}.

In this respect, the simplicity and the universality of the GEV
model can be exploited to characterize chaotic systems by focusing
on extreme values of suitable time series, rather then examining
the distribution of all states visited by the system in phase space.
Different model variants (both in boundary conditions and in model
structure) and other observables and will be considered
in future research.

\begin{acknowledgments}
  The authors wish to thank Stefano Pittalis for useful conversations.
  This work has been supported by MIUR PRIN Grant
  "Gli estremi meteo-climatici nell'area mediterranea:
  propriet\`a statistiche e dinamiche", Italy, 2003.
\end{acknowledgments}

\newpage

\appendix

\section{Classical Theory of Extreme Value Distributions}
\label{app:theory}

Let $X_1,\ldots,X_n$ be a sequence of independent and identically
distributed random variables (i.i.d.r.v.) where $F_X$ is the
common distribution function (d.f.). The classical theory of
Extreme Values deals with the statistical behavior of the random
variables
\begin{equation}
  \label{eq:maxima}
  M_n=\max\{X_1,\ldots,X_n\},
\end{equation}
which is the \emph{maximum} of the first $n$ variables (an
analogous theory for the minima is developed similarly, since
$\min\{X_1,\ldots,X_n\}=-\max\{X_1,\ldots,X_n\}$). Under the
assumptions of statistical independence and distributional
equality of the $X_i$, we known that
\begin{equation}
  P\{M_n\leq x\}= P\{X_1\leq x,\ldots, X_n \leq x\}
   =P\{X_1 \leq x\}\cdot \ldots \cdot P\{X_n \leq x\}
   = F^n_X.
\end{equation}
However, in most practical applications this property is useless
because, typically $F_X$ is unknown. Moreover, the limit of
$F^n_X$ is degenerate, since it is concentrated on the point
$x_{+}=sup\{x\, : \, F(x)<1\}$:
\begin{equation}
  \lim_{n\rightarrow +\infty}= F^n_X(x)=
  \begin{cases}
    0 &  x<x_{+},\\
    1 &  x\geq x_{+}.
  \end{cases}
\end{equation}
This difficulty is avoided by assuming the existence of two
sequences of constants, $\{\sigma_n>0\}$ and $\{\mu_n\}$, such
that $M_n^*=\frac{M_n-\mu_n}{\sigma_n}$, rather than $M_n$, has a
nondegenerate limit distribution $G(x)$:
\begin{equation}
  \label{eq:hypG}
  P\{M_n^*<x\}\stackrel{w}{\longrightarrow}G(x)
\end{equation}
for each continuity point $x$ of $G$.
\begin{thr}[\textbf{Extremal Types Theorem}]
  \label{thm:gnedenko}
  If there exist sequences of constant $\{\sigma_n>0\}$ and $\{\mu_n\}$
  such that the limit in~\eqref{hypG} exists, then the d.f. $G(x)$
  belongs to one of the following three parametric forms, called
  \emph{Extreme Value Distributions}:
  \begin{align}
    \textrm{Type I (Gumbel):}\quad G_1(x)= & \exp\left\{ -\exp \left[ -\left(
          \frac{x-\mu}{\sigma} \right) \right] \right\},
    \quad -\infty<\,x\,<+\infty
    \label{eq:gumbel}
    \\
    \textrm{Type II (Fr\'echet):}\quad G_2(x)= &
    \begin{cases}
      0 &   x \leq \mu, \\
      \exp\left \{-\left(\frac{x-\mu}{\sigma} \right)^{-\xi} \right \} &
      x>\mu, \quad \xi>0
    \end{cases}
    \label{eq:frechet}
    \\
    \textrm{Type III (Weibull):}\quad G_3(x)= &
    \begin{cases}
        \exp\left \{-\left[-\left(\frac{x-\mu}{\sigma}
            \right)\right]^{ \xi} \right \}  &   x < \mu, \quad \xi<0 \\
        1& x\geq \mu
    \end{cases}
    \label{eq:weibull}
  \end{align}
  with scale parameter $\sigma>0$, location parameter $\mu\in\mathbb{R}$
  and, for the types II and III, the shape parameter $\xi\neq 0$
  (for type I it is assumed $\xi=0$).
\end{thr}
\noindent This result was first proved by Fisher and
Tippett~\cite{FT28} and then it was extended by
Gnedenko~\cite{Gne43}. The strength of this theorem is the fact
that it is a \emph{universal} property, since it holds regardless
of the parent distribution $F_X$. Notice that:
  \begin{itemize}
  \item \thrref{gnedenko} does not guarantee the convergence
    in distribution for the variable $M_n^*$.
    In fact it \emph{assumes} it (compare~\eqref{hypG}).
    There are distributions for which the convergence requested
    in~\eqref{hypG} does not hold, see~\cite{LLR83}.
  \item
    The value of $x_{+}$ is finite only for the Weibull distribution,
    whereas the Fr\'echet and Gumbel densities decay polynomially and
    exponentially as $x\to+\infty$, respectively.
  \item
    The \emph{domain of attraction} $D(G_i)$, where $G_i$ is one
    of~\eqref{gumbel}-\eqref{weibull}, is defined as the
    set of all d.f. $F(x)$ such that one has convergence to $G_i(x)$
    in~\eqref{hypG}.
    Various criteria give necessary and sufficient conditions to determine
    what is the domain of attraction of each of the extremal
    distributions~\eqref{gumbel}-\eqref{weibull}.
    However, the limit type for a given $F(x)$ only depends
    on the \emph{upper tail} of $F(x)$. See~\cite{LLR83}.
  \item
    The hypotheses of \thrref{gnedenko} can be relaxed
    to the case of stationary stochastic processes with
    \emph{weak long-range} dependence
    (at extreme levels)~\cite[Chap. 5]{Col01}, which is of
    particular importance in our case.
  \end{itemize}
The Gumbel, Fr\'echet and Weibull families are unified into the
single GEV family of distribution functions, given in~\eqref{GEV}.
So in the sequel we denote by $G(x)$ the family in~\eqref{GEV}.
For positive values of the shape parameter $\xi$, the Fr\'echet
family is obtained from~\eqref{GEV} and, similarly, for negative
values we have Weibull. The Gumbel distribution is the limit for
$\xi \rightarrow 0$ of $G(x)$:
\begin{equation}
  \lim_{\xi \to 0}\,G(x)=\exp \left \{ - \exp\left[ -\left(
        \frac{x-\mu}{\sigma} \right) \right]\right\}.
\end{equation}
This highlights another strength of the GEV model in concrete
applications: the limit type is \emph{inferred from the data} by
estimating the parameter $\xi$. This removes the necessity of an
initial and arbitrary choice of the limit type when using
models~\eqref{gumbel}-\eqref{weibull}.

\newpage
\section{A Model for the Mid-Latitudes Atmospheric Circulation}
\label{app:model}

As mentioned in the introduction, the \emph{stochastic generator}
of the energy time series used in this paper is a model for the
baroclinic jet at mid-latitudes. The system is relaxed towards a
prescribed north-south temperature profile, where the gradient is
controlled by a parameter $T_E$. In fact, the parameter $T_E$
controls average baroclinicity of the system and is used to study
the relation with extreme values of the energy time series. Many
dynamical properties of the model depending on $T_E$ have been
analyzed before the analysis of extreme values presented in this
paper, providing a sort of \textit{road map}.
See~\cite{SM88,MTS90,LSV05}, to which we also refer for a detailed
derivation of the model and for discussion on the physics
involved. In this section, we confine ourselves to a brief sketch.

Starting point for the construction of the model is the two-level
quasi-geostrophic equation:
\begin{align}
  \label{eq:tau}
  \frac{\partial}{\partial t}\Delta_H\tau
  -\frac{2}{H_2^2}\frac{\partial}{\partial t}\tau
  +J\left(\tau,\Delta_H\phi+\beta y + \frac{2}{H_2^2}\phi\right)
  +J\left(\phi,\Delta_H\tau\right)=\nonumber\\
  \frac{2\nu_E}{H_2^2}\Delta_H\left(\phi-\tau\right)
  -\frac{2\kappa}{H_2^2}\Delta_H\tau
  +\frac{2\nu_N}{H_2^2}\left(\tau-\tau^\star\right),\\
  \label{eq:phi}
  \frac{\partial}{\partial t}\Delta_H\phi
  +J\left(\phi,\Delta_H\phi+\beta y\right)
  +J\left(\tau,\Delta_H\tau\right)=
  -\frac{2\nu_E}{H_2^2}\Delta_H\left(\phi-\tau\right).
\end{align}
Here $\tau$ and $\phi$ are the baroclinic and barotropic
components, respectively, of the streamfunction $\psi_1$ and
$\psi_3$ at the two levels:
\begin{equation}
  \tau=\frac{1}{2}\left(\psi_1-\psi_3\right),
  \qquad
  \phi=\frac12\left(\psi_1+\psi_3\right),
\end{equation}
$\Delta_H$ is the horizontal Laplacian, $1/H_2^2$ is the Froude
number, $\beta$ is the gradient of the Coriolis parameter,
$\nu_E$, $\kappa$ and $\nu_N$ parameterize the Ekman pumping at
the lower surface, the heat diffusion, and the Newtonian cooling,
respectively.

The system is driven for the baroclinic component by the term in
$(\tau-\tau^*)$ in~\eqref{tau}, which forces a relaxation to the
radiative equilibrium $\tau^*$ with a characteristic time scale of
$1/\nu_N$. We take
\begin{equation}
  \label{deftau}
  \tau^\star=\frac{R}{f_0}\frac{T_E}{4}\cos\left(\frac{\pi
      y}{L_y}\right),
\end{equation}
so that $T_E$ is the forced temperature difference between the low
and the high latitude border of the domain. In this sense, the
parameter $T_E$ is responsible for average baroclinicity of the
system and is the control parameter we vary to test changes in the
extreme value statistics.

The fields $\phi$ and $\tau$ are expanded in Fourier series in the
longitudinal direction $x$. Moreover, in order to avoid wave-wave
nonlinear interactions, only the terms of order $n=1$ and $n=6$
are retained (see~\cite{LSV05} for details). This yields
\begin{align}
  \label{phifourier2}
  &\phi\left(x,y,t\right)=
  -\int_0^y{U\left(z,t\right)\rm{d}z}
  +A\exp{\left(\textrm{i} \chi x\right)}+\textrm{c.c.}\\
  \label{taufourier2}
  &\tau\left(x,y,t\right)=
  -\int_0^y{m\left(z,t\right)\rm{d}z}
  +B\exp{\left(\textrm{i} \chi x\right)}+\textrm{c.c.}.
\end{align}
By substitution into~\eqref{tau}-\eqref{phi}, one obtains
\begin{align}
  \label{eq:dotA}
  &\begin{aligned}
    \dot{A}_{yy}-\chi^2\dot{A}
&+\left(\textrm{i} \chi U+\frac{2\nu_E}{H_2^2}\right)A_{yy}
    -\left(\textrm{i}  \chi^3U+\textrm{i}  \chi U_{yy}+\frac{2\nu_E}{H_2^2}\chi^2-\textrm{i}  \chi\beta\right)A
    \\
    &+\left(\textrm{i}  \chi m-\frac{2\nu_E}{H_2^2}\right)B_{yy}
    -\left(\textrm{i}  \chi^3m+\textrm{i}  \chi m_{yy}-\frac{2\nu_E}{H_2^2}\chi^2\right)B
    =0,
  \end{aligned}\\
  \label{eq:dotB}
  &\begin{aligned}
    \dot{B}_{yy}-\chi^2\dot{B}-&\frac{2}{H_2^2}\dot{B}
    +\left(\textrm{i}   \chi U+\frac{2\nu_E}{H_2^2}+\frac{2\kappa}{H_2^2}\right)B_{yy}\\
   &-\left(\textrm{i}  \chi^3U+\textrm{i}  \chi U_{yy}+\frac{2\nu_E}{H_2^2}\chi^2-\textrm{i}  \chi\beta
    +\frac{2\kappa}{H_2^2}\chi^2+\frac{2\nu_N}{H_2^2}+\frac{2}{H_2^2}\textrm{i}  \chi U\right)B\\
    &+\left(\textrm{i}  \chi m-\frac{2\nu_E}{H_2^2}\right)A_{yy}
    -\left(\textrm{i}  \chi^3m+\textrm{i}  \chi m_{yy}-\frac{2\nu_E}{H_2^2}\chi^2
     -\frac{2}{H_2^2}\textrm{i}  \chi m\right)A
    =0,
  \end{aligned}\\
  \label{eq:dotU}
  &\dot{U}+\frac{2\nu_E}{H_2^2}(U-m)+2\chi\imag(AA^*_{yy}+BB^*_{yy})=0,\\
  \label{eq:dotm}
  &\begin{aligned}
    \dot{m}_{yy} - \frac{2}{H_2^2}\dot{m} &+\frac{2\kappa}{H_2^2} m_{yy}
    -\frac{2\nu_E}{H_2^2}(U-m)_{yy}
    -\frac{2\nu_N}{H_2^2}(m-m^*)\\
    &+\frac{4}{H_2^2}\chi \imag(A^*B)_{yy}
    +2\chi\imag(AB^*_y+BA^*_y)_{yyy}
    =0,
  \end{aligned}
\end{align}
where the dot indicates time differentiation and $A^*$ denotes the
complex conjugate of $A$. This is a set of $6$ equations for the
real fields $A^1$, $A^2$, $B^1$, $B^2$, $U$, $m$, where $A^1$ and
$A^2$ are the real and imaginary parts of $A$ and similarly for
$B$. Rigid walls are taken as boundaries at $y=0,L_y$, so that all
fields have vanishing boundary conditions.

A system of ordinary differential equations is obtained
from~\eqref{dotA}-\eqref{dotm} by means of a pseudospectral
(collocation) projection, involving a Fourier half-sine expansion
of the fields of the form
\begin{align}
  \label{eq:weight1}
  &A^i=\sum_{j=1}^{JT}A^i_j\sin\left(\frac{\pi j y}{L_y}\right), \hspace{5pt}i=1,2,\\
  \label{eq:weight2}
  &B^i=\sum_{j=1}^{JT}B^i_j\sin\left(\frac{\pi j y}{L_y}\right), \hspace{5pt}i=1,2,\\
  \label{eq:weight3}
  &U=\sum_{j=1}^{JT}U_j\sin\left(\frac{\pi j y}{L_y}\right),\\
  \label{eq:weight4}
  &m=\sum_{j=1}^{JT}m_j\sin\left(\frac{\pi j y}{L_y}\right).
\end{align}
The resulting system is the generator of the time series used in
this paper for extreme value analysis. In particular, as
\emph{observable} (that is, as function of the state space
yielding the time series) we choose the total energy $E(t)$ of the
system, obtained by integration in the $(x,y)$-domain of the
energy density:
\begin{equation}
  \label{eq:energydensity}
  e\left(x,y,t\right)=\frac{\delta
    p}{g}\left[\frac{1}{2}\left(\vec{\nabla}\psi_1\right)^2+
    \frac{1}{2}\left(\vec{\nabla}\psi_3\right)^2+
    \frac{1}{2H_2^2}\left(\psi_1-\psi_3\right)^2\right].
\end{equation}
Here the factor $\delta p/g$ is the mass per unit surface in each
level, the first two terms inside the brackets describe the
kinetic energy and the last term describes the potential energy.
We emphasize that in the expression~\eqref{energydensity} the
potential energy term is half of what reported in~\cite{Ped87},
which contains a trivial algebraic mistake .

It turns out that the order $JT=32$ in the
expansion~\hbox{~\eqref{weight1}-\eqref{weight4}} is sufficiently
high to have an earth-like chaotic regime characterized by
intermediate dimensionality in suitable ranges of the parameter
$T_E$. By chaotic, we mean that the dynamics takes place on a
strange attractor with internally generated noise. By earth-like
we mean that the time-dependent Fourier coefficients
in~\eqref{weight1}-\eqref{weight4}, as well as the total energy
and mean zonal wind, have unimodal probability densities. The
mentioned chaotic range is $T_E>T_E^{crit}$, where
$T_E^{crit}=8.75$ approximately. For lower values of $T_E$, the
Hadley equilibrium (stationary solution) is stable and is
therefore the unique attractor. Again see~\cite{SM88,MTS90,LSV05}
for a complete discussion. Throughout this work, we consider
$JT=8$ , $16$, $32$, and $64$ and the considered parameter range
is $10\le T_E\le 50$ with integer steps of 2.

\clearpage

\begin{table}[!h]
  \centering
  \begin{tabular}{|c||c|c|c|}
    \hline
    n& $\gamma_1$& $T_E^{b}$& $\gamma_2$\\
    \hline \hline
    1000 & $1.7310 \pm 0.0007$ & 18& $1.6019 \pm 0.0017$\\
    300  & $1.7311 \pm 0.0013$ & 18& $1.6017 \pm 0.0030$\\
    100  & $1.7219 \pm 0.0018$ & 17& $1.5982 \pm 0.0018$\\
    \hline
  \end{tabular}
  \caption{
    Power law fits of the location parameter $\mu$ as a function of $T_E$
    of the form $\mu \propto T_E^{\gamma}$,
    performed in two adjacent intervals of $T_E$.
    The number of used annual extremes is $n$ and $T_E^{b}$
    is the value of $T_E$ separating the two intervals.
    Compare with figure \figref{powermu}.}
  \label{tab:powermu}
\end{table}

\begin{table}[!h]
  \centering
  \begin{tabular}{|c||c|c|c|c|}
    \hline $n$ & $T_E^{b1}$ & $\gamma_1$ & $T_E^{b2}$ & $\gamma_2$\\
    \hline \hline
    1000 & 15 & $3.011 \pm 0.076$ & 22 & $2.140 \pm 0.025$\\
    300  & 14 & $3.236 \pm 0.115$ & 22 & $2.114 \pm 0.047$\\
    100  & 14 & $2.944 \pm 0.157$ & 24 & $2.040 \pm 0.093$\\
    \hline
  \end{tabular}
  \caption{
    Same as \tabref{powermu} for the scale parameter $\sigma$.
    Here the fits $\sigma=T_e^{\gamma_1}$ and $\sigma=T_e^{\gamma_2}$
    hold for $T_E$ such that $T_E^{b1}\le T_E\le T_E^{b2}$
    and $T_E^{b2}\le T_E\le 50$, respectively.
    No power law fit is found for $T_E<T_E^{b1}$.
    Compare with \figref{powersigma}.}
  \label{tab:powersigma}
\end{table}

\begin{table}[htb]
  \centering
  \begin{tabular}{|c||c|c|c|}
    \hline
    JT& $\gamma_1$& $T_E^{b}$& $\gamma_2$\\
    \hline \hline
    64 &  $1.7346 \pm 0.0008$ & 15& $1.6027 \pm 0.0005$\\
    32  & $1.7310 \pm 0.0007$ & 18& $1.6019 \pm 0.0005$\\
    16  & $1.7027 \pm 0.0007$ & 18& $1.5982 \pm 0.0007$\\
    8   & $1.6794 \pm 0.0006$ & 22& $1.5977 \pm 0.0011$\\
    \hline
  \end{tabular}
  \caption{Power law fits of the location parameter $\mu$ as a function
    of $T_E$ of the form $\mu\propto T_E^{\gamma}$.
    $JT$ indicates the spectral resolution (number of Fourier modes)
    of the baroclinic model and $T_E^{b}$ is the value of $T_E$
    dividing the two considered intervals, see text for details.
  }\label{tab:muJT}
\end{table}

\begin{table}[htb]
  \centering
  \begin{tabular}{|c||c|c|c|c|}
    \hline $JT$ & $T_E^{b1}$ & $\gamma_1$ & $T_E^{b2}$ & $\gamma_2$\\
    \hline \hline
    64  & 18 & $2.514 \pm 0.046$ & 32 & $2.067 \pm 0.055$\\
    32  & 15 & $3.011 \pm 0.076$ & 22 & $2.140 \pm 0.025$\\
    16  & 15 & $2.821 \pm 0.045$ & 26 & $2.150 \pm 0.033$\\
    8   & 17 & $2.675 \pm 0.065$ & 26 & $2.149 \pm 0.033$\\\hline
  \end{tabular}
  \caption{
    Same as \tabref{muJT} for the scale parameter
    $\sigma\propto T_E^{\gamma}$.
    The interval $[T_E^{b1},T_E^{b2}]$ is the range of validity of the
    the first power-law, having exponent $\gamma_1$.
    The point dividing the two considered intervals is $T_E^{b2}$.
    No power law is detected for $T_E<T_E^{b1}$.
  }\label{tab:sigmaJT}
\end{table}

\clearpage

\newpage

\begin{thebibliography}{99}

\bibitem{Buz86}
  A. Buzzi, A. Speranza:
  A theory of deep cyclogenesis in the lee of the Alps. Part II:
  Effects of finite topographic slope and height,
  \textit{J. Atmos. Sci.}, \textbf{43} (1986), 2826--2837.

\bibitem{Buz94}
  A. Buzzi, M. Fantini, P. Malguzzi, and F. Nerozzi:
  Validation of a limited area model in cases of Mediterranean
  cyclogenesis: Surface fields and precipitation scores,
  \textit{Meteor. Atmos. Phys.}, \textbf{53} (1994), 137--153.

\bibitem{Cas88}
  E.~Castillo:
  Extreme Value Theory in Engineering,
  \textit{Academic Press}, 1988.

\bibitem{Char47}
  J.G. Charney:
  The Dynamics of Long Waves in a Baroclinic Westerly Current,
  \textit{J. Atmos. Sci.} \textbf{4} (1947), 136--162.

\bibitem{Col01}
  S.~Coles:
  \textit{An Introduction to Statistical Modelling of Extremes Values},
  Springer Series in Statistics,
  Springer-Verlag London, 2001.

\bibitem{Eady49}
  E.T. Eady:
  Long waves and cyclone waves,
  \textit{Tellus} \textbf{1} (1949), 33--52.

\bibitem{ER}
  J.-P.~Eckmann, D.~Ruelle:
  Ergodic theory of chaos and strange attractors,
  \textit{Rev. Mod. Phys.} \textbf{57} (1985), 617--655.

\bibitem{EKM97}
  P.~Embrechts, C.~Kl\"uppelberg, T.~Mikosch:
  \textit{Modelling Extremal Events for Insurance and Finance},
  1st ed., Stochastic Modelling and Applied Probability \textbf{33},
  Springer, 1997.

\bibitem{FT28}
  R.A.~Fisher, L.H.C.~Tippett:
  Limiting Forms of the Frequency Distribution of the Largest
  or Smallest Number of a Sample,
  \textit{Proc. Cambridge Phil. Soc.} \textbf{24} (1928), 108--190

\bibitem{Gal78}
  J.~Galambos:
  \textit{The Asymptotic Theory of Extreme Order Statistics},
  Wiley, New York, 1978.

\bibitem{Gne43}
  B.V.~Gnedenko:
  Sur la distribution limite du terme maximum d'une s\'erie
  al\'eatorie,
  \textit{Ann. Math} \textbf{44} (1943), 423--453.

\bibitem{HH80}
  I.M.~Held, A.Y.~Hou:
  Nonlinear Axially Symmetric Circulations
  in a Nearly Inviscid Atmosphere,
  \textit{J. Atmos. Sci.} \textbf{37} (1980), 515--533.

\bibitem{Hol92}
  J.R. Holton:
  \textit{An Introduction to Dynamic Meteorology}.
  Academic Press, San Diego, 1992.

\bibitem{IG96}
  R.~Ihaka, R.~Gentleman:
  R: A language for data analysis and graphics,
  \textit{Journal of Computational and Graphical Statistics},
  \textbf{5(3)} (1996), 299--314.

\bibitem{IPCC01}
  Intergovernmental panel on Climate Change,
  Report:
  \textit{Climate Change 2001: Impacts, Adaptation and Vulnerability},
  \texttt{www.ipcc.ch}.

\bibitem{IPCC02}
  Intergovernmental panel on Climate Change,
  Workshop Report
  \textit{IPCC Workshop on Changes in Extreme Weather and Climate Events}
  Beijing, China, 11--13 June, 2002,
  \texttt{www.ipcc.ch}.

\bibitem{IY93}
  T.~Iwashima, R.~Yamamoto:
  A statistical analysis of the extreme events:
  Long-term trend of heavy daily precipitation,
  \textit{Journal of the Meteorological Society of Japan}
  \textbf{71} (1993), 637--640.

\bibitem{Jef24}
  H.~Jeffreys:
  On the Formation of Waves by Wind,
  \textit{Proc. Roy. Soc. Lond.} \textbf{107} (1924), 189--206.

\bibitem{Jef25}
  H.~Jeffreys:
  On the Formation of Waves by Wind,
  \textit{Proc. Roy. Soc. Lond.}, \textbf{110A} (1925), 341--347.

\bibitem{Jen55}
  A.F.~Jenkinson:
  The frequency distribution of the annual maximum
  (or minimum) values of meteorological elements,
  \textit{Quart. J. Roy. Meteor. Soc.}
  \textbf{87} (1955), 158--171.

\bibitem{KK97}
  T.R.~Karl, R.W.~Knight:
  The Chicago heat wave: how likely is a recurrence?
  \textit{Bull. Amer. Meteor. Soc.}
  \textbf{78}, (1997) 1107--1119.

\bibitem{KK98}
  T.R.~Karl, R.W.~Knight:
  Secular trend of precipitation amount,
  frequency, and intensity in the United States.
  \textit{Bull. Amer. Meteor. Soc.}
  \textbf{79}, (1998) 231--242.

\bibitem{KKEQ96}
  T.R.~Karl, R.W.~Knight, D.R.~Easterling, R.G.~Quayle:
  Indices of climate change for the United States,
  \textit{Bull. Amer. Meteor. Soc.}
  \textbf{77}, (1996) 279--292.

\bibitem{KB92}
  R.W.~Katz, B.G.~Brown:
  Extreme events in a changing climate:
  Variability is more important than averages,
  \textit{Climatic Change} \textbf{21}, (1992) 289--302.

\bibitem{KPN02}
  R.W.~Katz, M.B.~Parlange, P.~Naveau:
  Statistics of extremes in hydrology,
  \textit{Adv. Water Resour.}, \textbf{25} (2002), 1287--1304.

\bibitem{KtK03}
  A.M.G.~Klein~Tank, G.P.~K\"onnen:
  Trends in Indices of Daily Temperature and Precipitation
  Extremes in Europe,
  \textit{J. Climate} \textbf{16} (2003), 1946--1999.

\bibitem{KPC99}
  K.E.~Kunkel, R.A.~Pielke~Jr., S.A.~Changnon:
  Temporal fluctuations in weather and climate extremes that cause
  economic and human health impacts: A review,
  \textit{Bull. Amer. Meteor. Soc.} \textbf{80} (1999), 1077-­1098.

\bibitem{KAE99}
  K.E.~Kunkel, K.~Andsager, D.R.~Easterling:
  Long-term trends in extreme precipitation events
  over the conterminous United States,
  \textit{J. Climate} \textbf{12} (1999), 2515-­2527.

\bibitem{LLR83}
  G.~Lindgren, M.~R.~Leadbetter, H.~Rootz\'en:
  \textit{Extremes and Related Properties of Random Sequences and
    Processes},
  Springer-Verlag, New York (1983).

\bibitem{LDE02}
  P.~Lionello, F.~Dalan, E.~Elvini:
  Cyclones in the Mediterranean Region:
  the present and the doubled CO2 climate scenarios,
  \textit{Clim. Res.} \textbf{22} (2002), 147--159.

\bibitem{Lor55}
  E.N.~Lorenz:
  Available potential energy and the maintenance of the
  general circulation,
  \textit{Tellus} \textbf{7} (1955), 157--167.

\bibitem{Lor60}
  E.N.~Lorenz:
  Generation of available potential energy and the
  intensity of the general circulation,
  in \textit{Dynamics of Climate},
  R.L. Pfeffer ed., Pergamon, Tarrytown (1960), 86--92.

\bibitem{Lor67}
  E.N.~Lorenz:
  \textit{The Nature and Theory of the General
    Circulation of the Atmosphere},
  World Meteorol. Organ., Geneva, 1967.

\bibitem{Luc02}
  V.~Lucarini:
  Towards a definition of climate science,
  \textit{Int. J. Environment and Pollution}
  \textbf{18} (2002), 409--414.

\bibitem{LSV05}
  V.~Lucarini, A.~Speranza, R.~Vitolo:
  Geometrical Properties of the Attractor of a Model
  of Intermediate Complexity of the Mid-Latitudes Atmospheric Circulation,
  preprint \verb|ArXiv, DOI:physics/0511208| (2005).

\bibitem{MTS90}
  P.~Malguzzi, A.~Trevisan, A.~Speranza:
  Statistic and Predictability for an intermediate dimensionality
  model of the baroclinc jet,
  \textit{Annales Geophysicae} \textbf{8} (1990), 29--36.

\bibitem{Mar03}
  M.~Margules:
  Die energie der St\"{u}rme,
  \textit{Jahrb. Zentralanst. Meteor. Wien} \textbf{40} (1903), 1--26.

\bibitem{MWMRH99}
  S.J.~Mason, P.R.~Waylen, G.M.~Mimmack, B.~Rajaratnam, J.M.~Harrison:
  Changes in extreme rainfall events in South Africa,
  \textit{Climatic Change} \textbf{41} (1999), 249--257.

\bibitem{Nor94}
  W.D.~Nordhaus:
  \textit{Managing the Global Commons. The Economics of Climate Change},
  MIT Press, Cambridge (MA), 1994.

\bibitem{PRT05}
  O.~Perrin, H.~Rootzen, R.~Taessler:
  A discussion of statistical methods for estimation
  of extreme wind speeds,
  \textit{Theoretical and Applied Climatology}, to appear (2005).

\bibitem{Ped87}
  J.~Pedlosky:
  \textit{Geophysical Fluid Dynamics}, 2nd ed.,
  Springer-Verlag, New York, 1987.

\bibitem{Peix}
  J.P.~Peixoto, A.H.~Oort:
  \textit{Physics of Climate}, Am. Inst. of Phys., College Park, 1992.

\bibitem{PSNSHLTPL99}
  N.~Plummer, M.~James~Salinger, N.~Nicholls, R.~Suppiah, K.J.~Hennessy,
  R.M.~Leighton, B.~Trewin, C.M.~Page, J.M.~Lough:
  Changes in climate extremes over the Australian region
  and New Zealand during the twentieth century,
  \textit{Climatic Change} \textbf{42} (1999), 183­-202.

\bibitem{Phi54}
  N.A.~Phillips:
  Energy transformations and meridional circulations
  associated with simple baroclinic waves in a two-level,
  quasi-geostrophic model,
  \textit{Tellus} \textbf{6} (1954), 273--286.

\bibitem{Spe85}
  A. Speranza, A. Buzzi, A. Trevisan, P. Malguzzi:
  A theory of deep cyclogenesis in the lee of the Alps. Part I.
  Modification of baroclinic instability by localized topography,
  \textit{J. Atmos. Sci.}, \textbf{42} (1985), 1521--1535.

\bibitem{SM88}
  A.~Speranza, P.~Malguzzi:
  The statistical properties of a zonal jet in a baroclinic atmosphere:
  a semilinear approach. Part I: two-layer model atmosphere,
  \textit{J. Atmos. Sci.} \textbf{48} (1988), 3046--3061.


\bibitem{SH98}
  R.~Suppiah, K.J.~Hennessy:
  Trends in total rainfall, heavy rain events,
  and number of dry days in Australia, 1910-1990,
  \textit{Intern. J. of Climatology} \textbf{18(10)} (1998), 1141--1164.

\bibitem{ZZL03}
  X.~Zhang, F.W.~Zwiers, G.~Li:
  Monte Carlo Experiment on the Detection of Trends in Extreme
  Values,
  \textit{J. Climate} \textbf{17}, (2003) 1945--1952.

\bibitem{ZK98}
  F.W.~Zwiers, V.V.~Kharin:
  Changes in the Extremes of the Climate Simulated by CCC GCM2
  under CO$_2$ Doubling,
  \textit{J. Climate} \textbf{11}, (1998) 2200--2222.

\bibitem{ZK00}
  F.W.~Zwiers, V.V.~Kharin:
  Changes in the Extremes in an Ensemble of Transient Climate Simulations
  with a Coupled Atmosphere­-Ocean GCM
  \textit{J. Climate} \textbf{13}, (2000) 3760--3788.

\end{thebibliography}
\end{document}